\documentclass[aps,prl,letterpaper,preprint,amsmath,amssymb]{revtex4}
\usepackage{graphicx}
\usepackage{dcolumn}
\usepackage{bm}
\begin{document}

\newcommand{\rf}[1]{(\ref{#1})}

\newcommand{\bfomega}{ \mbox{\boldmath{$\omega$}}}

\title{Self-similarity of complex networks}

\author{Chaoming Song$^1$, Shlomo Havlin$^2$, and Hern\'an A. Makse$^1$}

\affiliation{
$^1$ Levich Institute and Physics Department,
City College of
New York,
New York, NY 10031, US\\
$^2$ Minerva Center and Department of Physics,
Bar-Ilan University, Ramat Gan 52900, Israel}

\date{Nature 433, 392-395 (2005)}

\begin{abstract}
Complex networks have been studied extensively due
to their relevance to many real
systems
as diverse as the
World-Wide-Web (WWW), the Internet, energy landscapes,
biological and
social networks
\cite{ab-review,mendes,vespignani,newman,amaral}.
A large number of real networks are called
``scale-free'' because
they show a power-law distribution of
the number of links per node
\cite{ab-review,barabasi1999,faloutsos}.
However, it is widely believed that complex networks are not
{\it length-scale} invariant or self-similar. This conclusion originates
from the ``small-world'' property of these networks,
which implies
that the number of nodes increases exponentially with the
``diameter'' of the network \cite{erdos,bollobas,milgram,watts},
rather than the power-law relation
expected for a self-similar
structure. Nevertheless, here we
present a novel approach to the analysis of such networks,
revealing that their structure is indeed self-similar. This result
is achieved by the application of a renormalization procedure
which coarse-grains the system into boxes
containing nodes within a given "size".
Concurrently, we identify a power-law relation between the number
of boxes needed to cover the network and the size of the box
defining
a finite self-similar exponent.
These
fundamental properties,
which are shown for the WWW, social, cellular and  protein-protein
interaction networks, help to understand the emergence of
the scale-free property
in complex networks. They suggest a common
self-organization dynamics of diverse networks at different scales
into a critical
state and in turn bring together previously unrelated fields:
the statistical physics of complex networks with renormalization
group,
fractals and critical phenomena.

\end{abstract}

\maketitle

\newpage


Two fundamental properties of real complex networks have attracted
much attention recently: the small-world and the scale-free
properties.
Many naturally occurring networks are small world
since one can reach a given node from another one,
following the path with the smallest
number of links between the nodes, in a very small
number of steps. This
corresponds to the so-called ``six degrees of separation'' in social
networks \cite{milgram}. It is mathematically expressed by the slow
(logarithmic) increase of the average diameter of the network,
$\bar{\ell}$, with the total number of nodes $N$, $\bar{\ell} \sim \ln
N$, where $\ell$ is the {\it shortest} distance between two nodes
and defines the distance metric in complex networks
\cite{erdos,bollobas,watts,barabasi1999}.
Equivalently, we
obtain:
\begin{equation}
N \sim e^{\bar{\ell}/\ell_0},
\label{smallworld}
\end{equation}
where $\ell_0$ is a characteristic length.

A second fundamental property
 in the study of complex networks
arises with the
discovery that the probability distribution
of the number of links per node, $P(k)$ (also
known as the degree distribution),
can be represented by a
power-law (scale-free) with a degree exponent $\gamma$ usually in the
range $2<\gamma<3$
\cite{barabasi1999},
\begin{equation}
P(k) \sim k^{-\gamma}.
\label{scale-free}
\end{equation}

These discoveries have been confirmed in many empirical studies of
diverse networks
\cite{ab-review,mendes,vespignani,newman,barabasi1999,faloutsos}.

With the aim of providing a deeper understanding of
the underlying mechanism which leads to these common features one
needs to probe the patterns within the network structure in more
detail. The question of connectivity between groups
of interconnected nodes on different length-scales has received less
attention. Yet, a plethora of examples in Nature exhibits the
importance of collective behavior, from interactions between
communities within social networks, links between clusters of
web-sites of similar subjects, all the way to the highly modular
manner in which molecules interact to keep a cell alive.
Here we show that real complex networks, such as WWW, social,
protein-protein interaction networks (PIN) and cellular networks are
indeed constructed of self-repeating patterns on all length-scales, and
are therefore invariant or self-similar under a length-scale
transformation.

This result comes as a surprise since the exponential increase in
Eq. (\ref{smallworld}) has led to the general understanding that
complex networks are not self-similar, since self-similarity
requires a power-law relation between $N$ and $\ell$.

How can one reconcile the exponential increase in Eq.
(\ref{smallworld})  with self-similarity, or in other words
an underlying {\it
length}-scale-invariant topology? At the root of the self-similar
properties that we unravel in this study is a
scale-invariant renormalization procedure which we show to be
valid for dissimilar complex networks.

In order to demonstrate this concept we first
consider a self-similar network embedded in  Euclidean space, of
which a classical example would be a fractal percolation cluster
at criticality \cite{bunde-havlin}.
In order to unfold the self-similar properties of
such clusters we calculate the fractal dimension using  a ``box
counting'' method and a ``cluster growing'' method \cite{vicsek}.

In the first method we cover the percolation cluster with $N_B$
boxes of linear size $\ell_B$. The fractal dimension or box dimension
$d_B$ is then
given by \cite{feder}:

\begin{equation}
N_B \sim \ell_B^{-d_B},
\label{dh}
\end{equation}

In the second method, the network is not covered with boxes, instead
one seed node is chosen at random and a cluster of nodes
centered at the seed and separated by a minimum distance $\ell$
is calculated.
The procedure is then repeated by choosing many seed nodes
at random and the average ``mass'' of the resulting
clusters ($\langle M_c\rangle$, defined as the number of nodes in the cluster)
is calculated as a function of $\ell$
to obtain the following scaling:
\begin{equation}
\langle M_c\rangle  \sim \ell^{d_f},
\label{df}
\end{equation}
defining the fractal cluster dimension $d_f$ \cite{feder}.
Comparing Eq. (\ref{df}) and
(\ref{smallworld}) implies that $d_f=\infty$ for complex small-world
networks.

For a {\it homogeneous} network characterized by a {\it narrow}
degree distribution (such as a fractal percolation
cluster) the box covering method
of Eq. (\ref{dh}) and the cluster growing method of  Eq. (\ref{df})
are equivalent since every
node typically has the same number of links or neighbors.
Equation (\ref{df}) can then be derived from (\ref{dh}) and
$d_B=d_f$, and this relation has been regularly used.

The crux of the matter is to understand how one can
calculate a self-similar exponent (analogous to the fractal
dimension in Euclidean space)
in  complex
{\it inhomogeneous} networks with a
{\it broad} degree distribution such as Eq. (\ref{scale-free}).
Under these conditions Eqs. (\ref{dh}) and (\ref{df}) are not
equivalent as will be shown below.
The application of
the proper covering procedure in the box counting method,
Eq. (\ref{dh}), for complex networks unveils a set of self-similar
properties
such as a finite
self-similar exponent
and a new set of critical exponents for the
scale-invariant topology.

\begin{figure}
\centerline{
{\bf a}{ \resizebox{15cm}{!}{\includegraphics{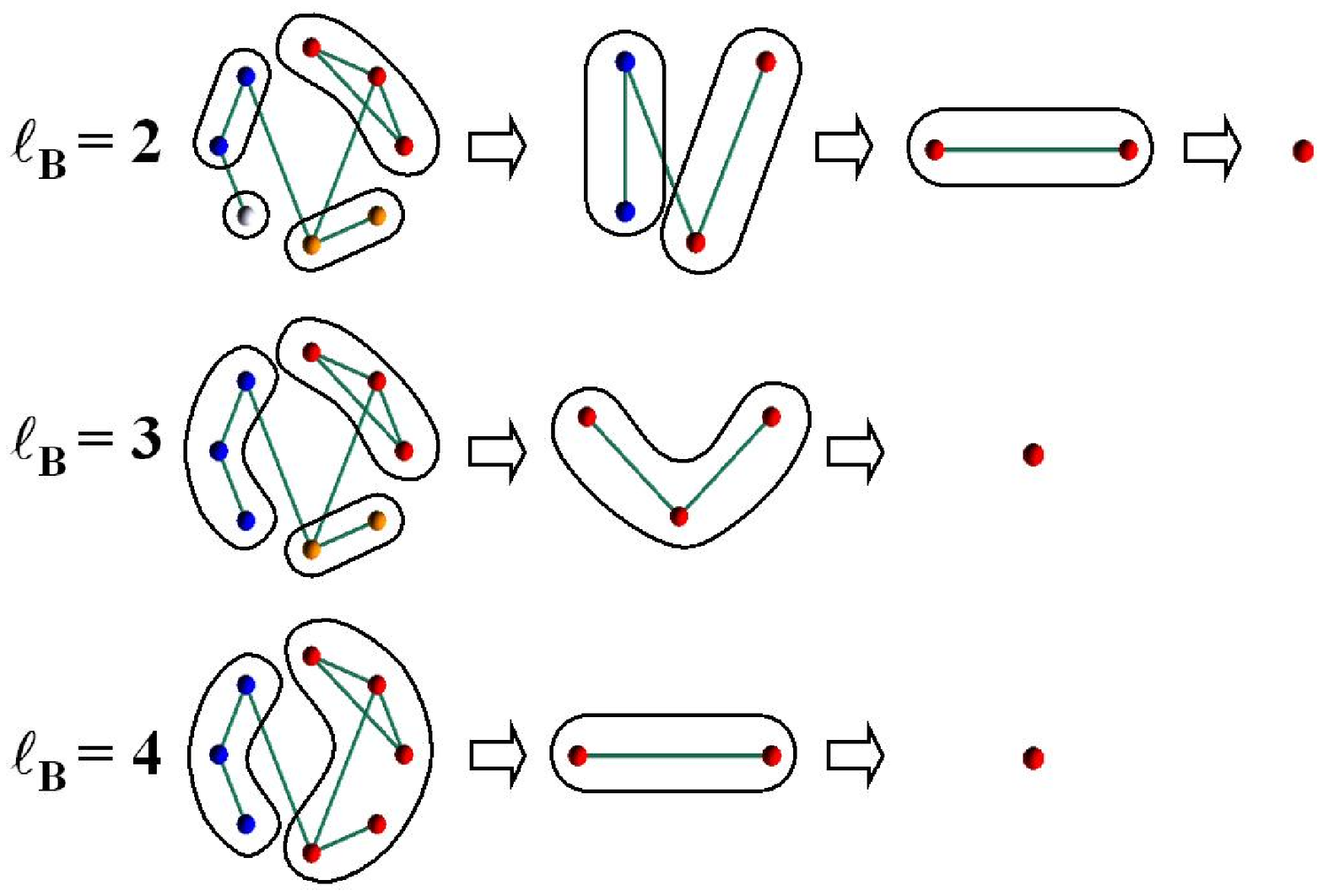}}}}
\centerline{
{\bf b}
{ \resizebox{15cm}{!}{\includegraphics{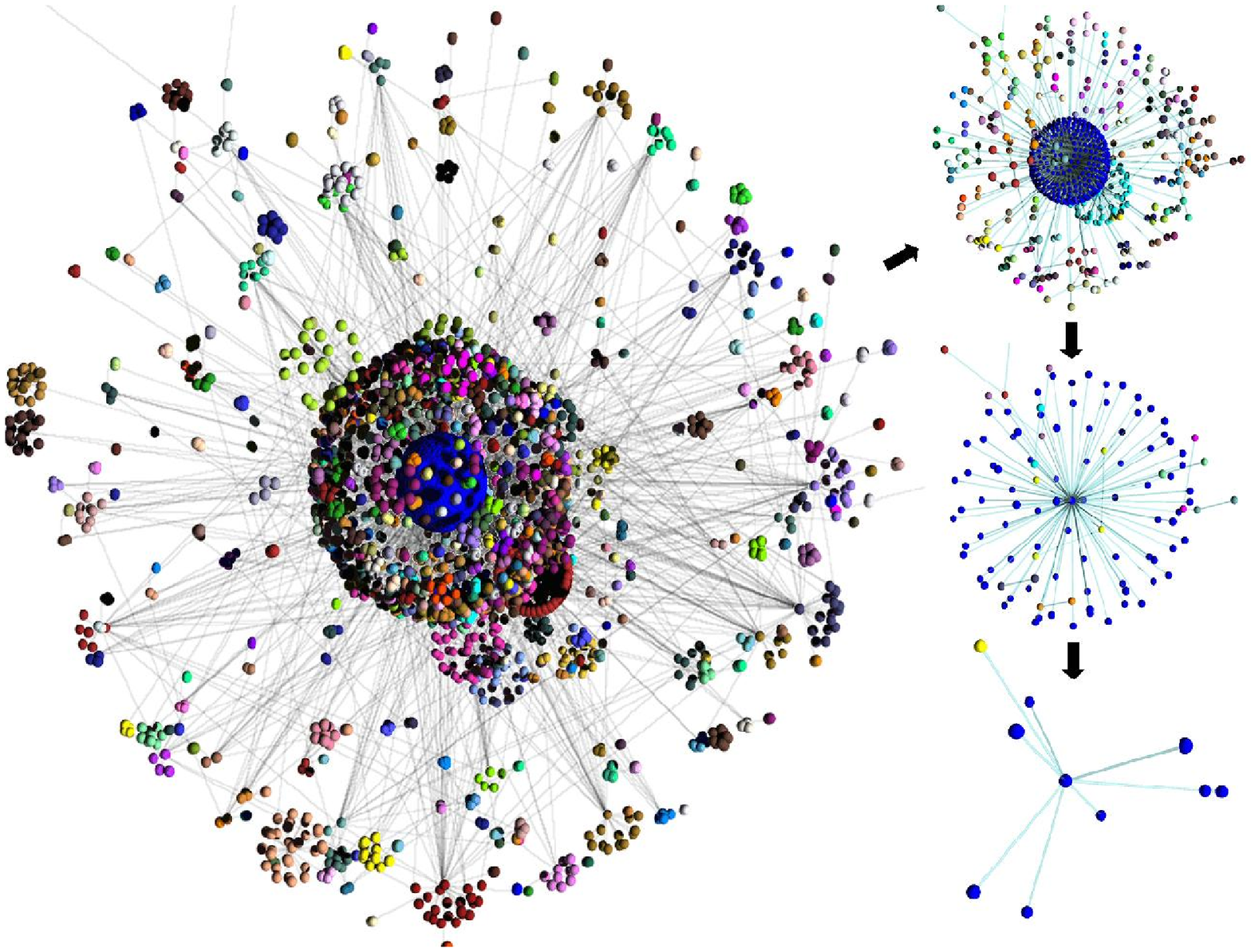}}}
}
\caption{}
    \label{web}
\end{figure}

Figure \ref{web}a illustrates the box covering method using a schematic
network composed of 8 nodes. For
each value of the box size $\ell_B$, we search for the
number of boxes needed to tile the entire network
such that each box contains nodes separated by a distance $\ell <
\ell_B$.

\begin{figure}
\centering
\hbox{
{\bf a}{ \resizebox{8cm}{!}{\includegraphics{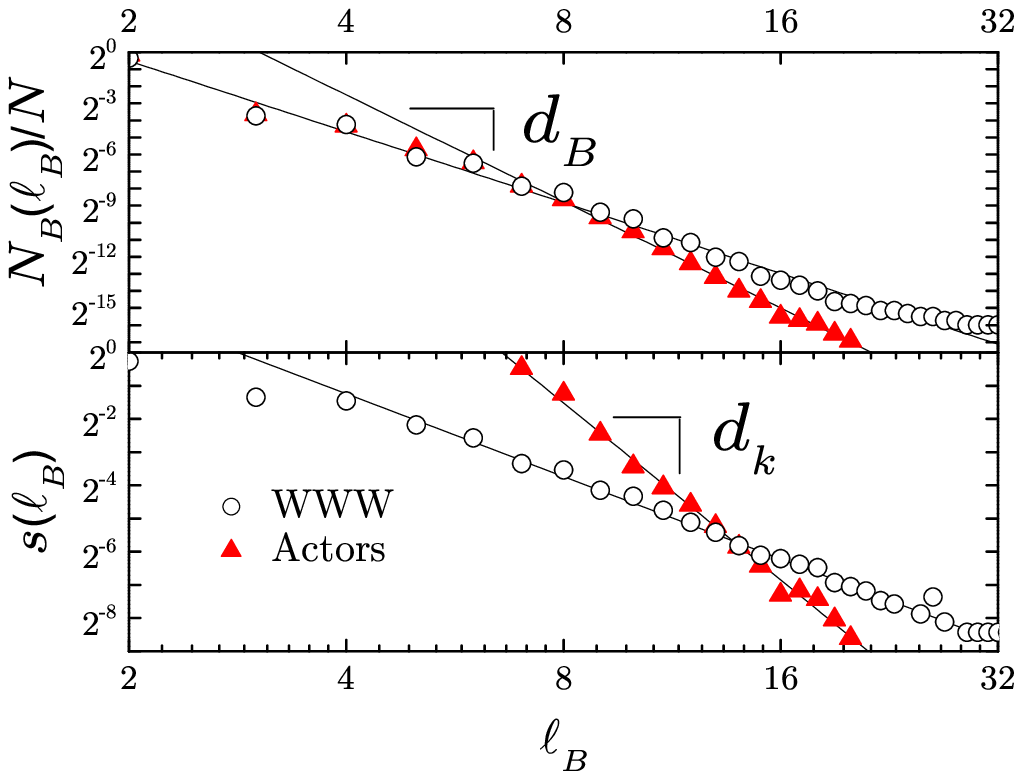}}}

\centering
{\bf b}{ \resizebox{8cm}{!}{\includegraphics{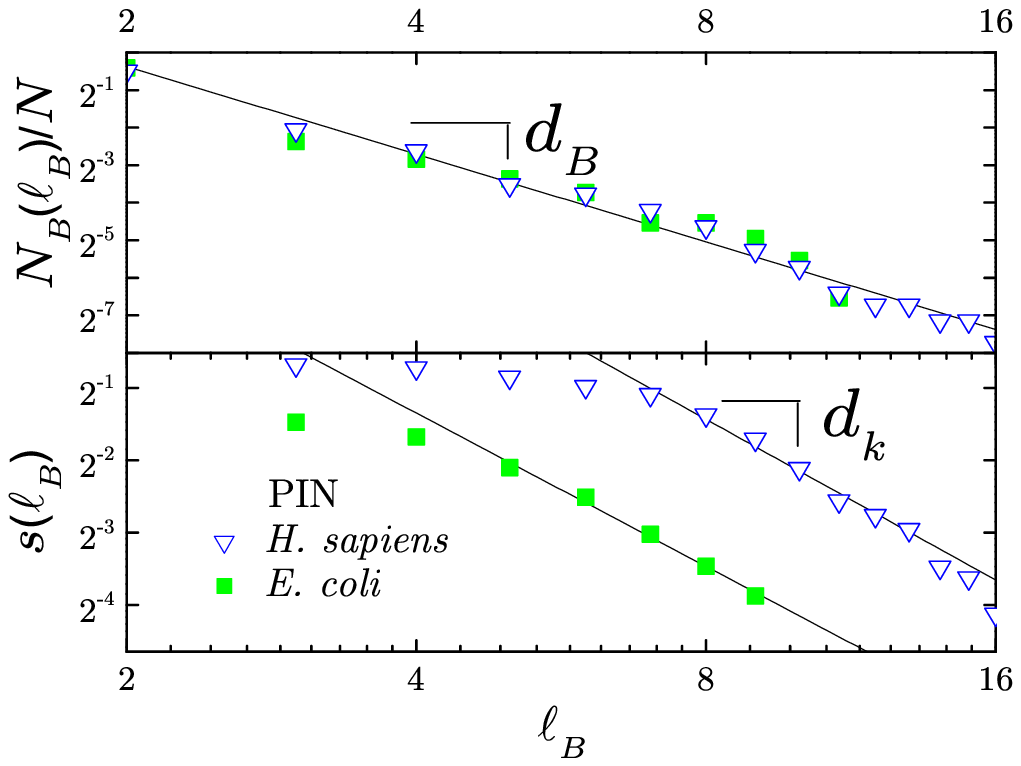}}}
}
\hbox{
\centering
{\bf c}{ \resizebox{8cm}{!}{\includegraphics{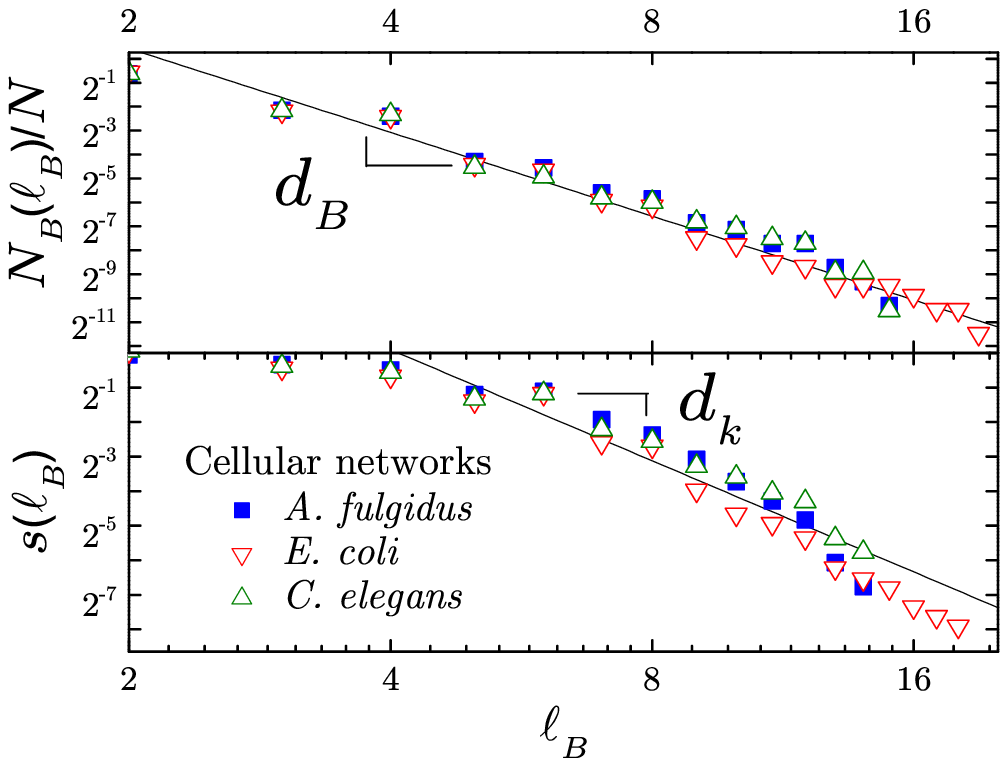}}}

\centering
{\bf d}{ \resizebox{8cm}{!}{\includegraphics{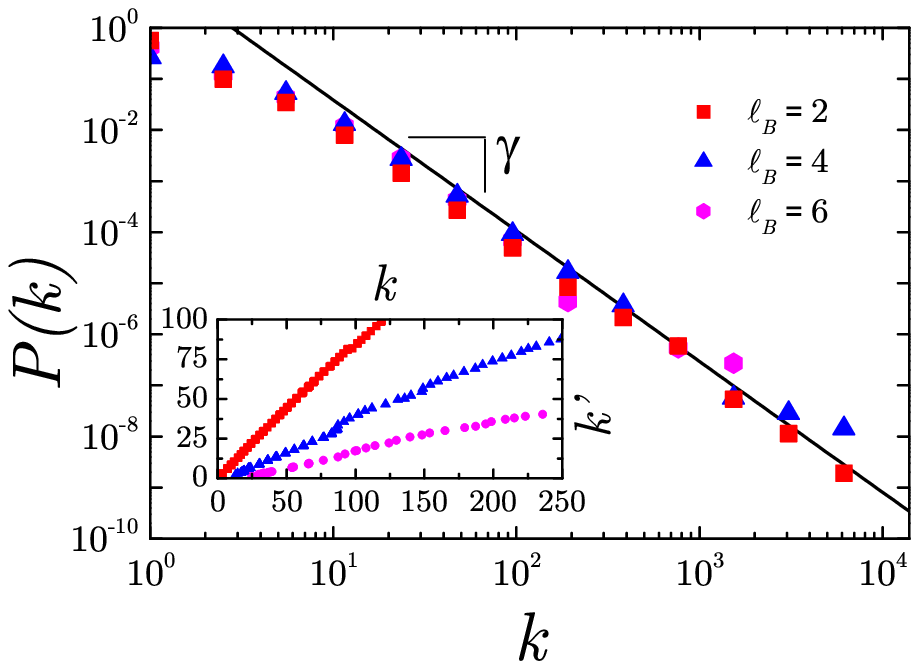}}}
}
  \caption{
}
\label{fractal}
\end{figure}

This procedure is applied to several different real networks: {\it (i)} a
part of the WWW composed of 325,729 web-pages which are connected
if there is a URL link from one page to another \cite{barabasi1999}
(http://www.nd.edu/$\sim$networks),
{\it (ii)} a social
network where the nodes are 392,340 actors linked if they were
cast together in at least one movie \cite{ba},
{\it (iii)} the biological
networks of protein-protein interactions found in {\it E. coli}
(429 proteins) and {\it H. sapiens} (human) (946 proteins)
linked if there is a physical binding
between them (database available via the
Database of Interacting Proteins \cite{xenarios,dip},
other PINs  are discussed in the Supplementary Materials),
and {\it (iv)} the cellular networks  compiled by \cite{cellular}
using a graph-theoretical representation of the whole biochemical pathways
 based on the WIT
 integrated-pathway genome database \cite{wit}
(http://igweb.integratedgenomics.com/IGwit)
of 43 species from
Archaea, Bacteria, and Eukarya. Here we show the results
for {\it A. fulgidus, E. coli} and {\it C. elegans} \cite{cellular}, while
the full database is analyzed in the Supplementary
Materials.
It has been previously determined that the WWW and actors networks
are small-world and scale-free, characterized by
Eq. (\ref{scale-free}) with $\gamma =
2.6$ and 2.2, respectively \cite{ab-review}. For the
PINs of {\it E. coli} and {\it H. sapiens} we find
$\gamma=2.2$ and $2.1$, respectively.
All cellular networks are scale-free with  average
exponent $\gamma=2.2$ \cite{cellular}.
We confirm these values and show the results for the WWW in Fig.
\ref{fractal}.

Figures \ref{fractal}a and \ref{fractal}b show the results of $N_B(\ell_B)$
according to Eq. (\ref{dh}). They reveal the
existence of self-similarity in the WWW, actors, and
{\it E. coli} and {\it H. sapiens}
protein-protein interaction networks with
self-similar exponents
$d_B=4.1$, $d_B=6.3$ and $d_B=2.3$  and $d_B=2.3$, respectively.
The cellular networks are shown in Fig. \ref{fractal}c and have
$d_B=3.5$.

\begin{figure}
\centering
{\bf a}
{ \resizebox{12cm}{!}{\includegraphics{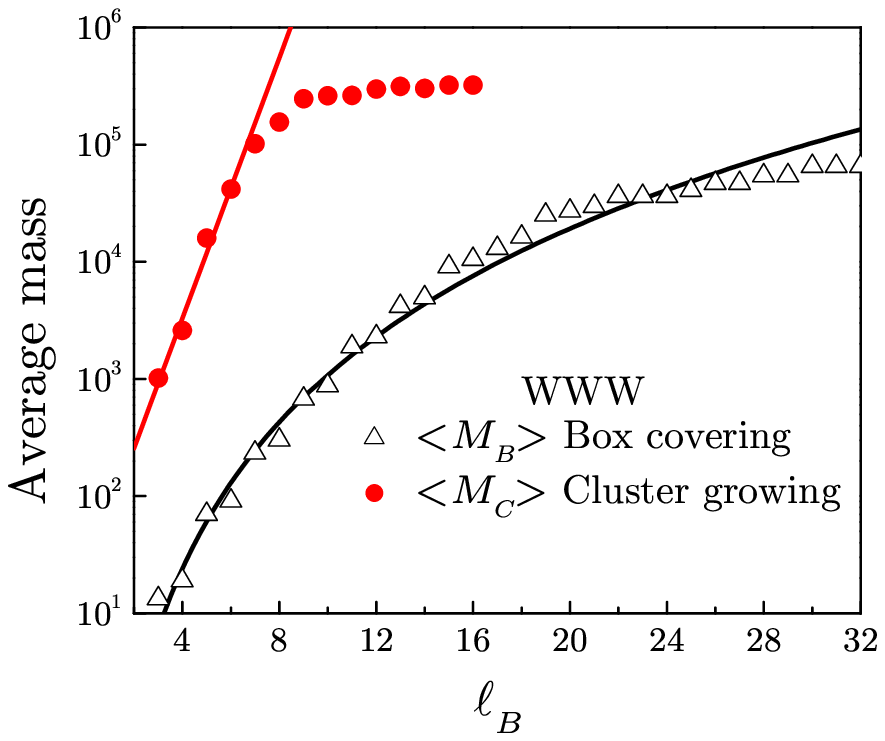}}}

{\bf b}
{ \resizebox{14cm}{!}{\includegraphics{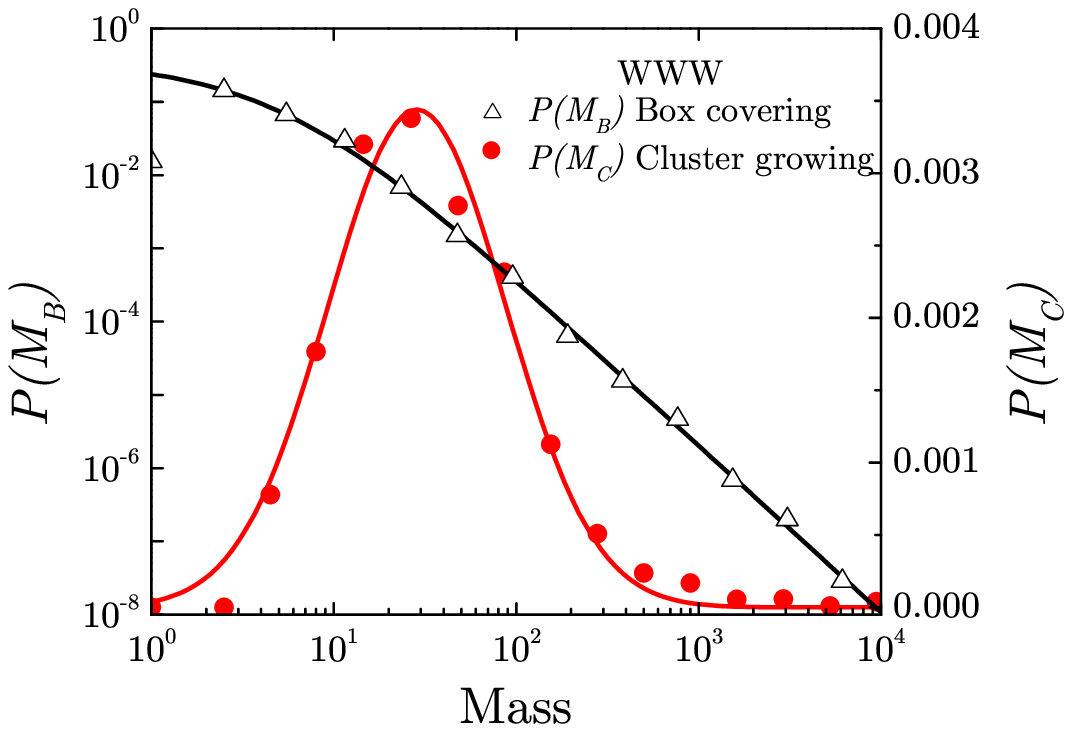}}}
\caption{}
\label{mass}
\end{figure}

We now elaborate on the apparent
contradiction between the two definitions of  self-similar exponents
in complex networks.
After performing a renormalization at a
given $\ell_B$, we calculate the mean mass of the boxes covering the network,
$\langle M_B (\ell_B)\rangle$,
to obtain
\begin{equation}
\langle M_B (\ell_B)\rangle  \equiv N / N_B(\ell_B) \sim \ell_B^{d_B},
\label{mass_b}
\end{equation}
which is corroborated by direct measurements for all the networks and shown
in Fig. \ref{mass}a for the WWW.

%

On the other hand, the average performed in the cluster growing
method (for this calculation
we average over single boxes without tiling the system)
gives rise to an exponential
growth of the mass
\begin{equation}
\langle M_c(\ell_B)\rangle  \sim e^{\ell_B/\ell_1},
\label{mass_c}
\end{equation}
with $\ell_1\approx 0.78 $
in accordance with the small-world effect Eq. (\ref{smallworld}),
as seen in Fig. \ref{mass}a.

The topology of scale-free
networks is dominated by several highly connected
hubs--- the nodes with the largest degree---
implying that most of the nodes are connected to the hubs via one
or very few steps.
Therefore the average performed in the cluster growing  method
is biased; the
hubs are overrepresented in Eq. (\ref{mass_c})
since almost every node is a neighbor of a hub.
By choosing the seed of the clusters at random,
there is a very large probability of including the hubs in the clusters.
On the other hand the box covering method is a global tiling of the
system providing a flat average over all the nodes, i.e.
each part of the network is covered with an equal
probability. Once a hub (or any node) is covered, it cannot be covered again.
We conclude that
Eqs. (\ref{dh}) and (\ref{df}) are not equivalent
for inhomogeneous networks with topologies dominated by hubs with a large degree.

The biased sampling of the randomly chosen nodes is clearly demonstrated
in Fig. \ref{mass}b.
We find that the
probability distribution of the mass of the boxes
for a given $\ell_B$ is very broad and can be approximated by
a power-law: $P_{\ell_B}(M_B) \sim M_B^{-2.2}$
in the case of WWW and $\ell_B=4$.
On the other hand, the probability distribution of
$M_c$ is very narrow and can be fitted by a log-normal
distribution (see Fig. \ref{mass}b).
In the box covering method there are many boxes with very large and very small
masses in contrast to the peaked distribution in the
cluster growing method, thus showing the biased nature of the latter method
in inhomogeneous networks.
This biased average leads to the exponential growth of the mass
in Eq. (\ref{mass_c}) and it also explains why the average
distance is logarithmic with $N$ as in Eq. (\ref{smallworld}).

The box counting method provides a powerful tool for further
investigations of the network properties as it enables a
renormalization procedure, revealing that the
self-similar properties
and the scale-free degree distribution
persists irrespective of the amount of coarse-graining of the
network.


Subsequent to the first step of assigning the nodes to the boxes
we create a new renormalized network by replacing each box by a
single node. Two boxes are then connected,  provided that there
was at least one link between their constituent nodes.
The second column of the panels in Fig. \ref{web}a shows this step
in the renormalization procedure for the schematic network, while Fig.
\ref{web}b shows the results for the same procedure applied to the
entire WWW for $\ell_B=3$.

The renormalized network gives rise to a new probability
distribution of links, $P(k')$, which is invariant
under the renormalization:
\begin{equation}
P(k)\rightarrow P(k') \sim (k')^{-\gamma}.
\label{pk}
\end{equation}
Figure \ref{fractal}d supports the validity of this scale
transformation by showing a data collapse of all distributions
with the same $\gamma$ according to (\ref{pk}) for the WWW.



Further
insight arises from relating the scale-invariant properties (\ref{dh})
to the scale-free degree distribution (\ref{scale-free}).
Plotting (see inset in Fig.
\ref{fractal}d for the WWW) the number of links
$k'$ of each node in the renormalized network
versus the maximum number of links $k$ in each box
of the unrenormalized network
exhibits a scaling law
\begin{equation}
k \rightarrow k' = s(\ell_B) k.
\label{s}
\end{equation}
This equation defines the scaling transformation in the
connectivity distribution. Empirically we find that the scaling
factor $s(<1)$ scales with  $\ell_B$ with a new exponent $d_k$ as
\begin{equation}
s(\ell_B) \sim \ell_B^{-d_k},
\label{sl}
\end{equation}
shown in Fig. \ref{fractal}a for the WWW and actor networks (with $d_k=2.5$
and $d_k = 5.3$, respectively), in
Fig. \ref{fractal}b for the protein networks ($d_k=2.1$ for {\it E. coli} and
$d_k=2.2$ for {\it H. sapiens}) and in Fig. \ref{fractal}c
for the cellular networks with $d_k=3.2$.

Equations (\ref{s}) and (\ref{sl}) shed light on how families of
hierarchical sizes are linked
together. The larger the families, the fewer links exist.
Surprisingly the same power-law relation exists for
large and small families represented by
Eq. (\ref{scale-free}).


From Eq. (\ref{pk}) we obtain $n(k) dk = n'(k') dk'$,
where $n(k) = N P(k)$ is the number of nodes with links $k$
and $n'(k') = N' P(k')$ is the number of nodes with links $k'$ after
the renormalization ($N'$ is the total number of nodes in
the renormalized network).
Using Eq. (\ref{s})
we obtain $n(k) = s^{1-\gamma} n'(k)$.
Then, upon renormalizing
a network with $N$ total nodes
we obtain a smaller number of nodes $N'$ according to
$N' = s^{\gamma-1} N$.
Since the total number of nodes in the renormalized network is the
number of boxes needed to cover the unrenormalized network
at any given $\ell_B$, we have $N' =N_B(\ell_B)$. Then, from
Eqs. (\ref{dh}) and (\ref{sl}) we obtain
the relation between the three indexes
\begin{equation}
\gamma = 1 + d_{B} / d_{k}.
\label{relation}
\end{equation}

Equation (\ref{relation}) is confirmed for all the networks analyzed
here (see Supplementary Materials).
In all cases the
calculation of $d_B$ and $d_k$ and
Eq. (\ref{relation})
gives rise to the same $\gamma$ exponent as
that obtained in the direct calculation of the degree
distribution. The significance of this result is that the
scale-free properties characterized by $\gamma$ can be related to
a more fundamental length-scale invariant property,
characterized by the two new indexes $d_{B}$ and $d_{k}$.

Summarizing, we elucidate a fundamental property of a wide variety of
complex networks: that
of a scale-invariant topology.
Concepts first introduced for the study of critical phenomena
in statistical physics are shown to be valid here in the characterization
of a different class of phenomena: the topology of
complex networks.
One could
envisage a great deal of fundamental information being understood
by the
application of renormalization techniques to this kind of
complex system.
For instance,  networks with similar degree distributions
are characterized by different
self-similar exponents, thus indicating that they may belong to
different
``universality classes''.
It is as though
each node (ranging from web-pages
in the WWW,  to people in social networks, to proteins and substrates
in cellular networks)
were connected to other nodes under
a single self-organizing principle
according to which
groups of nodes of all sizes
self-organize too;
such that everything links with everything
else, governed by
one universal dynamics  in Nature \cite{schopenhauer}.

Acknowledgments: This work was supported by the National Science Foundation.



{\bf Acknowledgements}. SH wishes to thank the Israel Science Foundation
for support. This work is supported by the
National Science Foundation, DMR-0239504.

\newpage
\clearpage

FIG \ref{web}. {The renormalization procedure to complex networks.
\bf a,} Demonstration of the method
for different $\ell_B$ and different stages in a network demo.
The first column depicts the original network.
We tile the system with boxes of size $\ell_B$
(different colors correspond to different boxes).
 All nodes in
a box are connected by a minimum distance smaller than the given $\ell_B$.
For instance, in the case of
$\ell_B=2$, we identify four boxes which contain the nodes depicted
with colors red, orange, white, and blue, each containing 3, 2, 1, and
2 nodes, respectively.
Then we replace each box by a single node; two renormalized nodes are
connected if there is at least one link
between the unrenormalized boxes.
 Thus we obtain the network
shown in the second column.
The resulting number of boxes needed to tile
the network, $N_B(\ell_B)$,
is plotted in Fig. \protect\ref{fractal}
versus $\ell_B$ to
obtain   $d_B$
as in Eq. (\protect\ref{dh}).
The renormalization procedure is applied again
and repeated until
the network is reduced to a single node
(third and fourth columns for different $\ell_B$).
{\bf b,} Three stages in the renormalization scheme applied to the entire WWW.
We fix the box size to  $\ell_B = 3$ and apply the
renormalization for four stages. This corresponds, for instance,
to the sequence for the network demo
depicted in the second row in part {\bf a} of this figure.
We color the nodes in the web according to the boxes to which
they belong.
The network is invariant under this renormalization as explained in
the legend of Fig. \ref{fractal}d and the Supplementary Materials.


\vspace{.5cm}

FIG \ref{fractal}. Self-similar scaling in complex networks.
{\bf a,}  Upper panel: Log-log plot of the $N_B$ vs $\ell_B$ revealing
the
self-similarity
of the WWW and actor network
according to Eq. (\ref{dh}). Lower panel: The scaling of $s(\ell_B)$
vs. $\ell_B$ according to Eq. (\ref{sl}).
The errors bars are of the order of the symbol size.
{\bf b,} Same as (a) but for two
protein interaction networks: {\it H. sapiens} and {\it E. coli}.
Results are analogous to (b) but with different scaling exponents.
{\bf c,} Same as (a) for the cellular networks of
{\it A. fulgidus}, {\it E. coli} and {\it C. elegans}.
{\bf d,}
Invariance of the
 degree distribution of the WWW under the renormalization
for different box sizes, $\ell_B$.
We show the data collapse of the degree distributions
demonstrating the self-similarity at different scales.
The inset shows the scaling of $k'=s(\ell_B) k$ for different
$\ell_B$, from where we obtain the scaling factor $s(\ell_B)$.
Moreover, we also apply the renormalization for a fixed
box size, for instance $\ell_B=3$ as shown in Fig. \ref{web}b for the WWW,
until the network is reduced to a few nodes
and find that $P(k)$ is invariant
under these multiple renormalizations as well, for several iterations
(see Supplementary Materials).


\vspace{.5cm}

FIG. \ref{mass}. Different  averaging techniques lead
 to qualitatively different results.
{\bf a,} Mean value of the box mass
in the box counting method, $\langle M_B\rangle$,
 and the cluster mass in the cluster growing method,
$\langle M_c\rangle$,
 for the WWW. The solid lines represent the power-law fit
for $\langle M_B\rangle$ and
the exponential fit for $\langle M_c\rangle$
according to Eqs. (\ref{mass_b}) and (\ref{mass_c}), respectively.
{\bf b,} Probability distribution of $M_B$ and $M_c$
for $\ell_B=4$ for the WWW. The curves are fitted by a power-law
and a log-normal distribution, respectively.

\newpage
\clearpage

\newpage
\clearpage

\newpage
\clearpage

\newpage
\clearpage

\centerline{\bf SUPPLEMENTARY MATERIALS}

\vspace{1cm}

\section{\bf The  box covering method}

\begin{figure}
\centering
{\bf a}
{
\resizebox{8cm}{!}{\includegraphics{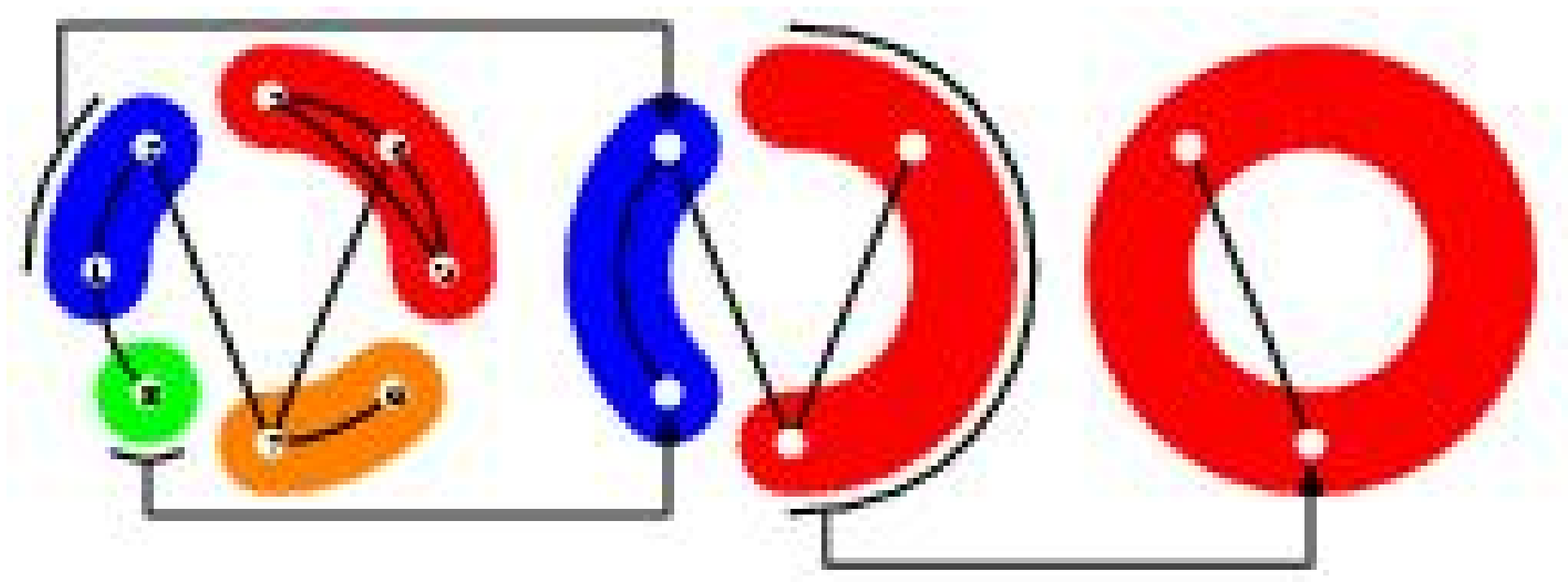}}
}

\centering
{\bf b}
{
\resizebox{3cm}{!}{\includegraphics{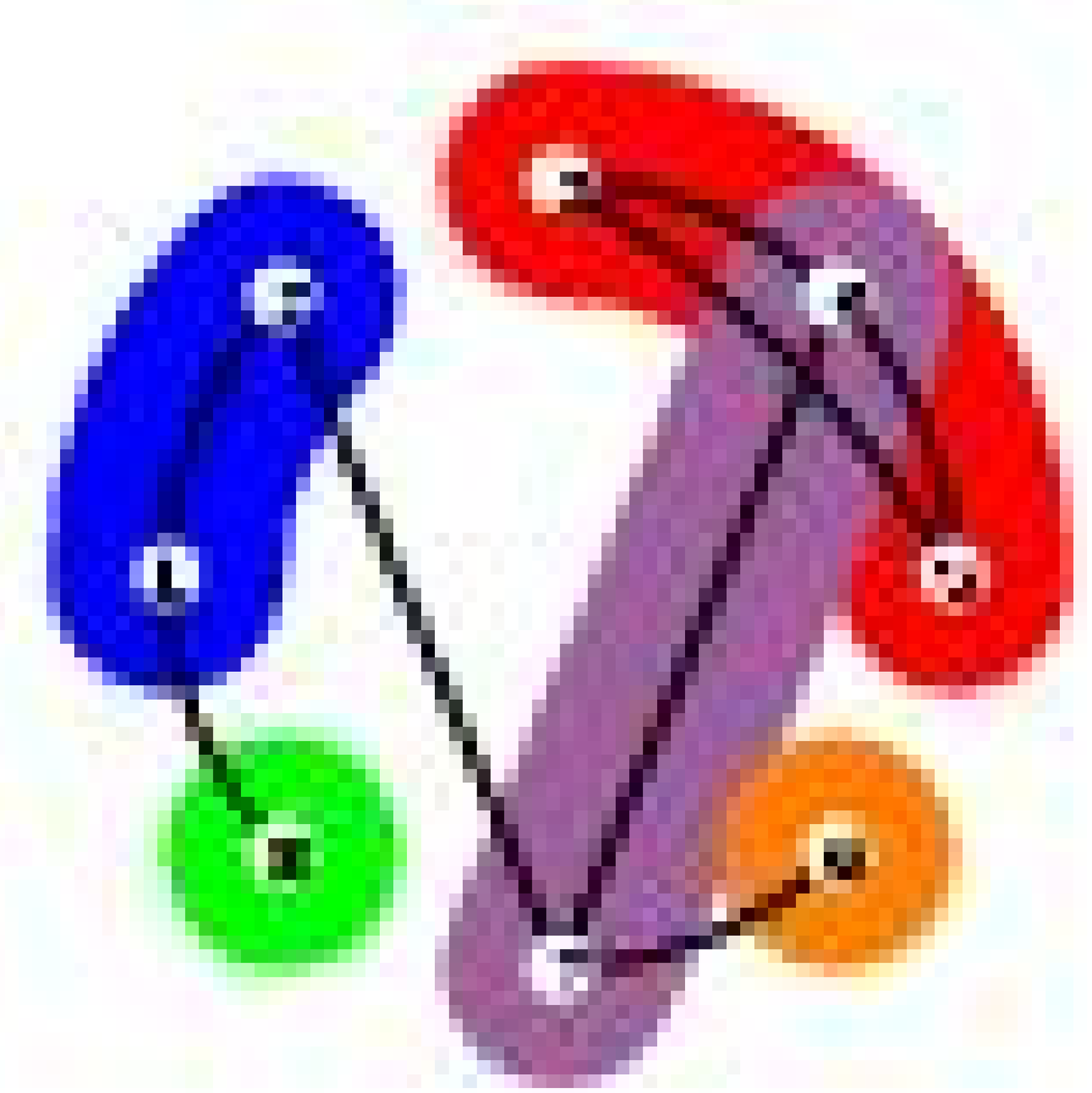}}
}
\caption{Details of the box covering method for
{\bf a, } $\ell_B=2$.
{\bf b,} A different covering for the same network as in (a) for
$\ell_B=2$. Different coverings give raise to the same
exponents
as explained in the text.}
\label{demo}
\end{figure}

Since the box covering method is central to the understanding of the
scale-invariant properties of networks, we describe it in more
detail here.
Figure \ref{demo}a shows the same network
as in Fig.
\ref{web}a for the case $\ell_B=2$.
We tile the system by first assigning nodes 1 and 2 to the
box colored in blue.  Notice that the maximum distance between the nodes
of a given box is $\ell_B-1$. Thus, node 8 would not be in the blue
box since its distance from node 2 is $\ell=2$ (even though its
distance from 1 is $\ell=1$).  Then we cover the nodes
6 and 7 with the orange box, and the nodes 3, 4, and 5
with the red box.
Finally, the
last node 8 is assigned to the green box.
The number of boxes to cover the network
is then $N_B=4$.

The renormalization is then
applied by replacing each box by a single node.  Thus, nodes 1 and 2
will be combined into a single node as indicated by the arrow
from the first panel to the second panel in Fig. \ref{demo}a.
This renormalized node is
connected with the orange and green boxes because there is a link
between nodes 2 and 7, and 1 and 8, respectively.
The same rule applies to the other
boxes.  The renormalized network is shown in the second panel.  The
system is then tiled again with boxes; in this case two boxes (blue and red)
are
needed to cover the entire network. The two boxes are then replaced by
nodes and a second renormalized network is obtained as shown in the
third panel.  Finally, the last two nodes belong to the same (red) box and
are replaced by a single
node.

This procedure is applied to the WWW in Fig. \ref{web}b.
The main panel corresponds to the first stage in the renormalization
of the web for $\ell_B=3$.
The procedure is applied again obtaining the remaining
panels in Fig. \ref{web}b until the web is reduced to a single box
in the last panel. The colors of the nodes corresponds to the boxes to which
they belong.

\begin{figure}
\centering
{
\resizebox{8cm}{!}{\includegraphics{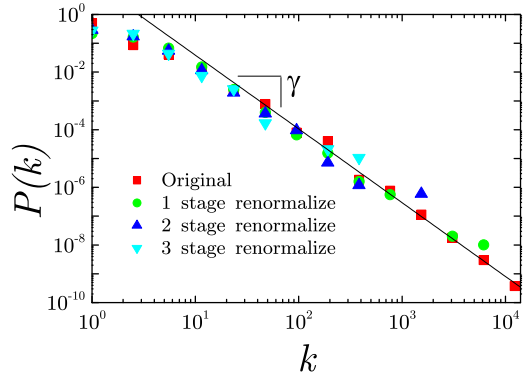}}
}
\caption{Invariance of the degree distribution of the WWW under
multiple renormalizations  done at fixed $\ell_B=3$. The stages
1, 2, and 3 correspond to the networks depicted in the first three stages in
Fig. \protect\ref{web}b.}
\label{pk1}
\end{figure}

In Fig. \ref{fractal}d we show the invariance of the degree
distribution $P(k)$ under the renormalization performed as a function
of the box size in the WWW. The other networks analyzed in this study
present the same invariant property.
It is important to mention that  the networks are also invariant
under multiple renormalizations applied for a fixed
box size $\ell_B$. This corresponds, for instance, to the stages depicted
in Fig. \ref{web}a in the second row for $\ell_B=3$ for the network demo.
Figure \ref{pk1} shows the invariance of $P(k)$ for the WWW
after several stages of the renormalization for a fixed $\ell_B=3$,
and it is the analogous of Fig. \ref{fractal}d for
different box size.
The stages 1, 2, and 3 correspond to the networks
depicted in the first 3 stages in Fig. \ref{web}b.

From  the above explanation it should be clear that there are many ways to tile
the network. For instance in Fig. \ref{demo}b we show another tiling.
In this case we assign nodes 4 and 7 together in a single box instead of
nodes 6 and 7 as in Fig. \ref{demo}a. This tiling
results in an extra box needed to cover node 6 and therefore
in a larger number of nodes to tile
the system, $N_B=5$.

While there are many ways to assign nodes to the boxes, we notice that
the rigorous mathematical definition
of Eq. (\ref{dh})
corresponds to the {\it minimum} number of boxes
needed to cover the network \cite{feder}.
This minimization does not have any consequence for the determination
of the fractal dimension in homogeneous clusters.  However, it may become
relevant when calculating the
self-similar exponent
of a complex network
with a {\it widely} distributed number of links.
Finding the minimum number of boxes to cover the network
is a hard optimization problem to solve, analogous to the
graph coloring problem in the NP-complete complexity class.
This minimization problem has to be solved by an exhaustive numerical search
since there is no numerical
algorithm to solve this kind of problems.

We have performed the search over a limited part of the phase-space
for the WWW
to obtain an estimation of the average and the minimum number of boxes
needed to tile the network for every value of $\ell_B$.
We find that the average value of the boxes is very close to the
estimated minimum number of boxes. Moreover,
we find that the minimization
is not relevant  and
any covering gives rise to the same  exponent.

\begin{figure}
\hbox{ \centering
{\bf a}
{ \resizebox{6cm}{!}{\includegraphics{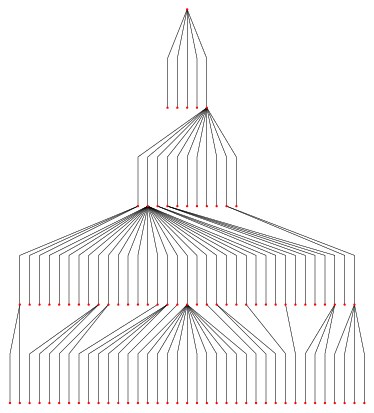}}}

\centering
{\bf b}
{ \resizebox{8cm}{!}{\includegraphics{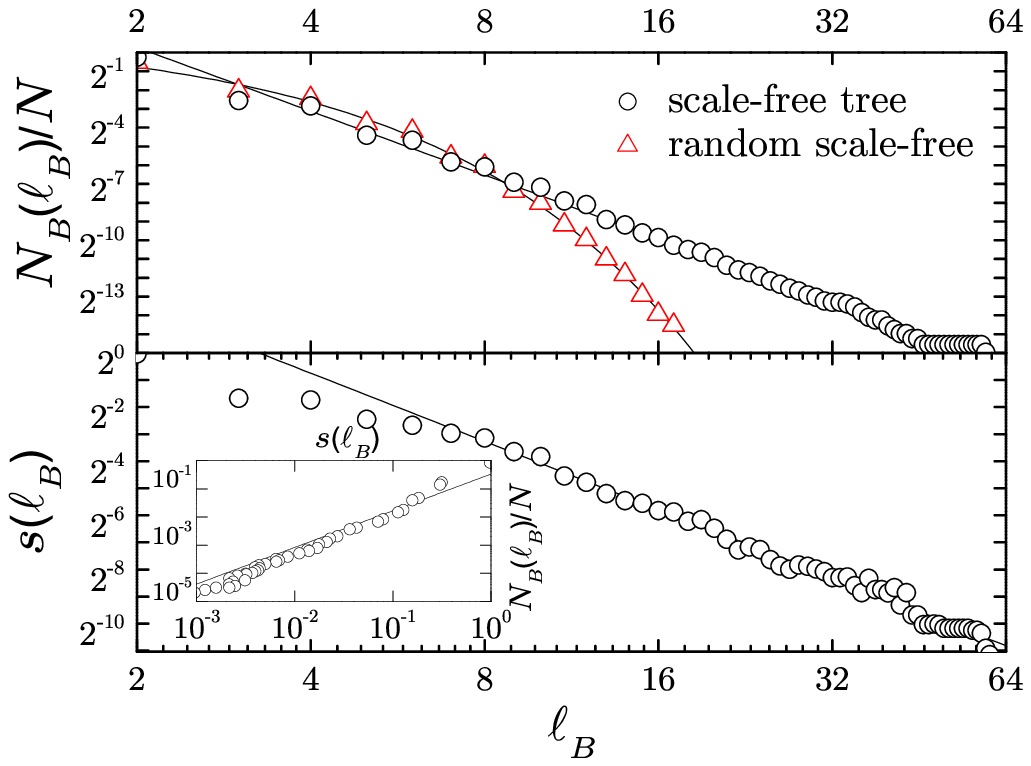}}}
}
\caption{The scale-free tree structure and the random scale-free model.
{\bf a,} Example of a scale-free tree structure.
Nodes with a power-law degree distribution are connected in a tree
structure without loops. {\bf b,} The log-log plot of $N_B$ vs
$\ell_B$ reveals a
self-similar
structure for the scale-free tree (upper panel)
while $s(\ell_B)$ scales as
in Eq. (\ref{sl}) (lower panel). In contrast the random scale-free
network where nodes (with a power-law distribution of links)
are connected at random shows a lack of self-similarity expressed in the
exponential decrease with $\ell_B$ in the upper panel.}
\label{tree}
\end{figure}

\section{\bf Scale-free tree structure}

The underlying meaning of the existence of scale-free networks
which are self-similar is yet to be deciphered, but some insight
can be gained by examining the simplest structure of a
known network of that kind: a {\it tree} network
which has been characterized  using field theoretical arguments
and fractal dimensions in
\cite{burda}.

The sequence of
renormalization steps depicted in Fig. \ref{web} suggests the
following scheme: one begins with a single node and then
constructs the network by applying the renormalization procedure
in a reversed fashion. This can be achieved by following the
procedure in Fig. \ref{web} for a specific value of $\ell_B$.

More specifically, a single node with a large number of links is
first connected to the next generation of nodes. For every node we
assign a number of links from a power-law distribution with a
given $\gamma$. The next layer of the tree is generated in the
same way. A tree structure with a power-law degree distribution
and self-similar topology emerges which is depicted in Fig. \ref{tree}a.

This is corroborated numerically in  Fig. \ref{tree}b where we
study a scale-free tree structure with 192,827 nodes and
$\lambda=2.3$, and we find
$d_B=3.4$ and $d_k = 2.5$. The parallels between the
features of such a simple structured network and those discussed
in this paper suggest that this simplified view may lie at the
core of more complex self-similar networks.

\begin{figure}
\hbox{
\centering
{\bf a}
{ \resizebox{7cm}{!}{\includegraphics{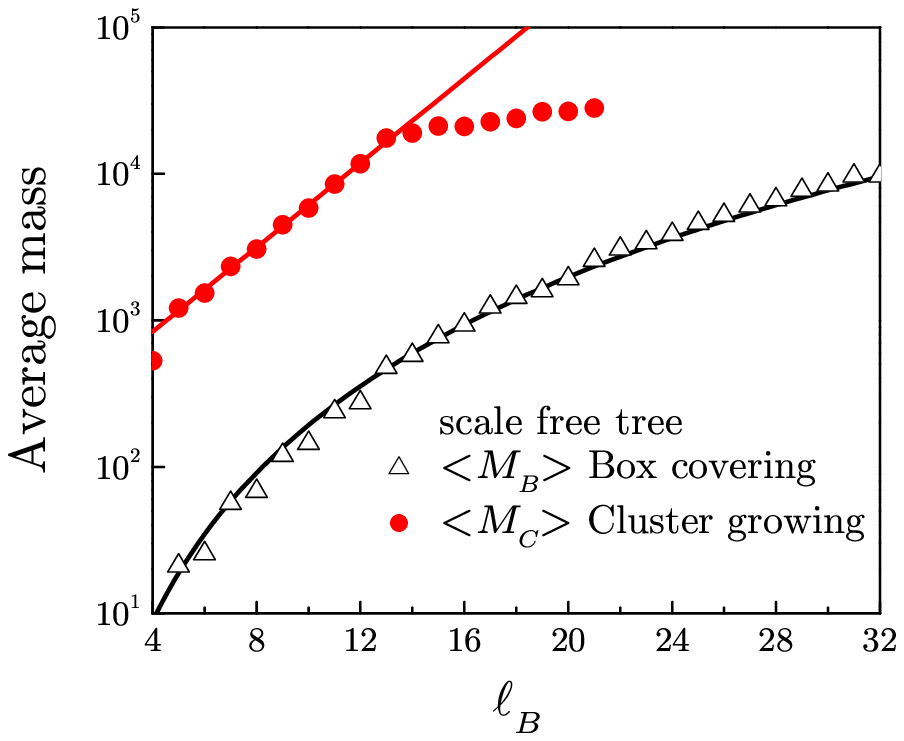}}}

{\bf b}
{ \resizebox{8cm}{!}{\includegraphics{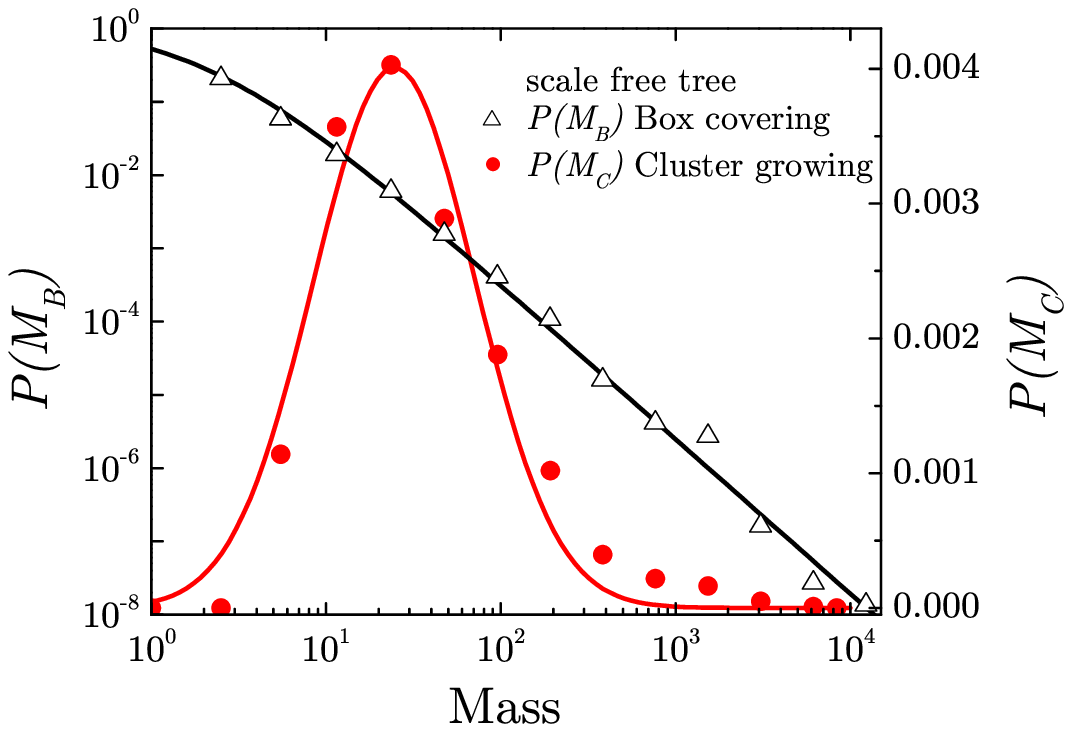}}}
}
\caption{
Results for the scale-free tree model.
{\bf a,} Mean value of the box mass
in the box counting,
$\langle M_B\rangle$, and mean value of the cluster mass
in the cluster growing method,
$\langle M_c\rangle$ versus $\ell_B$.
{\bf b,} Probability distribution of $M_B$ and $M_c$
for $\ell_B=5$. The results are in agreement with the finding
of real networks in Fig. \ref{mass}. A power-law distribution is
found for $M_B$ while a log-normal distribution is found
for $M_c$ as shown by the fits.
}
\label{mass_model}
\end{figure}

Moreover, we also calculate the average mass of the
boxes and the mass of the clusters
in the box covering method and the cluster covering method,
respectively, and we find the power law of Eq. (\ref{mass_b}) and
the exponential behaviour of Eq. (\ref{mass_c})
 (see Fig. \ref{mass_model}a)
 in
agreement with the results of the real networks
analyzed in the main manuscript, Fig. \ref{mass}a.
Figure \ref{mass_model}b shows the probability distribution
of $M_B$ (power-law) and $M_c$ (log-normal)
in agreement with previous results as well, Fig. \ref{mass}b.

\section{\bf  Internet }

\begin{figure}
\centering
{ \resizebox{10cm}{!}{\includegraphics{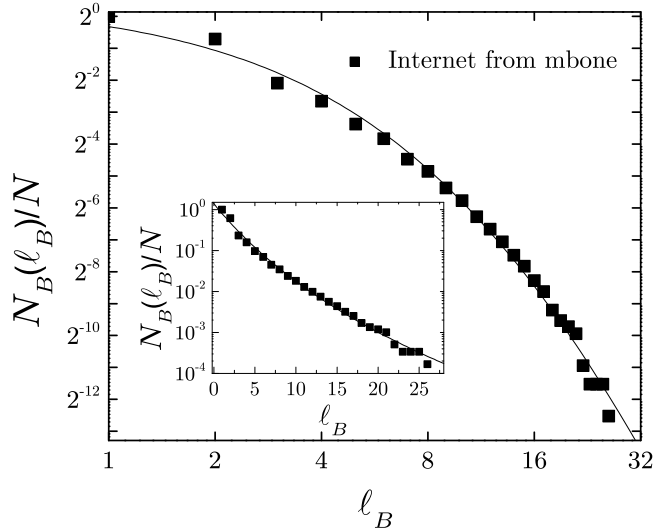}}}
\caption{Internet. Log-log plot of $N_B(\ell_B)$.
The solid line represents the modified power law fit, Eq. (\ref{modified}).
The inset shows a linear-log plot indicating
that the decay  is slower than exponential.}
\label{internet}
\end{figure}

It is interesting to note that not all complex networks
show the clear  self-similarity of the networks presented so far.
We analyze the Internet
composed of computers and routers linked by
physical lines such as
the database collected by the SCAN project
(the ``Mbone'',
www.isi.edu/scan/scan.html, we also analyze the database of
the Internet Mapping Project
\cite{lucent} and found similar results).
 Figure
\ref{internet} shows the result of $N_B(\ell_B)$. We fit
the curve with a modified power-law
\begin{equation}
 N_B(\ell_B) \sim (\ell_B+\ell_c)^{-d_B},
\label{modified}
\end{equation}
with $\ell_c=14.9$ representing a cut-off and
$d_B=8.5$,
suggesting a large self-similar exponent.
The decay of $N_B$ with $\ell_B$ is faster than a power-law and
slower than  exponential as shown in the inset of Fig. \ref{internet}.

Thus these networks
lack the clear  self-similar
structure found for the WWW, actors and the
biological networks.
However,
we find that the distribution of $P(M_B)$ remains a power law and
the degree distribution $P(k)$ is invariant under
the renormalization suggesting that some
self-similar properties might still be valid for the Internet.
We notice that Internet
maps are made by programs that use the IP protocol to trace the
connections between each registered node in the Internet.
These maps are incomplete since they map a few
routers from each domain and also due to the existence of firewalls.
Thus, the apparent lack of self-similarity might be due to incomplete
information of the network.


\section{ \bf Protein-protein interaction networks}

\begin{figure}
\centering
{ \resizebox{10cm}{!}{\includegraphics{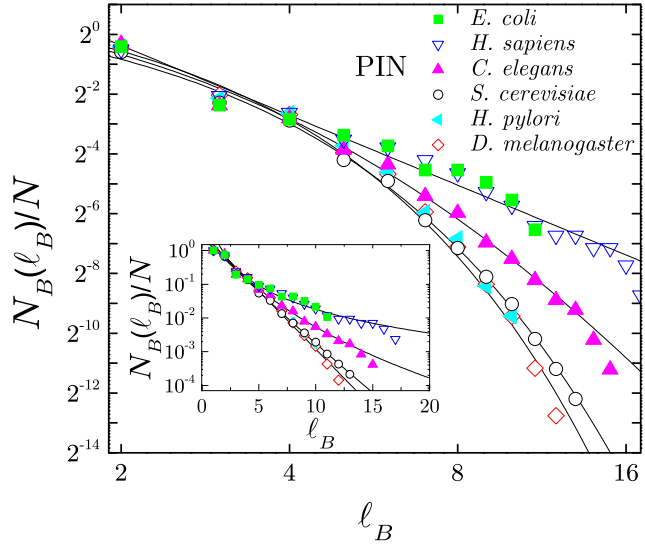}}}
\caption{Scaling for the protein-protein interaction networks.
Log-log plot of $N_B$ versus $\ell_B$ for different
protein-protein interaction networks. While {\it E. coli}
and {\it
H. sapiens} show  a clear  power law behavior, the other protein
networks show a modified power-law behaviour or a pure
exponential decay.
The inset shows a linear-log plot of   $N_B(\ell_B)$.
 }
\label{proteins}
\end{figure}

We also analyze the protein interaction networks of the fruit fly {\it
D. melanogaster} as given in \cite{dm}, the bacterium {\it H. pylori}
\cite{rain}, the baker's yeast {\it S. cerevisiae} \cite{sc},
and the nematode worm {\it C. elegans} \cite{ce},
 which are all available via the DIP database \cite{dip}.  Figure
\ref{proteins} shows the results of $N_B$ versus $\ell_B$
indicating that their behaviour is in
between a pure power-law decay and a
pure exponential.
As with the Internet data, we are able to fit the results
with Eq. (\ref{modified}) with
$\ell_c=7.2$ and $d_B=7.6$ for {\it C. elegans}.
For {\it H. pylori} and {\it D. melanogaster} the fit is a pure exponential
$N_B(\ell_B) \sim \exp(-\ell_B/\ell_e)$ with $\ell_e\approx 1$,
while for {\it S. cerevisiae} the data could be fitted
either by an exponential or by large values of $\ell_c$
and $d_B$ (note that
the exponential is the limit of Eq. (\ref{modified}) for
$\ell_c\to\infty$, $d_B\to\infty$ and $\ell_c/d_B=$ constant).
On the other hand, we observe that for small scales,
$N_B$ seems to display the same power law
found for {\it E. coli} and {\it
H. sapiens}.
The lack of clear self-similarity in these networks
might be due to the incompleteness of these
databases which are continuously being updated with newly discovered
physical interactions \cite{xenarios}.

\section{\bf Random scale-free network}

Next we introduce an example of a model
lacking self-similarity:
the random scale-free model.
This model consists of nodes to which a number of
links are assigned  with a power-law degree distribution and then
connected randomly. Such a network shows a small world effect and
a scale-free property but is not self-similar. We numerically
find that the number of boxes decays exponentially with the box
size (see Fig. \ref{tree}b).
Moreover, while Eq. (\ref{s}) is still valid in this case,
 the power law relation in  Eq. (\ref{sl}) is replaced by an
exponential law.
We conjecture that the reason for
this is a clustering of hubs;
by assigning randomly the connections between the nodes, two nodes
with a large number of links will have a large probability to be
connected.  This induces spatial correlations in the values of $k$
which may explain the breakdown of self-similarity. In contrast,
the simple tree-structure proposed above does not cluster the hubs
by construction.
A summary of our results is presented in Table \ref{table}.

\begin{figure}
\centering
{ \resizebox{10cm}{!}{\includegraphics{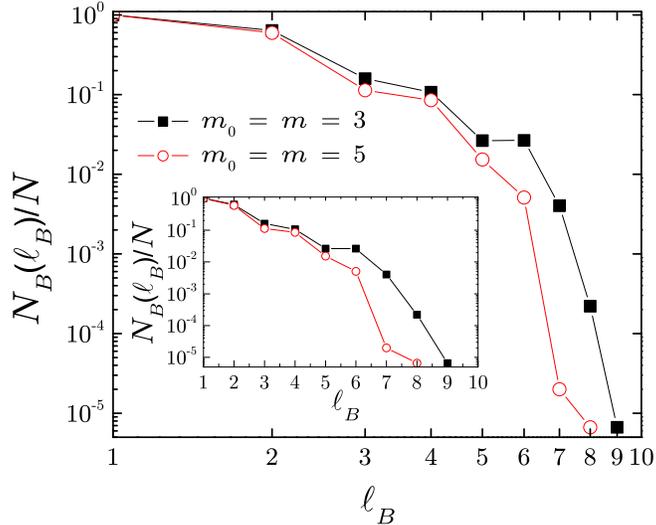}}}
\caption{Barab\'asi-Albert model of scale-free networks with preferential
attachment for 150,000 nodes and
$m=m_0=3$ and $m=m_0=5$. $m_0$ is the initial
number of nodes in the system and $m$ is the number of links
of a newly created node in the dynamical growth of the network \cite{ba}.
Log-log plot of $N_B$ versus $\ell_B$ showing the lack
of a power law behaviour. The inset shows a linear-log
plot indicating that $N_B$ decreases faster than exponential with
$\ell_B$.}
\label{abmodel}
\end{figure}

\begin{figure}
\centering
{ \resizebox{10cm}{!}{\includegraphics{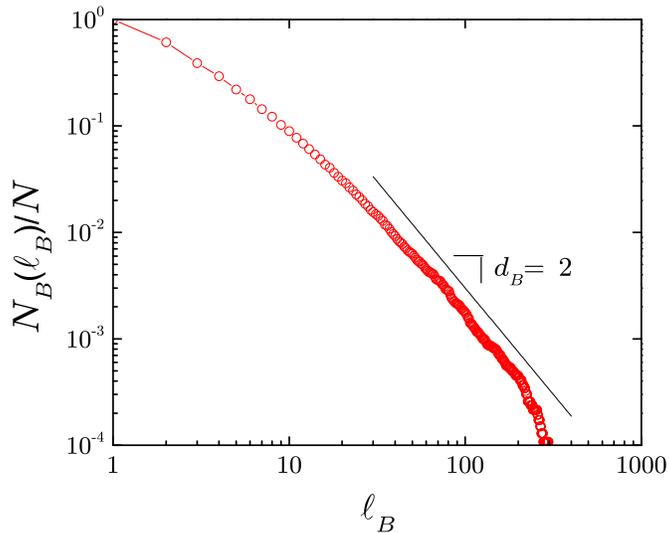}}}
\caption{Erd\"{o}s-R\'enyi random
graph at criticality.
Log-log plot of $N_B$ versus $\ell_B$ showing
the self-similar exponent $d_B=2$ which is obtained
for large distances.}
\label{erc}
\end{figure}

\section{\bf The Barab\'asi-Albert model and the Erd\"{o}s-R\'enyi random
graph at criticality}

We also analyzed the Barab\'asi-Albert model of complex networks
\cite{ba}  (which introduces the concepts of preferential attachment
to describe the dynamics of scale-free
networks). The results of $N_B(\ell_B)$
are shown in Fig. \ref{abmodel} for different parameters in the model
(see \cite{ba} for details) reveling that
the structure is not self-similar; $N_B$ seems to decrease faster than
exponential with $\ell_B$.

It is interesting to compare our results with the random
Erd\"{o}s-R\'enyi graph \cite{erdos,bollobas}
at the critical percolation threshold.
In this case the largest cluster
has self-similar properties and Eq. (\ref{mass_b}),
$\langle M_B (\ell_B)\rangle \sim \ell_B^{d_B}$,
 is valid with $d_B=2$ \cite{braunstein}. We corroborate this result
in Fig. \ref{erc} showing the scaling of the number of
boxes $N_B$ with the
box size $\ell_B$. However,
for this case the network is not small-world since Eq. (\ref{mass_c})
is not valid--- as well as Eq. (\ref{smallworld})--- but rather the
mean distance $\bar{\ell}$ scales as
$\langle M_c\rangle^{1/2}$,  i.e., a power-law
relation rather than the
logarithmic relation characteristic of small world networks.


\begin{table}
\begin{ruledtabular}
\begin{tabular}{lccccc}
Network&$d_B$&$d_k$& $1+d_B/d_k$ & $\gamma$    \\
 & & &  Eq. (\ref{relation}) &  Eq. (\ref{scale-free})  \\
\hline
WWW              & 4.1 & 2.5 & 2.6 & $2.6$ \\
Actor            & 6.3 & 5.3 & 2.2 & $2.2$ \\
{\it E. coli} (PIN)     & 2.3 & 2.1 & 2.1 & $2.2$ \\
{\it H. sapiens} (PIN)    & 2.3 & 2.2 & 2.0 & $2.1$ \\
43 cellular networks  & 3.5   & 3.2  & 2.1 & $2.2$ \\
Scale-free tree  & 3.4 & 2.5 & 2.4 & $2.3$ \\
\end{tabular}
\end{ruledtabular}
\caption{\label{table}
Summary of the exponents  obtained for the scale-invariant networks
studied in the manuscript.}
\end{table}


\newpage
\clearpage

\section{\bf Cellular networks}

The WIT database \cite{cellular}
(http://igweb.integratedgenomics.com/IGwit)
of cellular networks
considers the cellular functions divided according to
bioengineering principles containing datasets for
intermediate metabolism and bioenergetics (core metabolism),
information pathways,  electron transport, and  transmembrane transport.
The metabolic network is a subset of all reactions that take place in
the cell. Since this is the largest part of the network
we analyze it separately and compare
it with the full biochemical reaction network.
The data presented in Fig.  \ref{fractal}c represents the full
biochemical reaction networks of only three substrates.
Here we present results of  the 43 different substrates represented
in the database for the metabolic and full networks.
The following figures show the results of $N_B$ vs $\ell_B$.
Both the metabolic and full networks display the power law relationship
of self-similar networks with the same
exponent
(within error bars)
for all the organisms  considered (the metabolic networks show
a finite size effect due to their smaller size).
We find an average
$d_B=3.5$. The solid line in the figures
represent the average fit.
The values are
reported in Table \ref{table}.

\begin{center}
\begin{minipage}[c]{0.33\textwidth}
\centering Aquifex aeolicus\\\includegraphics[width=5.9cm]{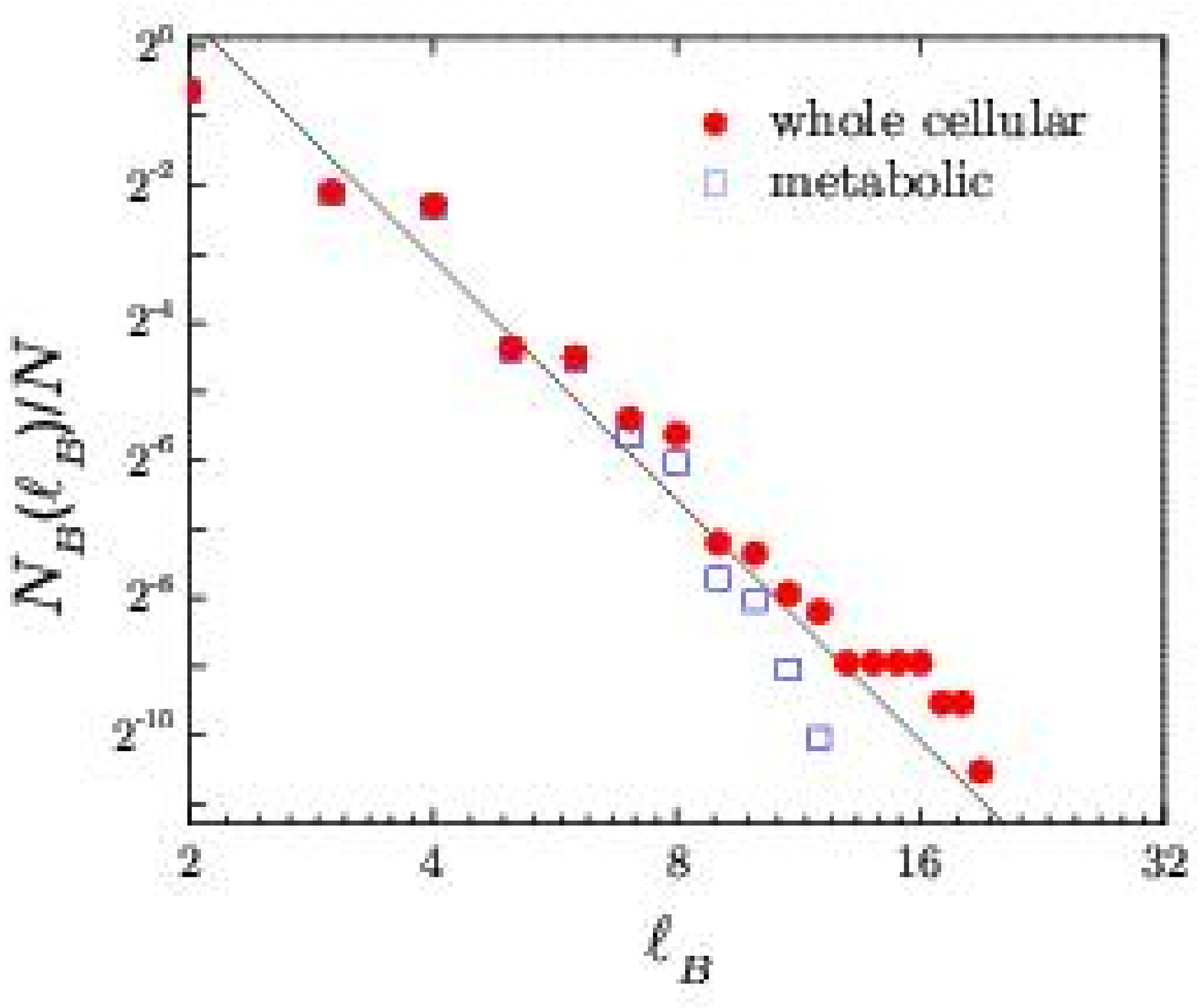}
\end{minipage}%
\begin{minipage}[c]{0.33\textwidth}
\centering Actinobacillus
actinomycetemcomitans\\\includegraphics[width=5.9cm]{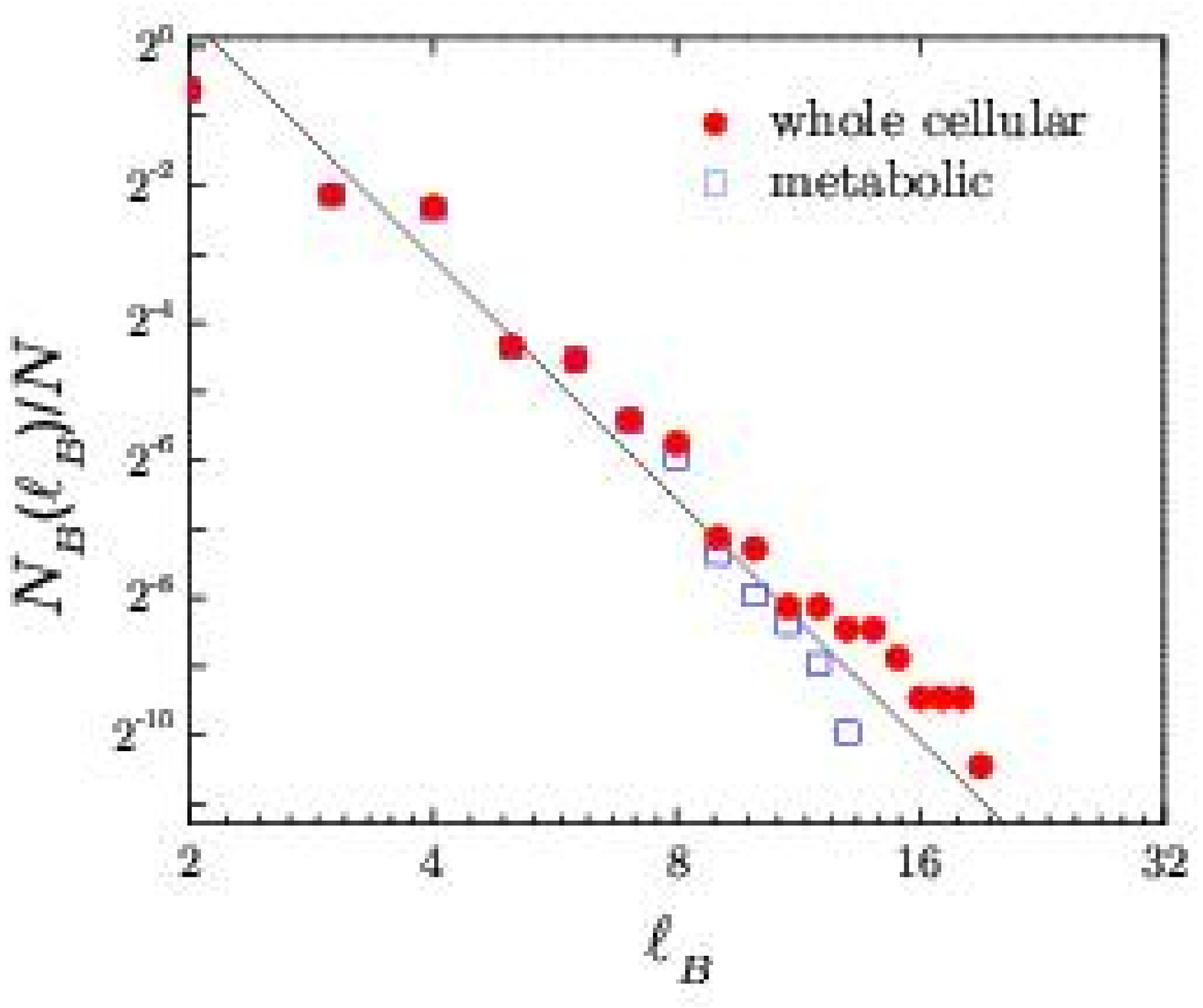}
\end{minipage}%
\begin{minipage}[c]{0.33\textwidth}
\centering Archaeoglobus
fulgidus\\\includegraphics[width=5.9cm]{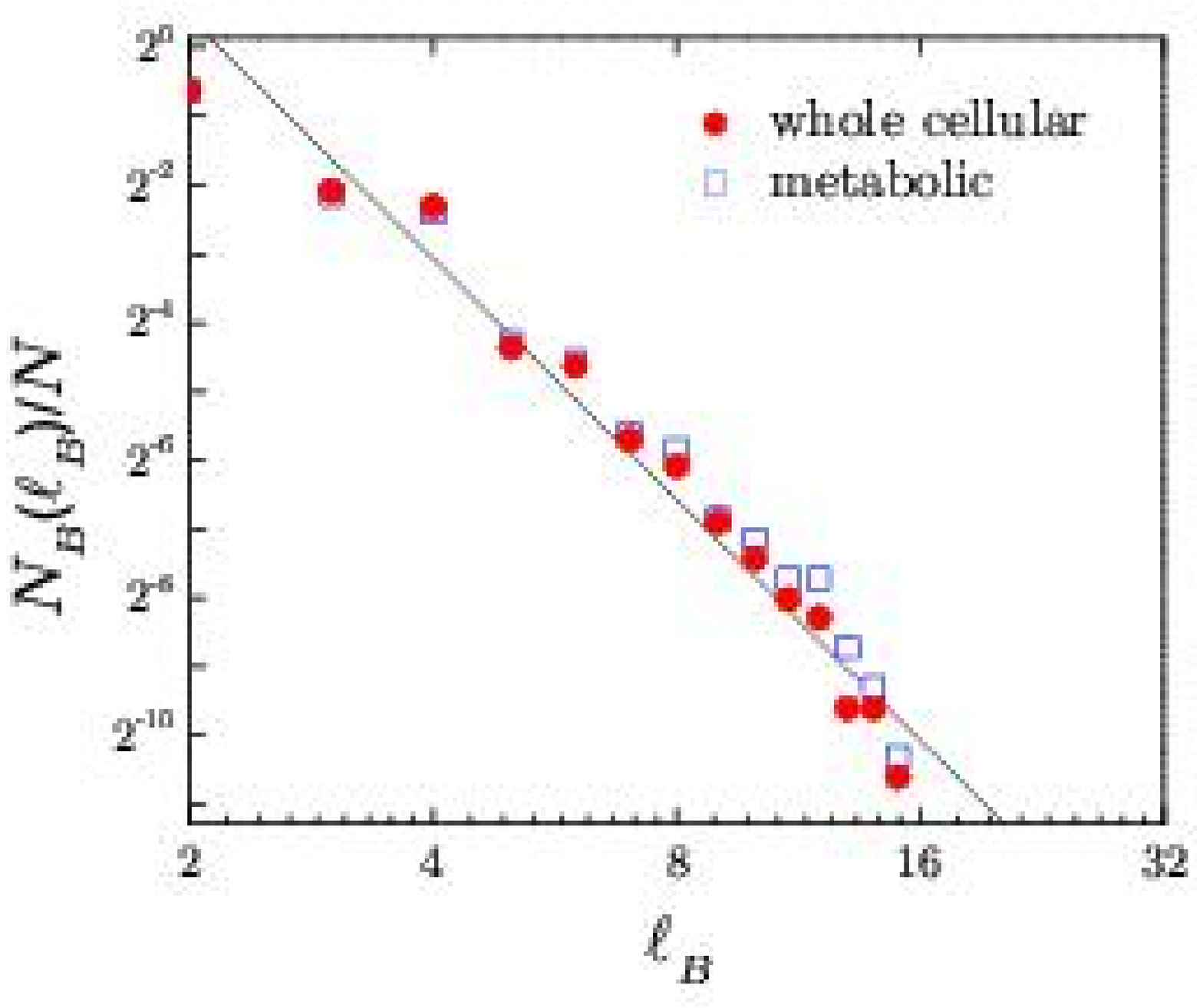}
\end{minipage}%
\end{center}

\begin{center}
\begin{minipage}[c]{0.33\textwidth}
\centering Aeropyrum pernix\\\includegraphics[width=5.9cm]{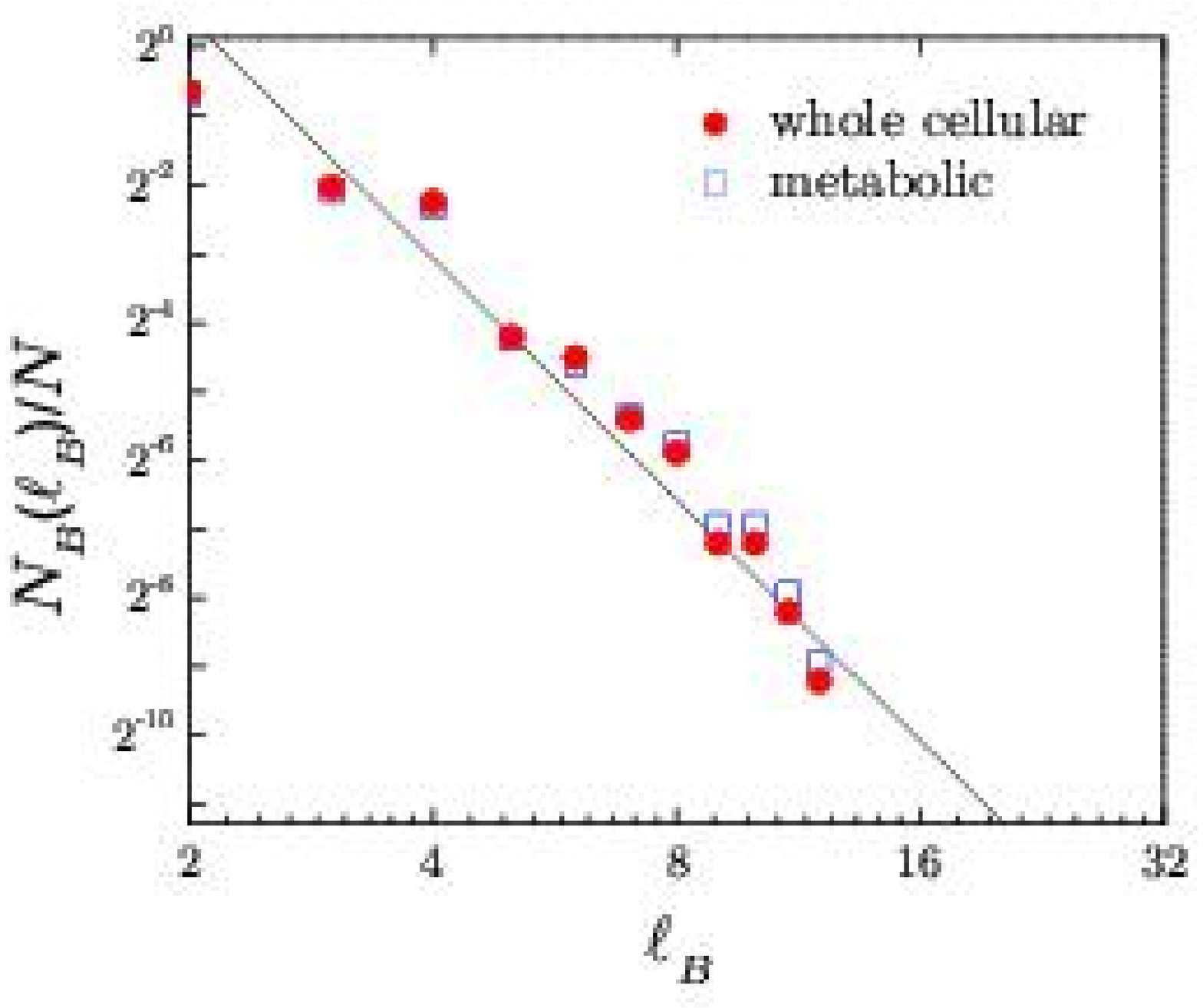}
\end{minipage}%
\begin{minipage}[c]{0.33\textwidth}
\centering Arabidopsis
thaliana\\\includegraphics[width=5.9cm]{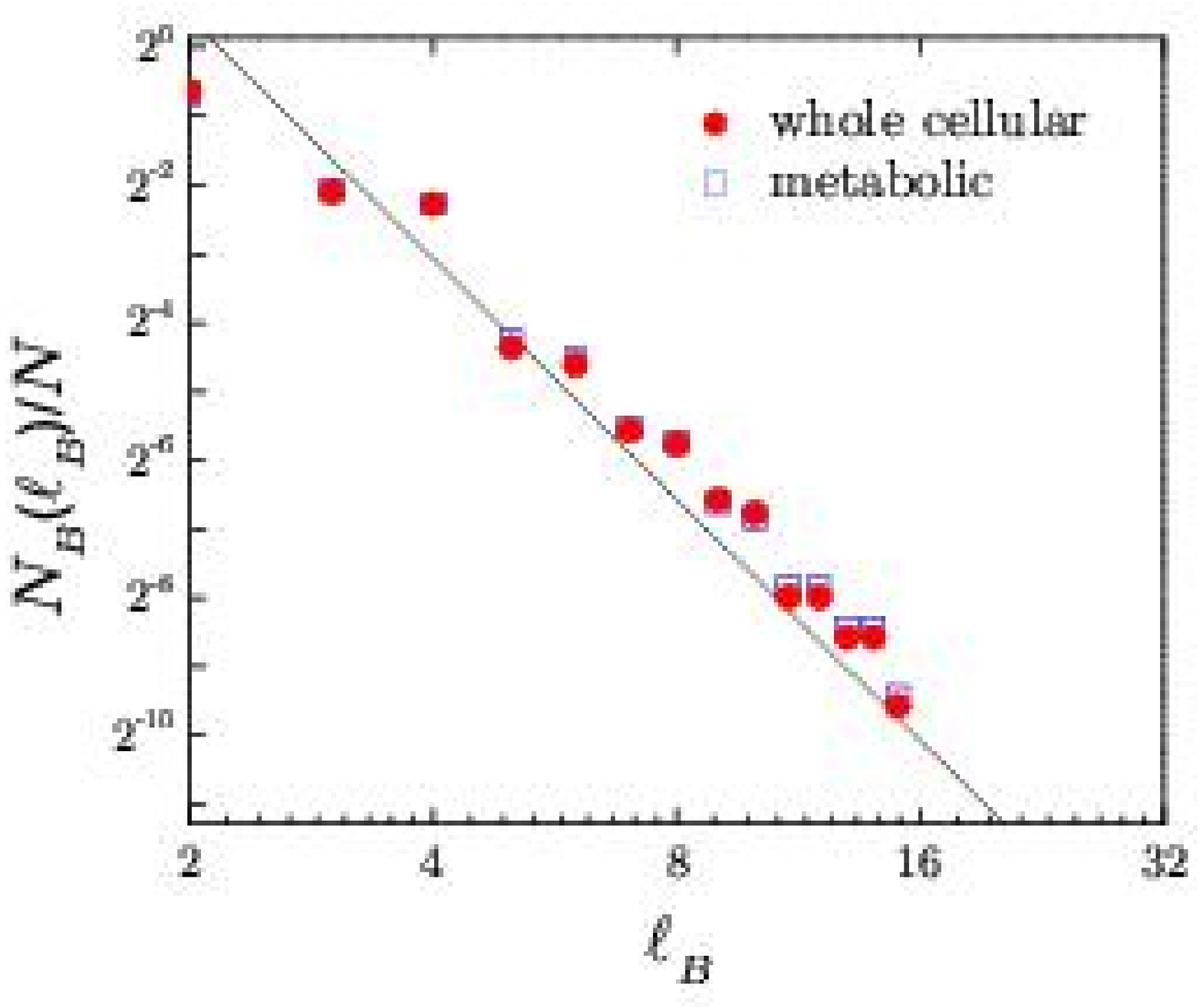}
\end{minipage}%
\begin{minipage}[c]{0.33\textwidth}
\centering Borrelia
burgdorferi\\\includegraphics[width=5.9cm]{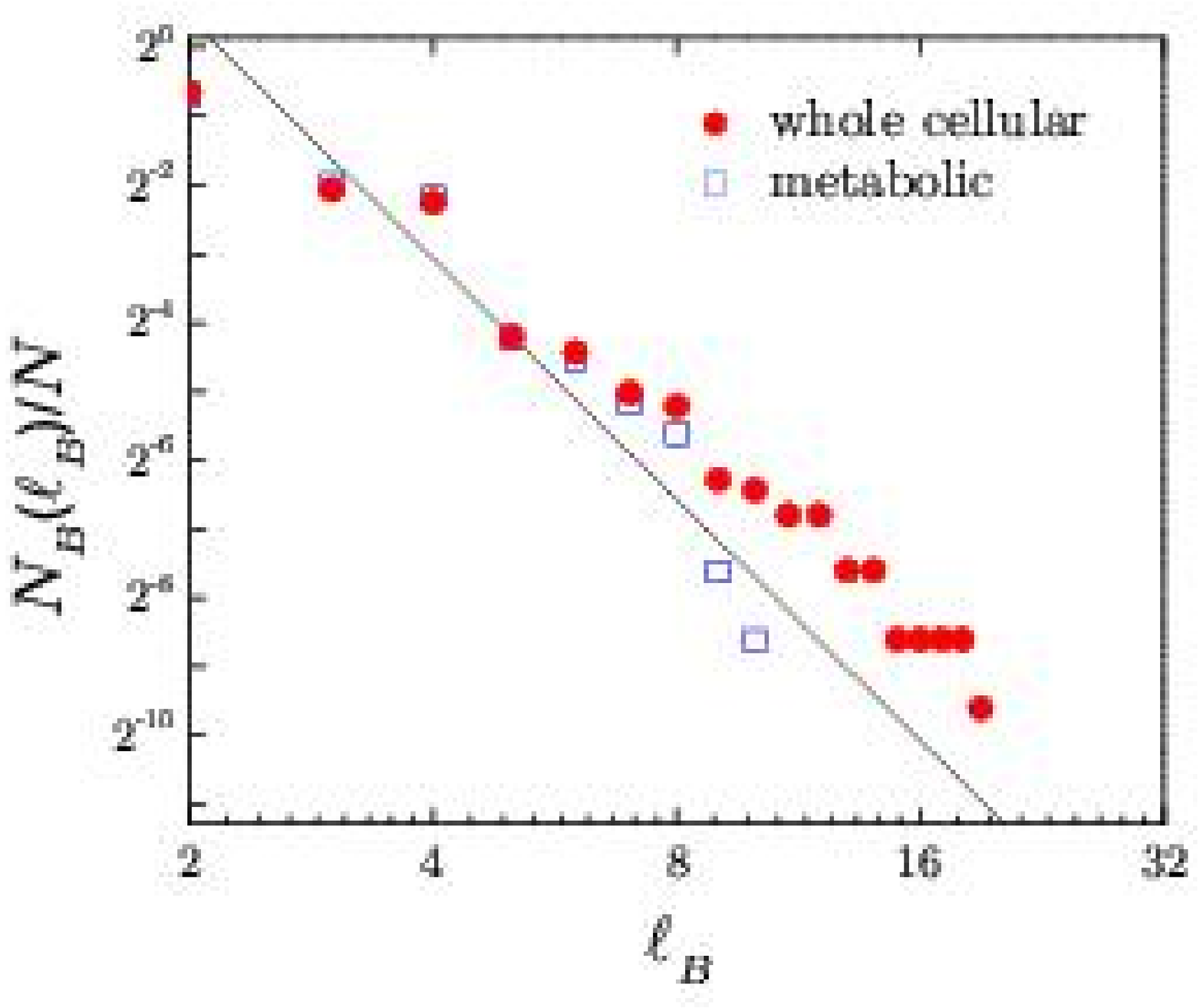}
\end{minipage}%
\end{center}

\begin{center}
\begin{minipage}[c]{0.33\textwidth}
\centering Bacillus
subtilis\\\includegraphics[width=5.9cm]{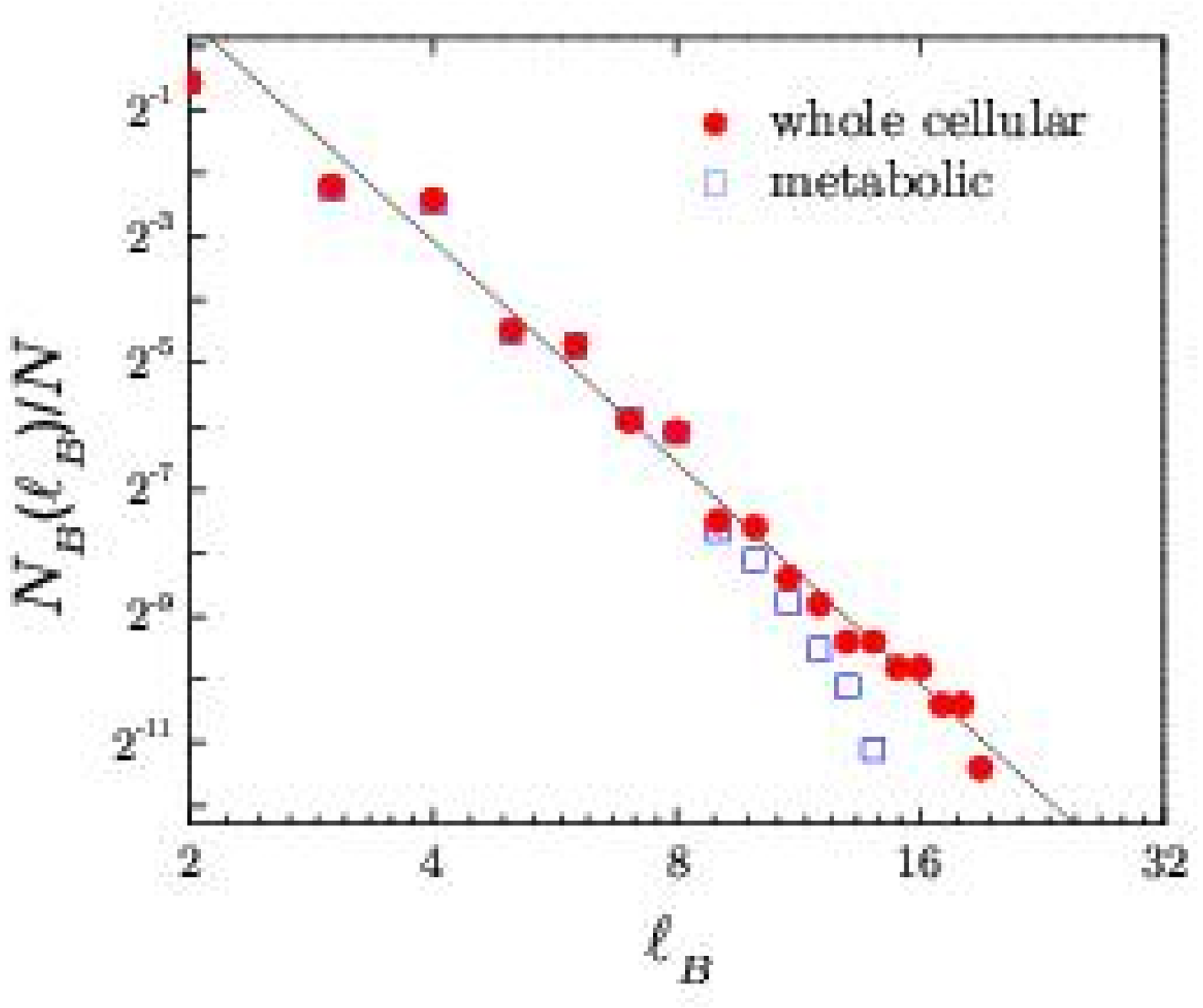}
\end{minipage}%
\begin{minipage}[c]{0.33\textwidth}
\centering Clostridium
acetobutylicum\\\includegraphics[width=5.9cm]{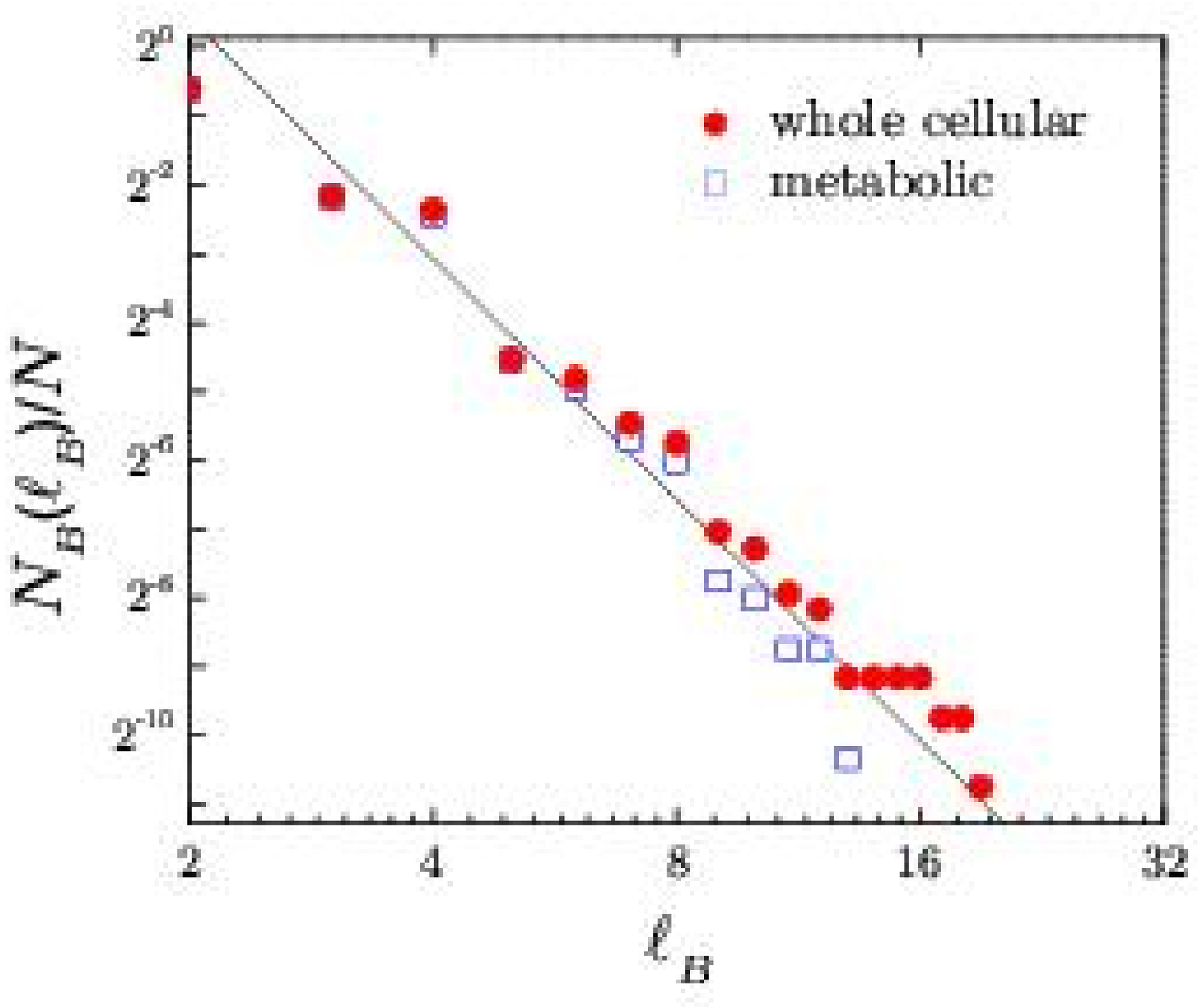}
\end{minipage}%
\begin{minipage}[c]{0.33\textwidth}
\centering Caenorhabditis
elegans\\\includegraphics[width=5.9cm]{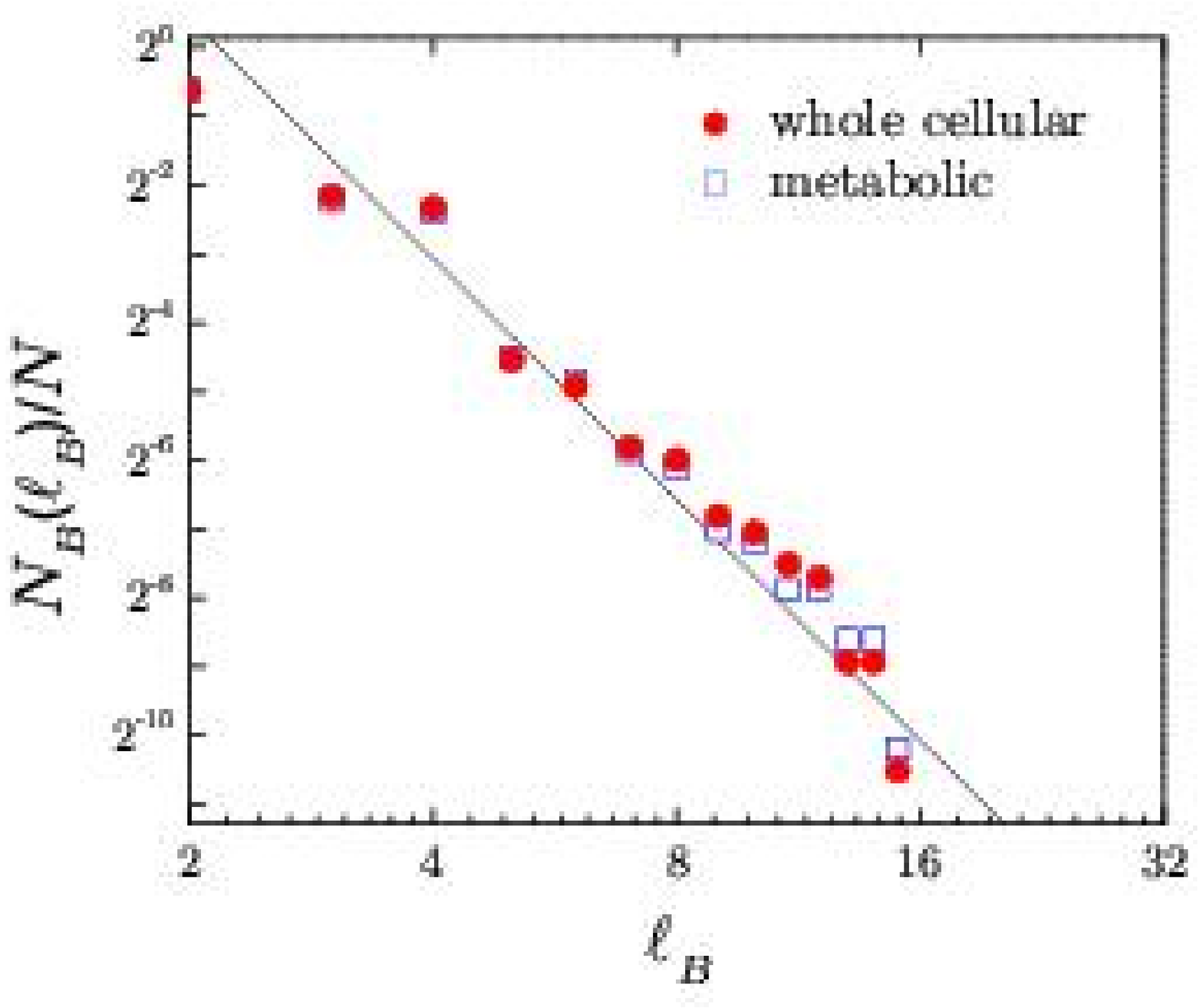}
\end{minipage}%
\end{center}

\begin{center}
\begin{minipage}[c]{0.33\textwidth}
\centering Campylobacter
jejuni\\\includegraphics[width=5.9cm]{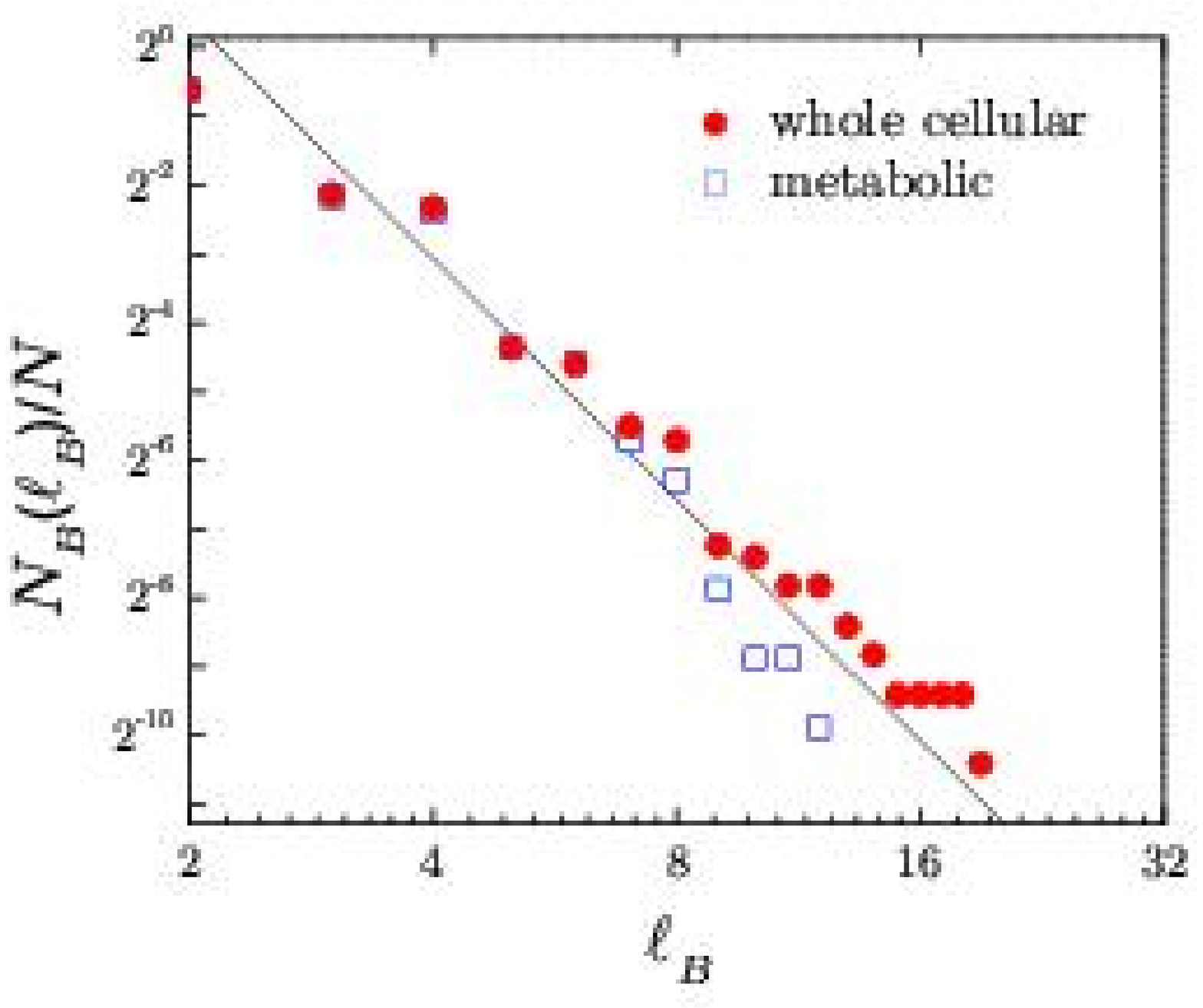}
\end{minipage}%
\begin{minipage}[c]{0.33\textwidth}
\centering Chlorobium
tepidum\\\includegraphics[width=5.9cm]{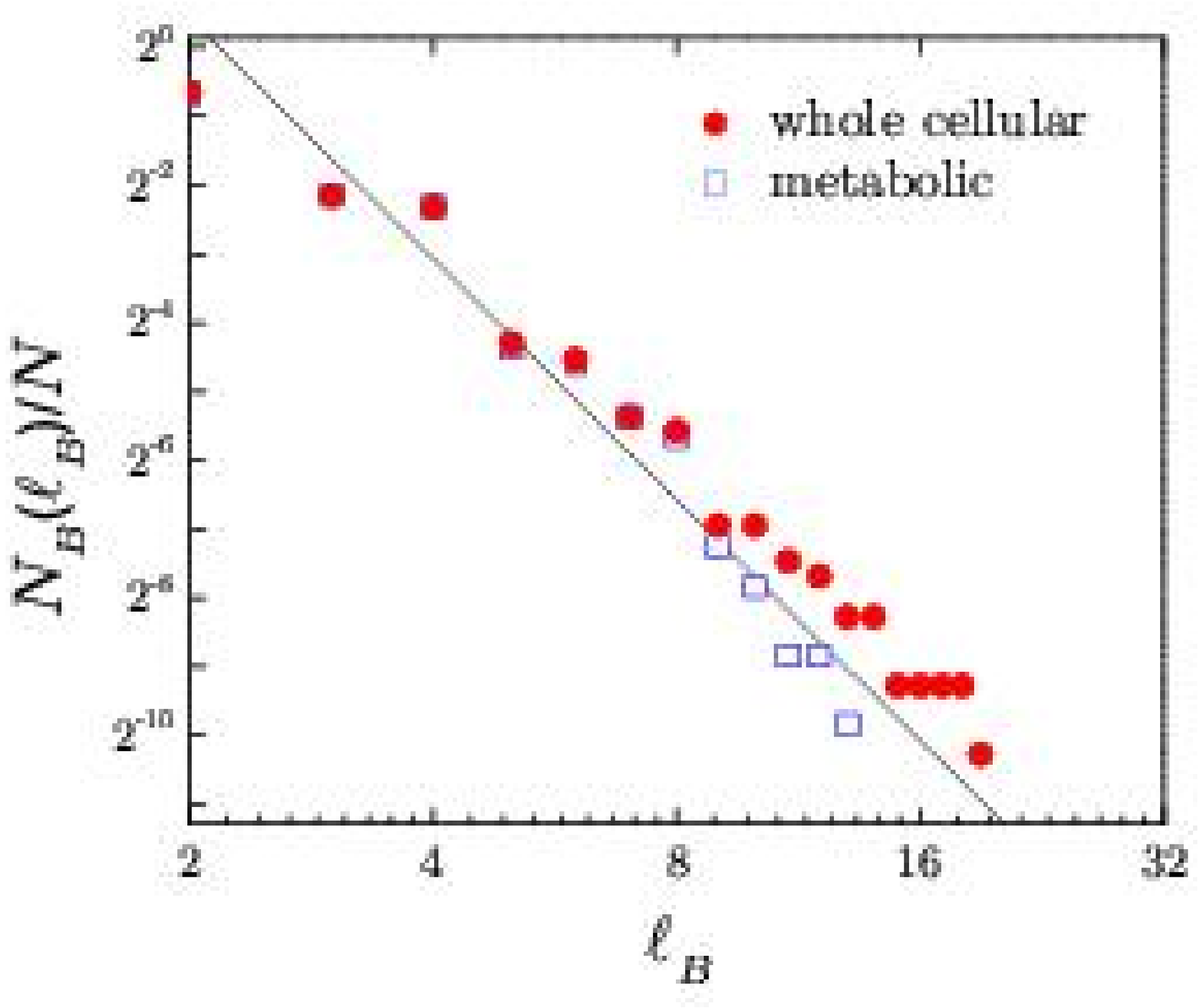}
\end{minipage}%
\begin{minipage}[c]{0.33\textwidth}
\centering Chlamydia
pneumoniae\\\includegraphics[width=5.9cm]{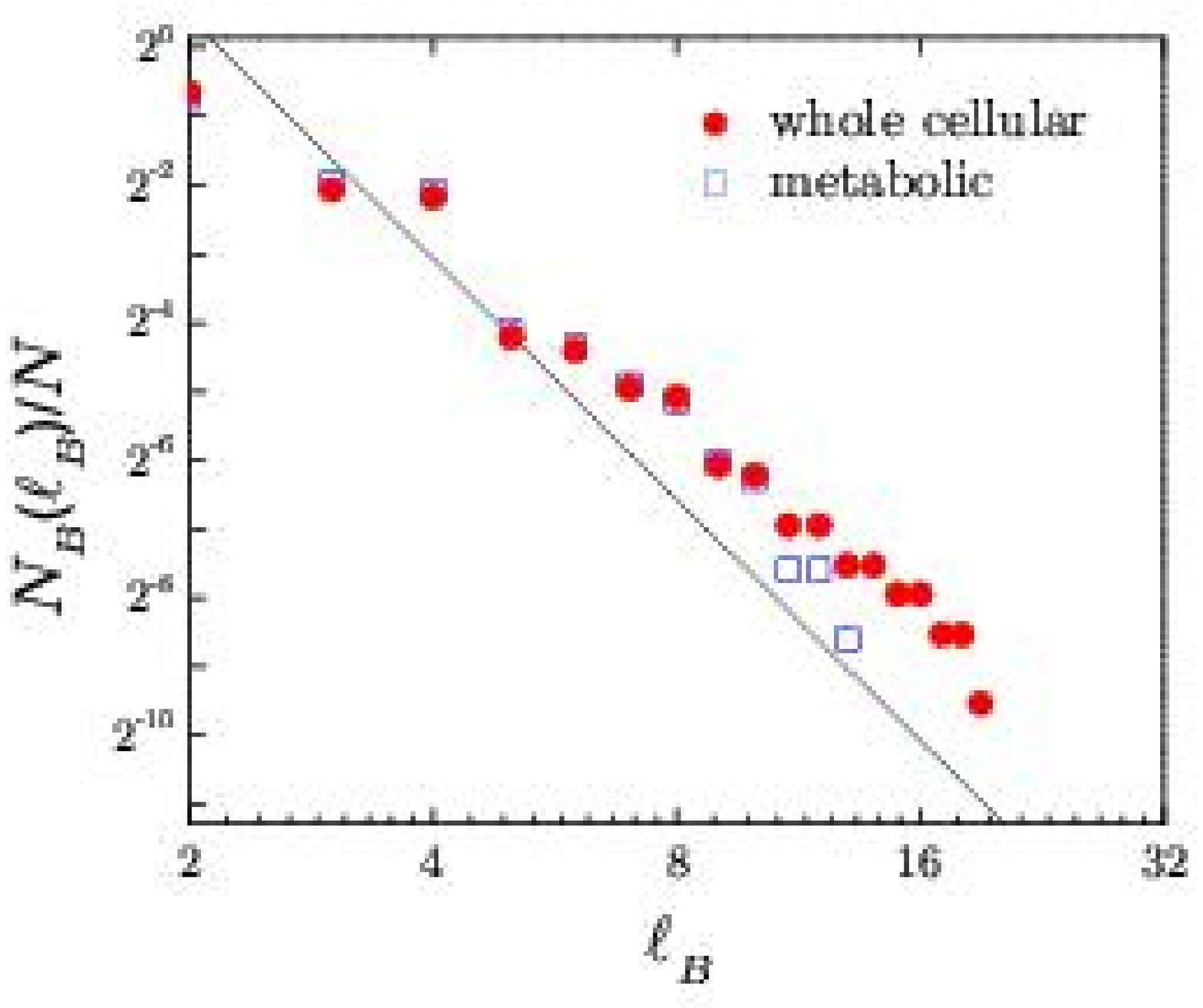}
\end{minipage}%
\end{center}

\begin{center}
\begin{minipage}[c]{0.33\textwidth}
\centering Chlamydia
trachomatis\\\includegraphics[width=5.9cm]{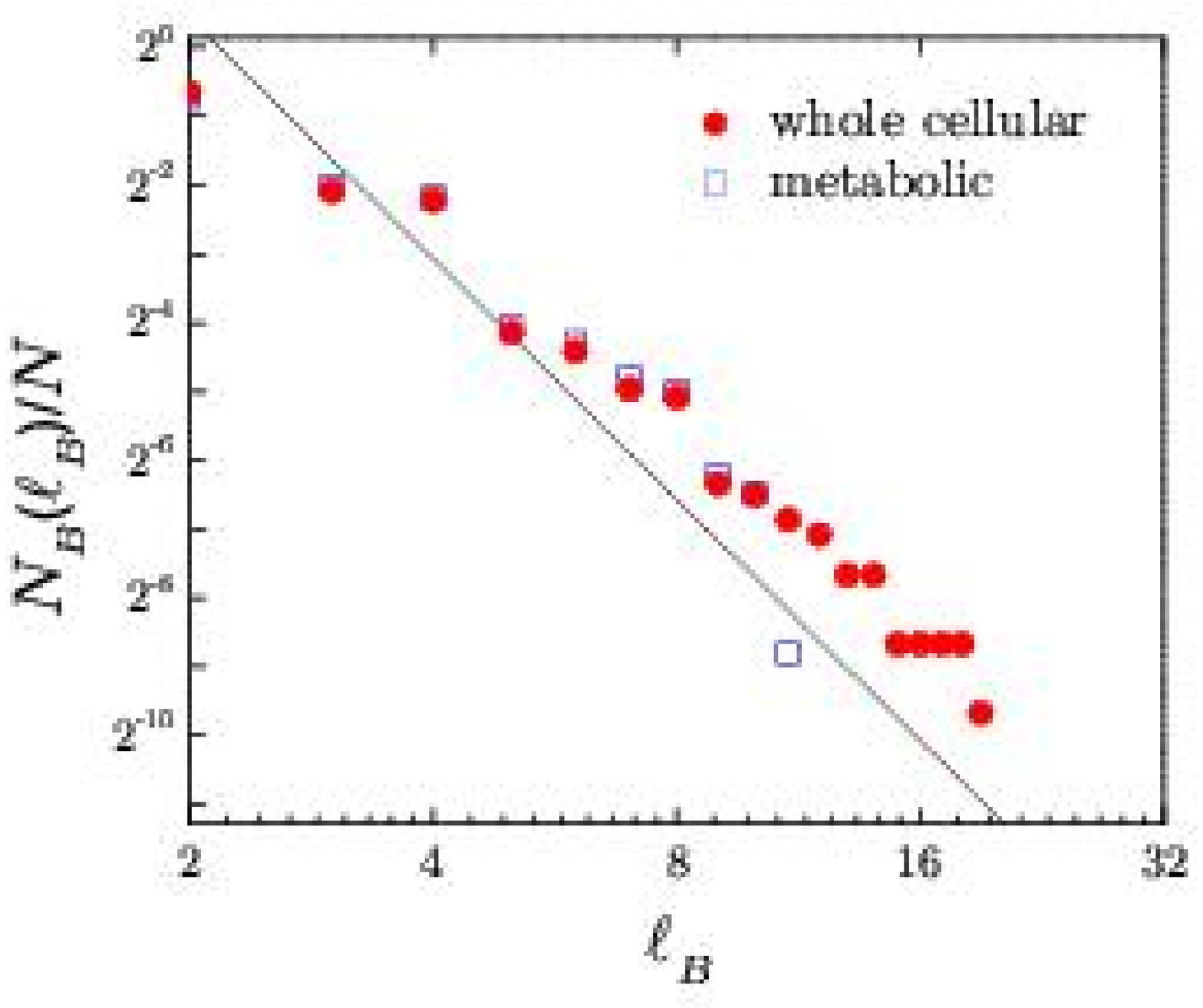}
\end{minipage}%
\begin{minipage}[c]{0.33\textwidth}
\centering Synechocystis
sp.\\\includegraphics[width=5.9cm]{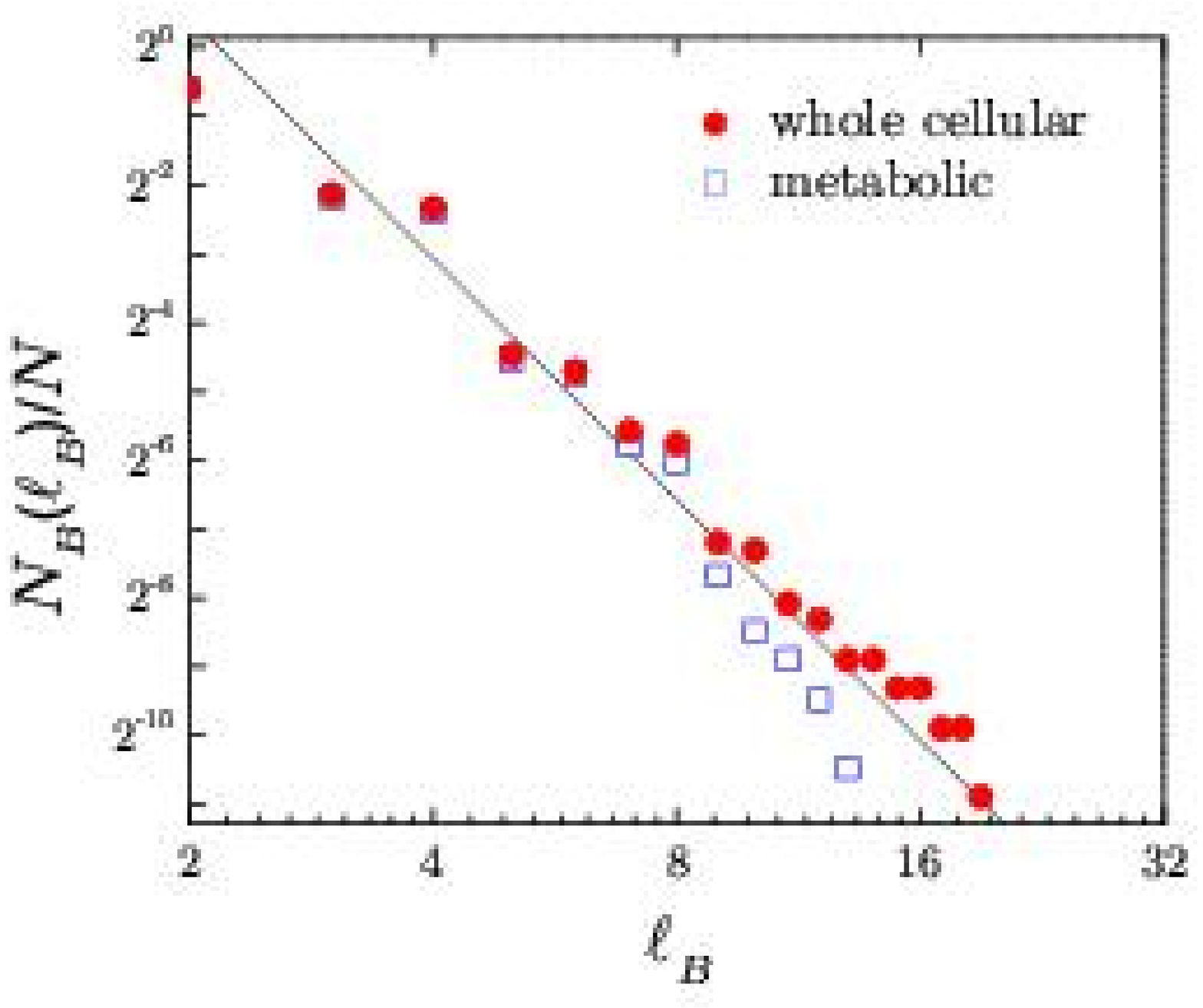}
\end{minipage}%
\begin{minipage}[c]{0.33\textwidth}
\centering Deinococcus
radiodurans\\\includegraphics[width=5.9cm]{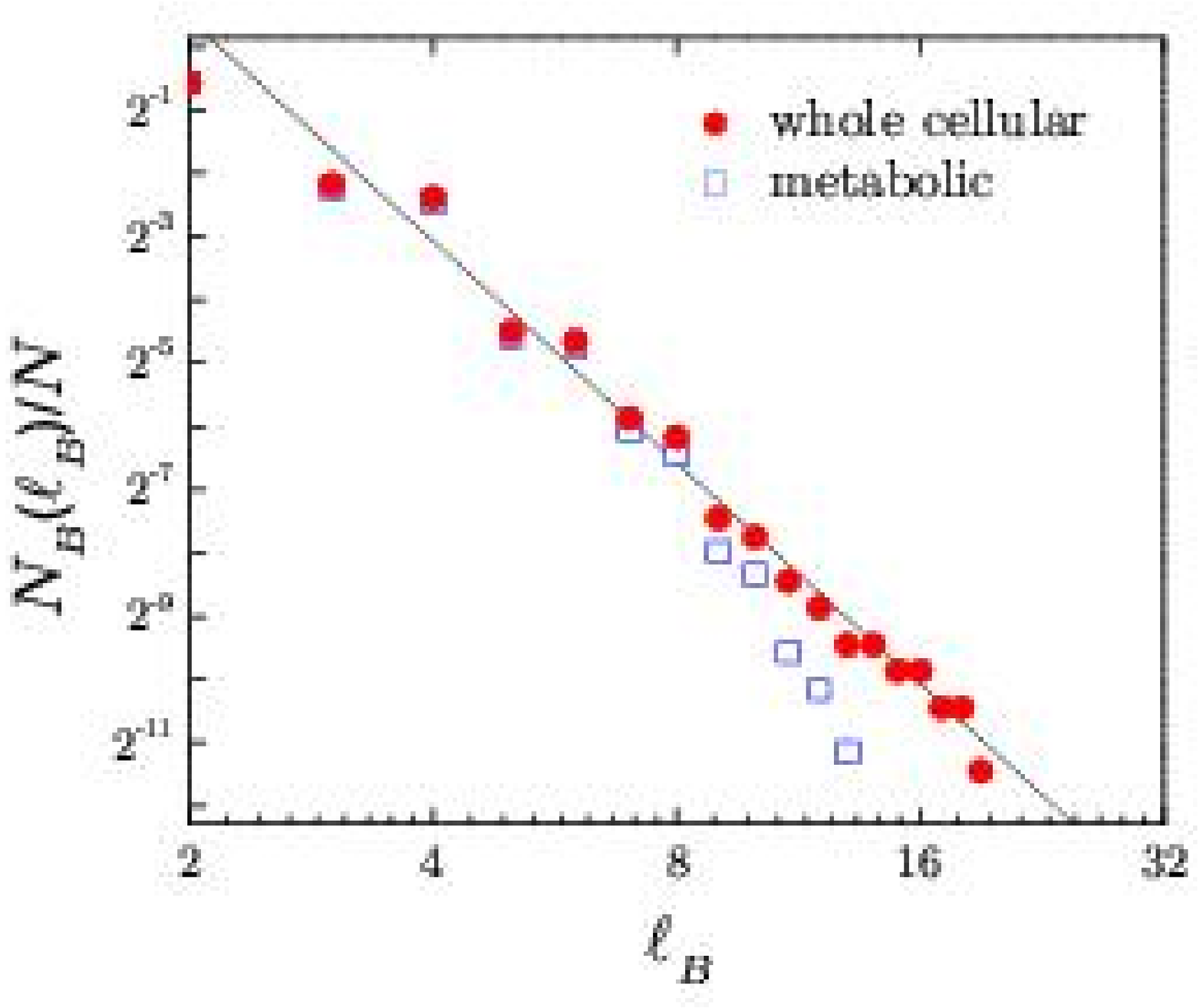}
\end{minipage}%
\end{center}

\begin{center}
\begin{minipage}[c]{0.33\textwidth}
\centering Escherichia coli\\\includegraphics[width=5.9cm]{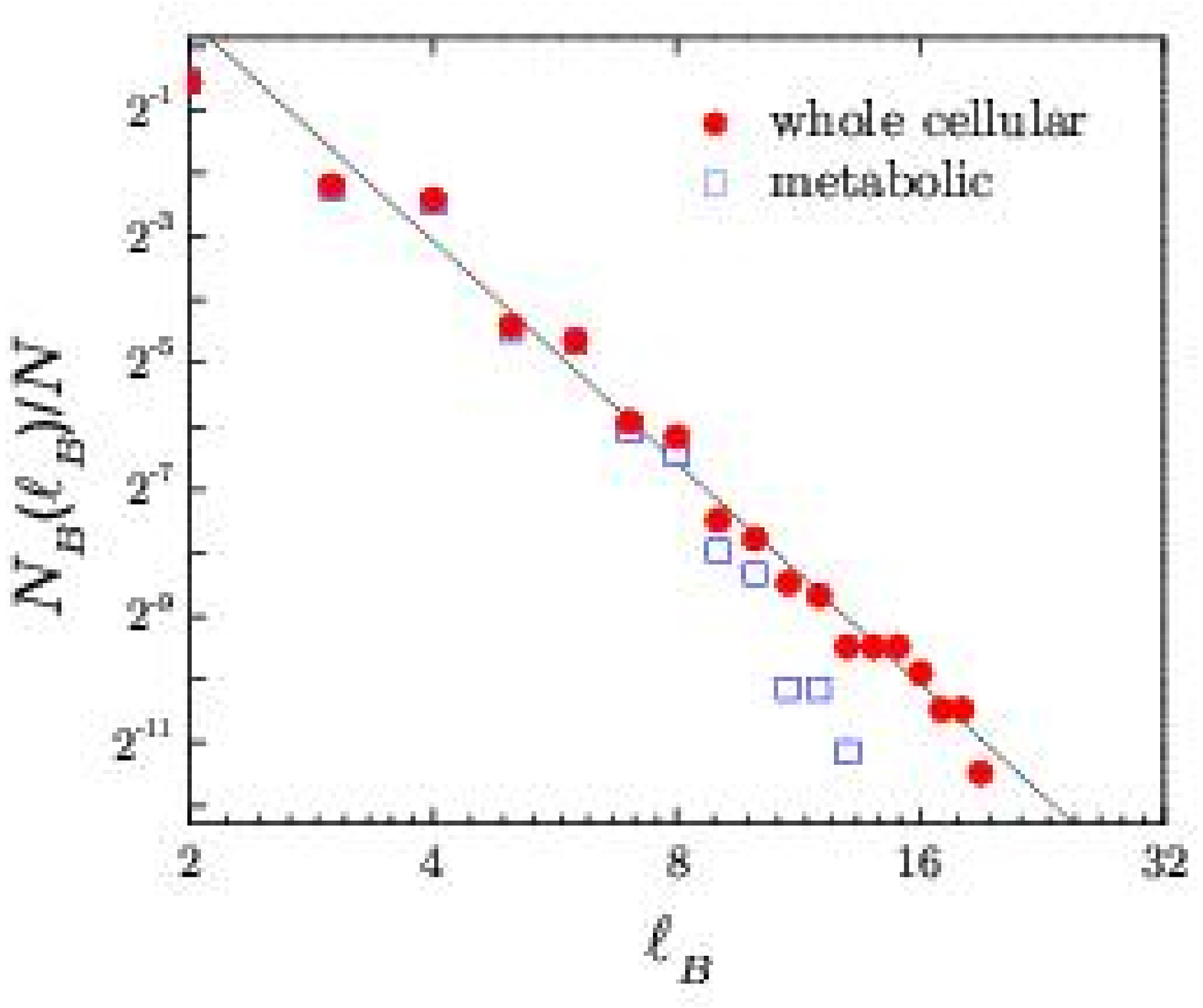}
\end{minipage}%
\begin{minipage}[c]{0.33\textwidth}
\centering Enterococcus
faecalis\\\includegraphics[width=5.9cm]{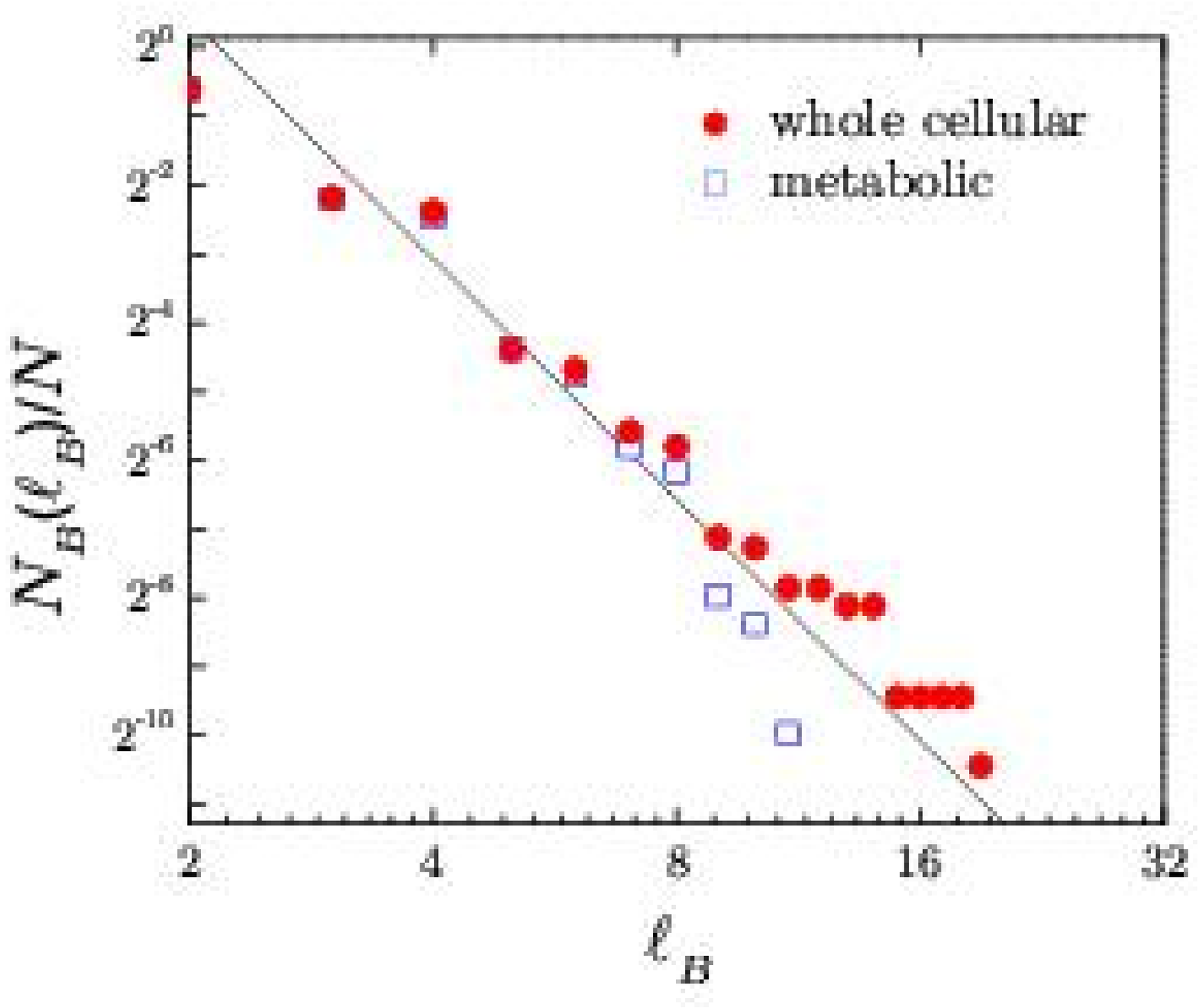}
\end{minipage}%
\begin{minipage}[c]{0.33\textwidth}
\centering Emericella
nidulans\\\includegraphics[width=5.9cm]{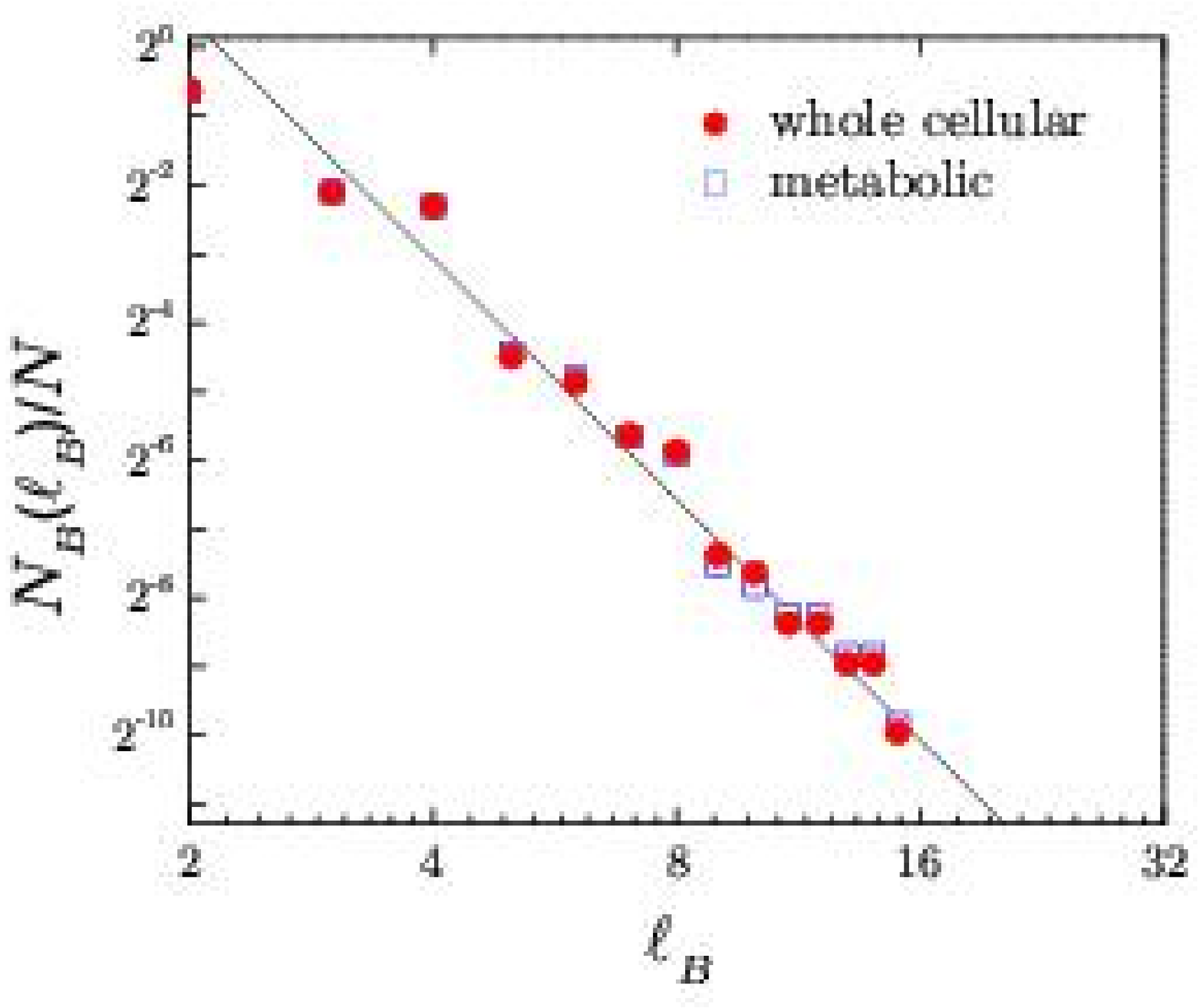}
\end{minipage}%
\end{center}

\begin{center}
\begin{minipage}[c]{0.33\textwidth}
\centering Haemophilus
influenzae\\\includegraphics[width=5.9cm]{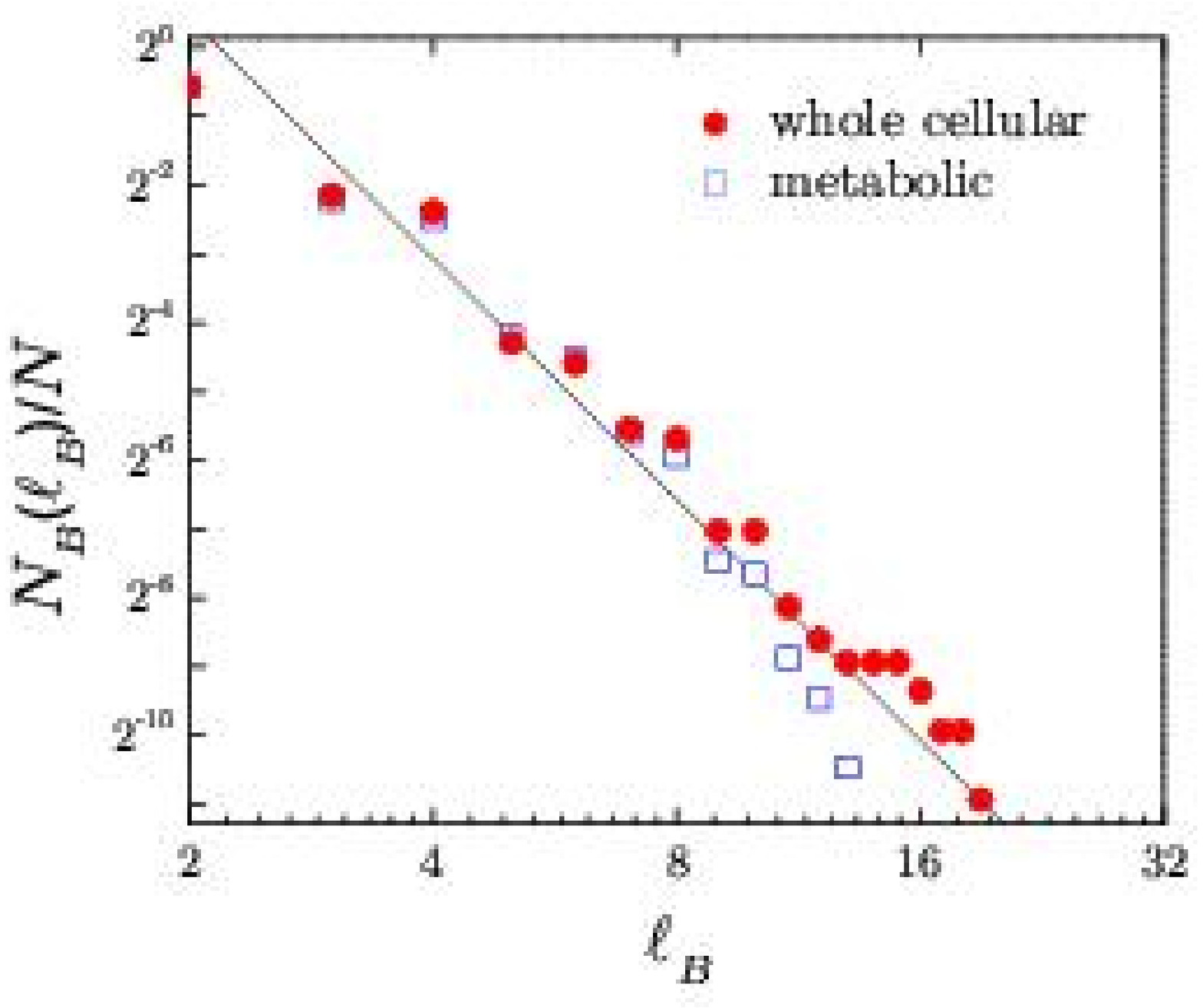}
\end{minipage}%
\begin{minipage}[c]{0.33\textwidth}
\centering Helicobacter
pylori\\\includegraphics[width=5.9cm]{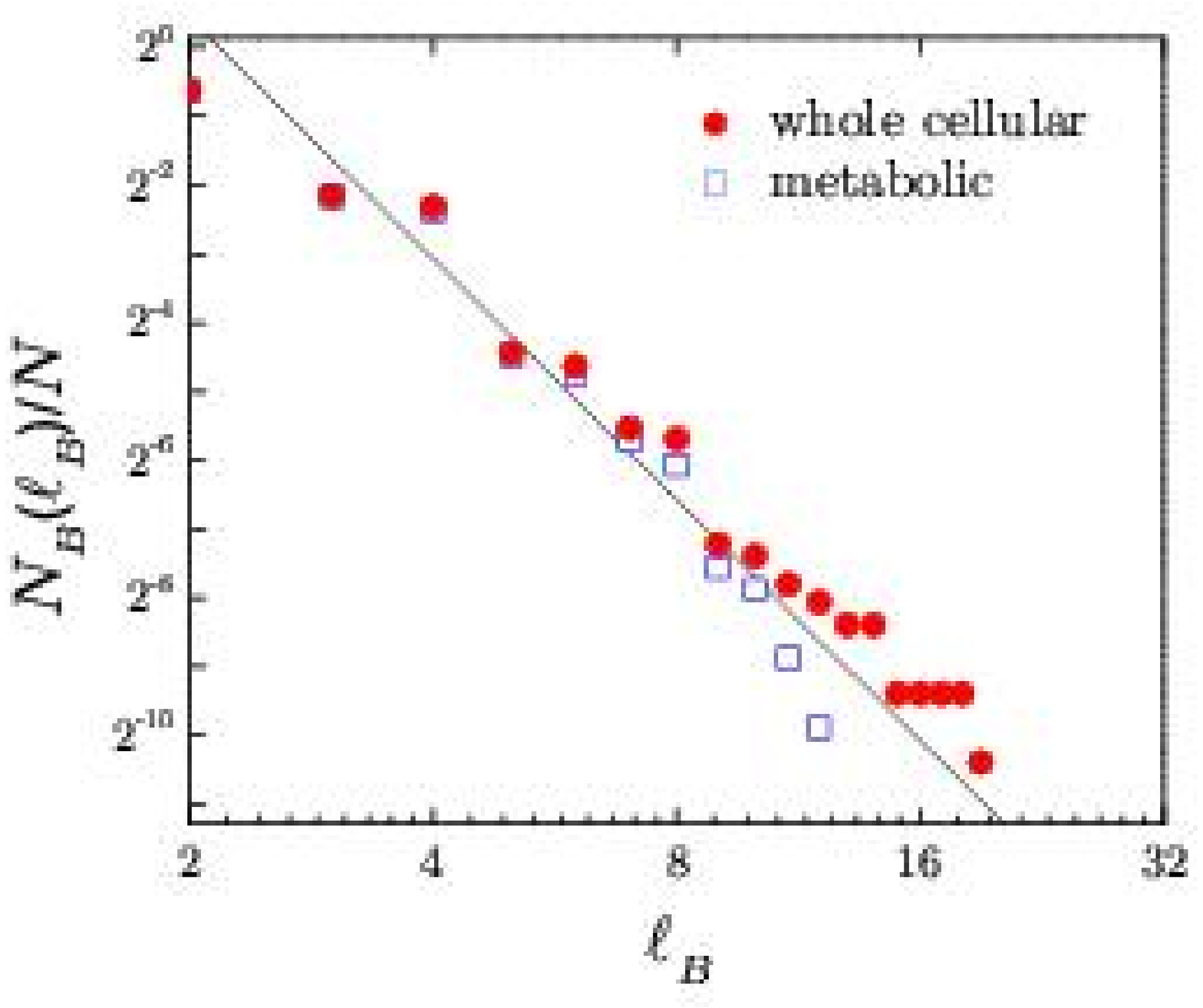}
\end{minipage}%
\begin{minipage}[c]{0.33\textwidth}
\centering Mycobacterium
bovis\\\includegraphics[width=5.9cm]{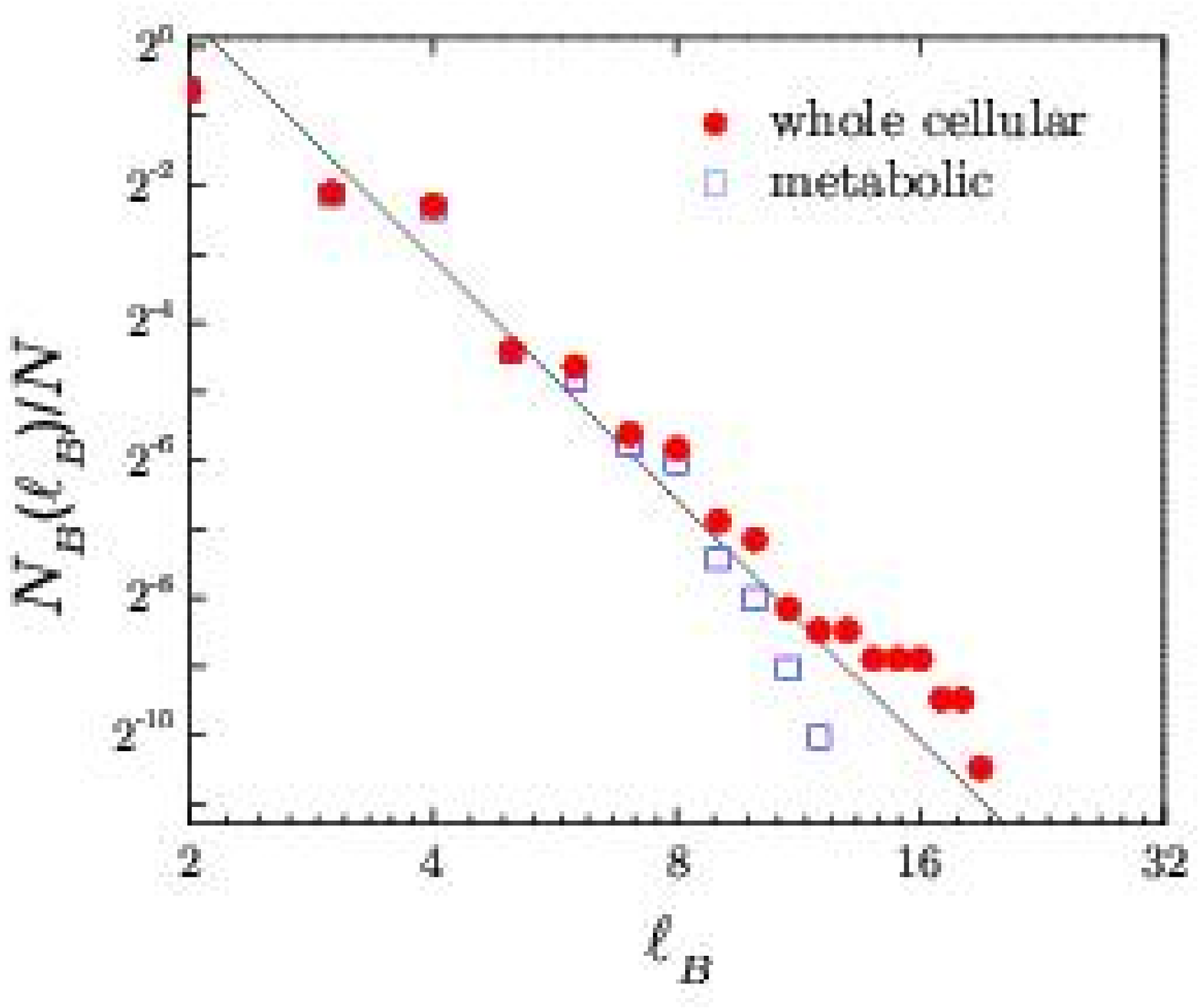}
\end{minipage}%
\end{center}

\begin{center}
\begin{minipage}[c]{0.33\textwidth}
\centering Mycoplasma
genitalium\\\includegraphics[width=5.9cm]{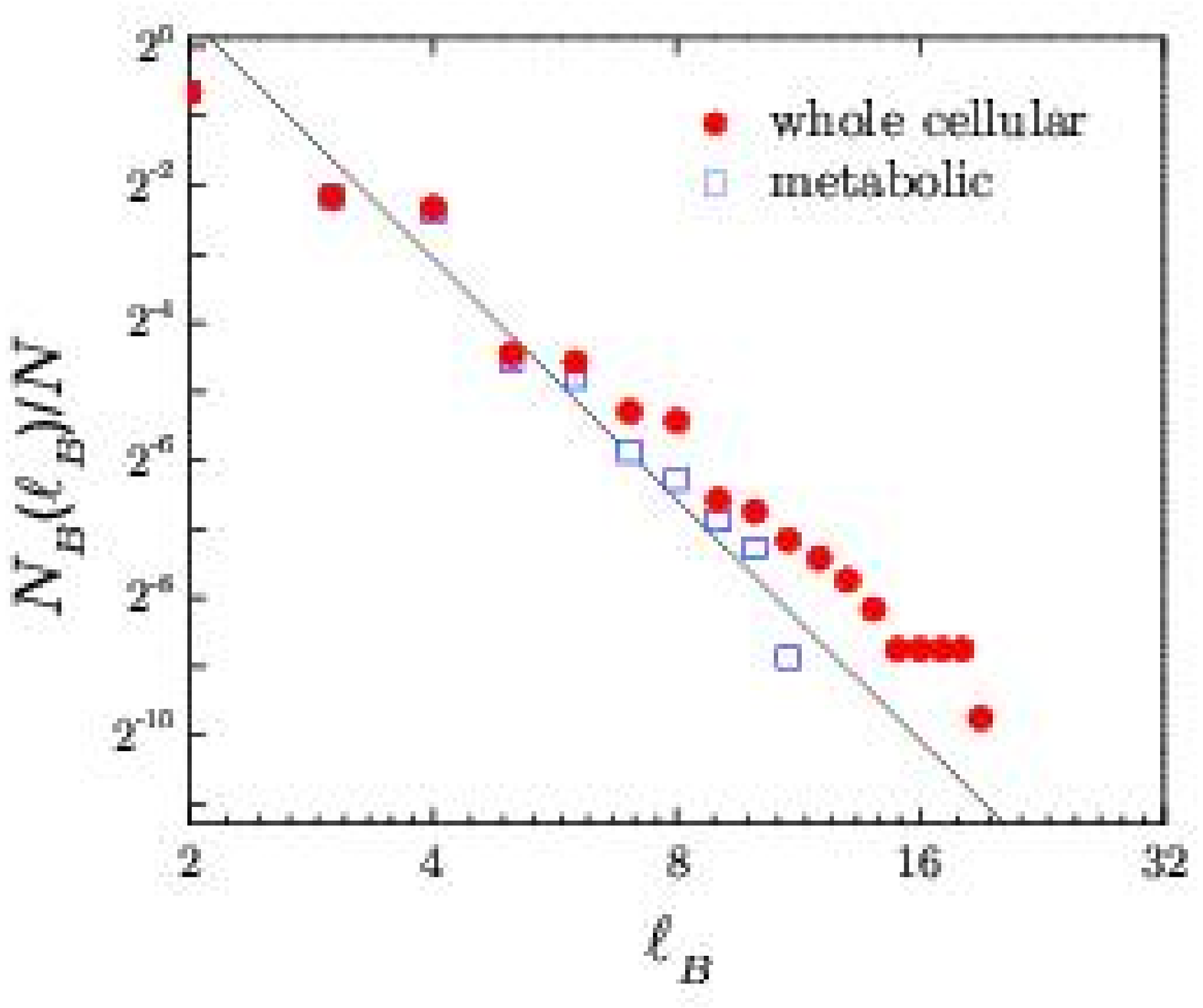}
\end{minipage}%
\begin{minipage}[c]{0.33\textwidth}
\centering Methanococcus
jannaschii\\\includegraphics[width=5.9cm]{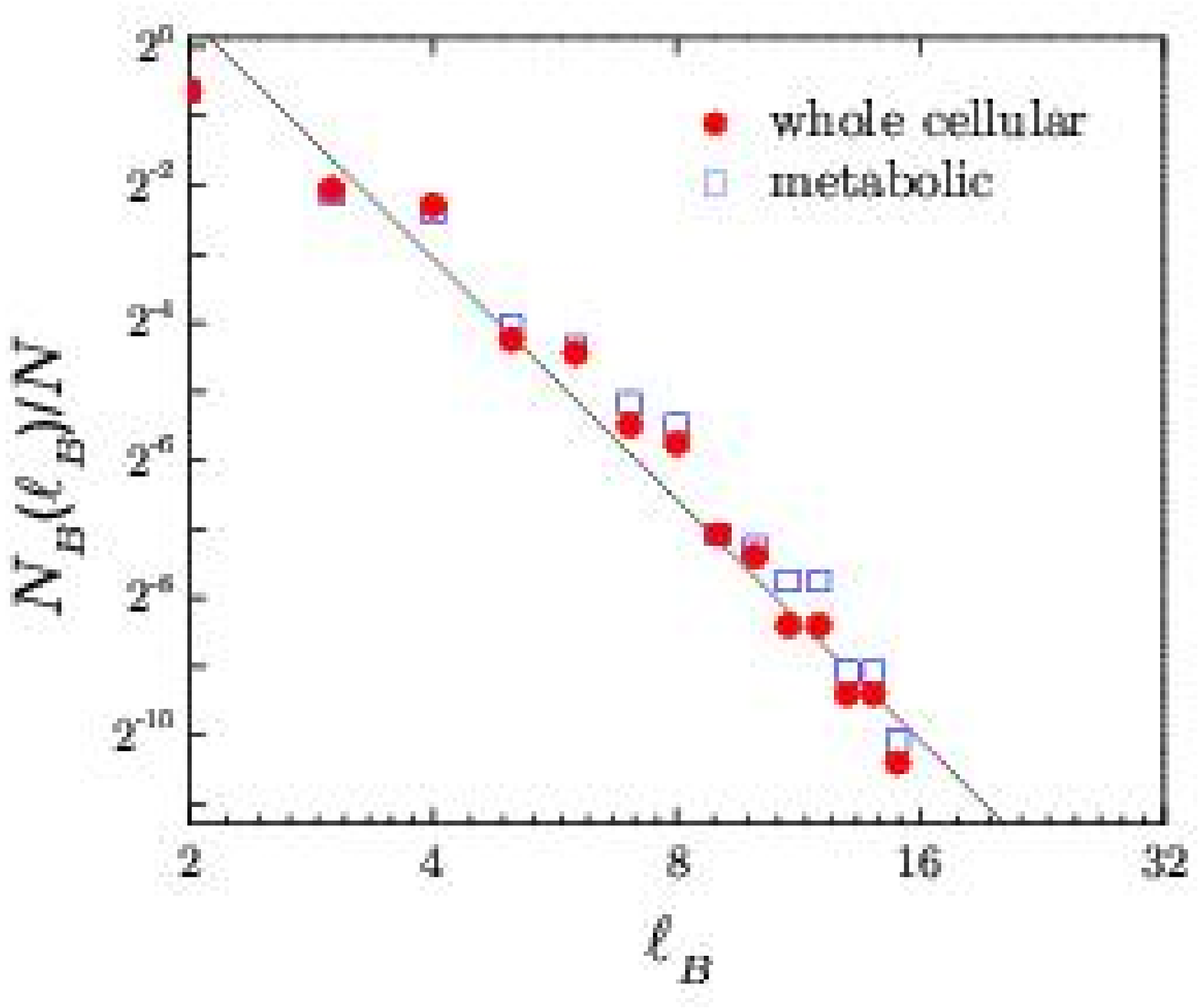}
\end{minipage}%
\begin{minipage}[c]{0.33\textwidth}
\centering Mycobacterium
leprae\\\includegraphics[width=5.9cm]{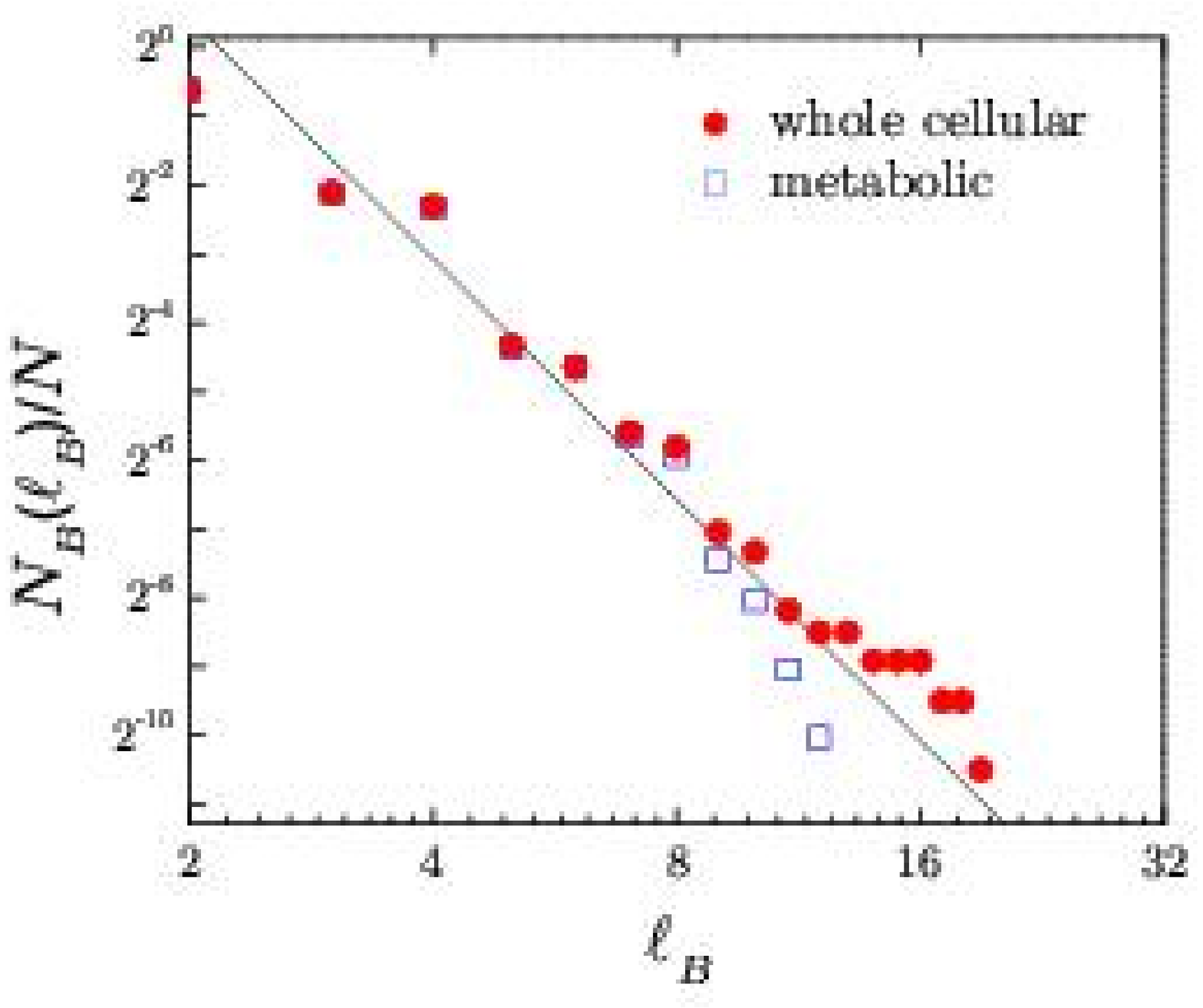}
\end{minipage}%
\end{center}

\begin{center}
\begin{minipage}[c]{0.33\textwidth}
\centering Mycoplasma
pneumoniae\\\includegraphics[width=5.9cm]{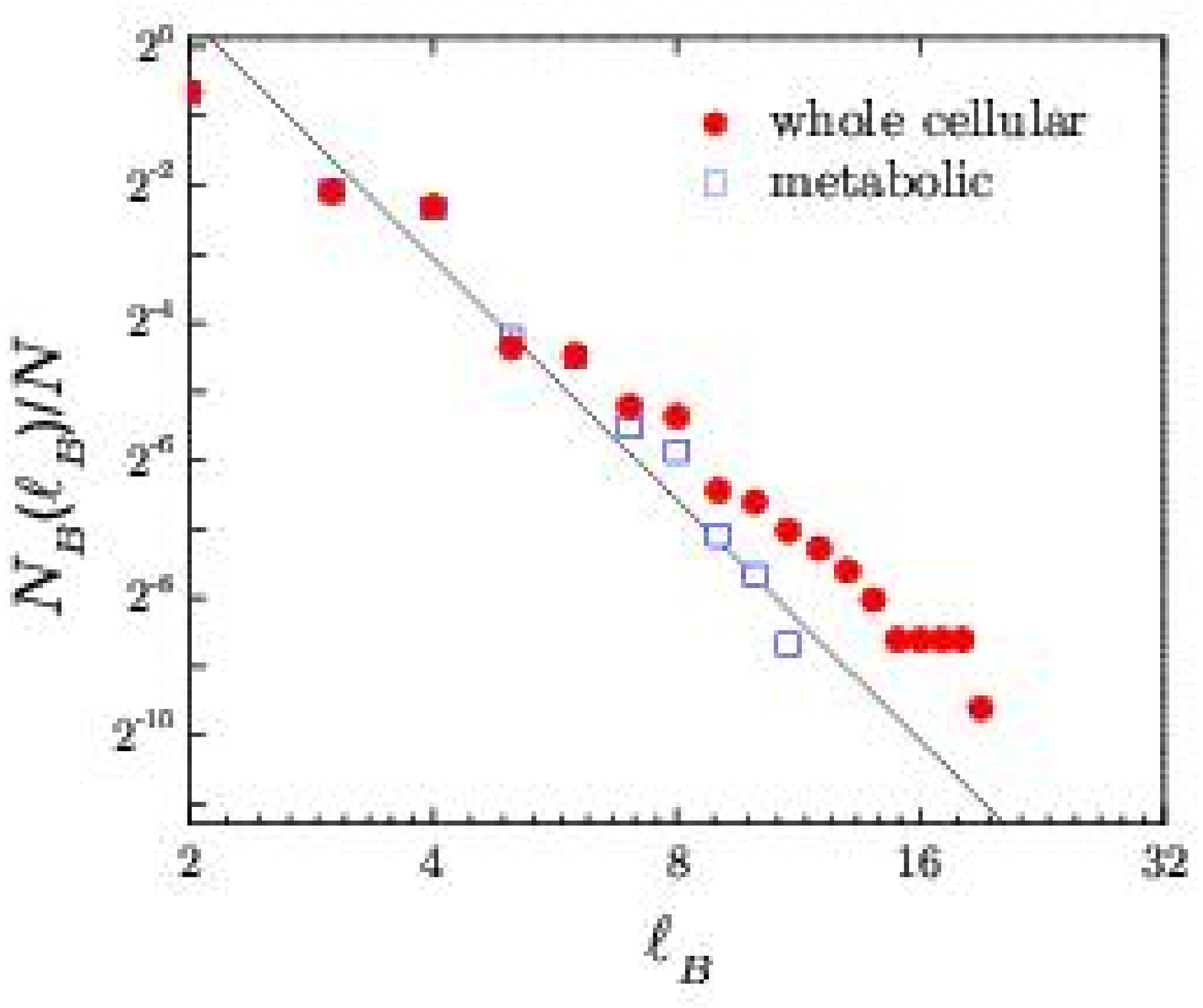}
\end{minipage}%
\begin{minipage}[c]{0.33\textwidth}
\centering Mycobacterium
tuberculosis\\\includegraphics[width=5.9cm]{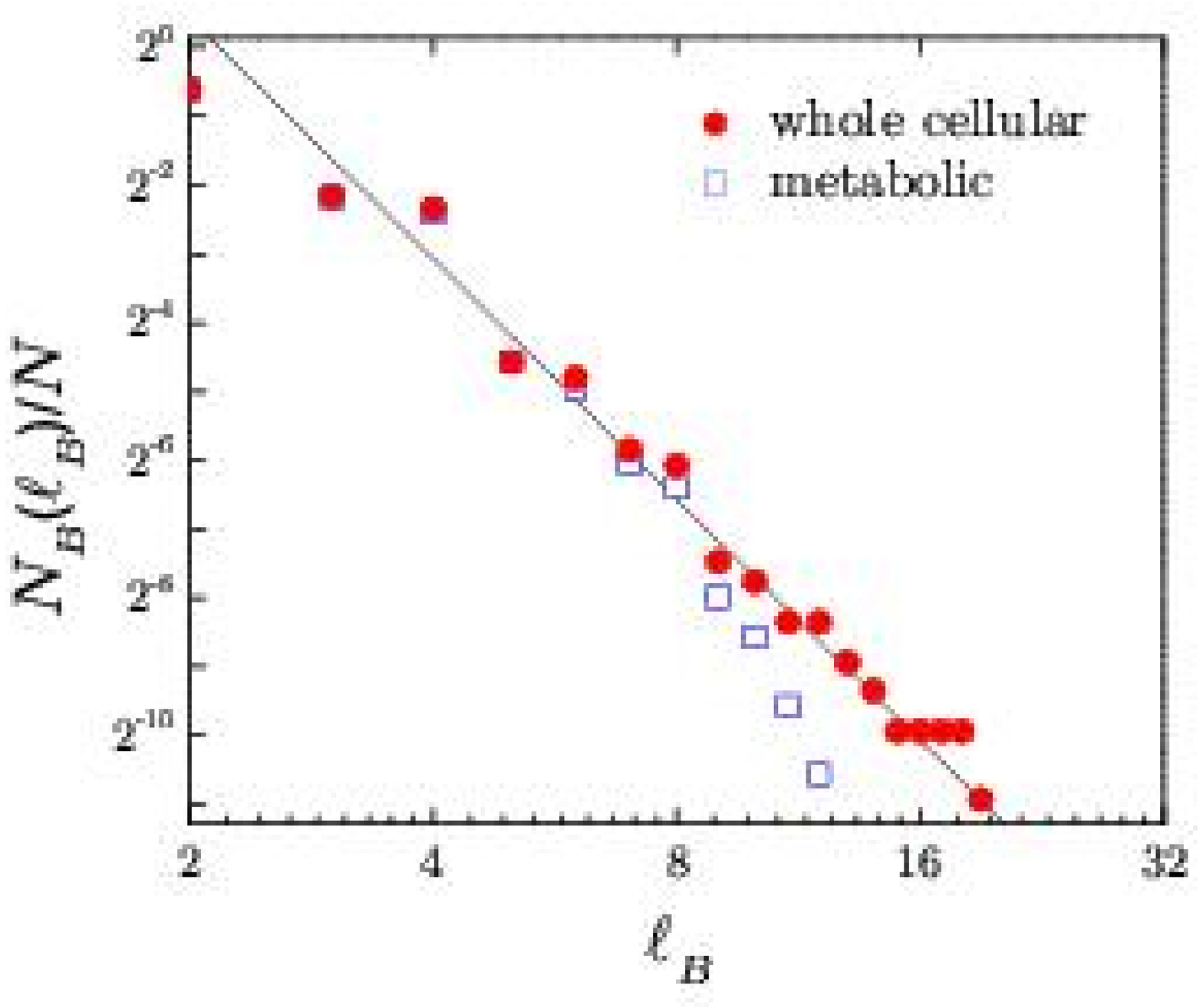}
\end{minipage}%
\begin{minipage}[c]{0.33\textwidth}
\centering Neisseria
gonorrhoeae\\\includegraphics[width=5.9cm]{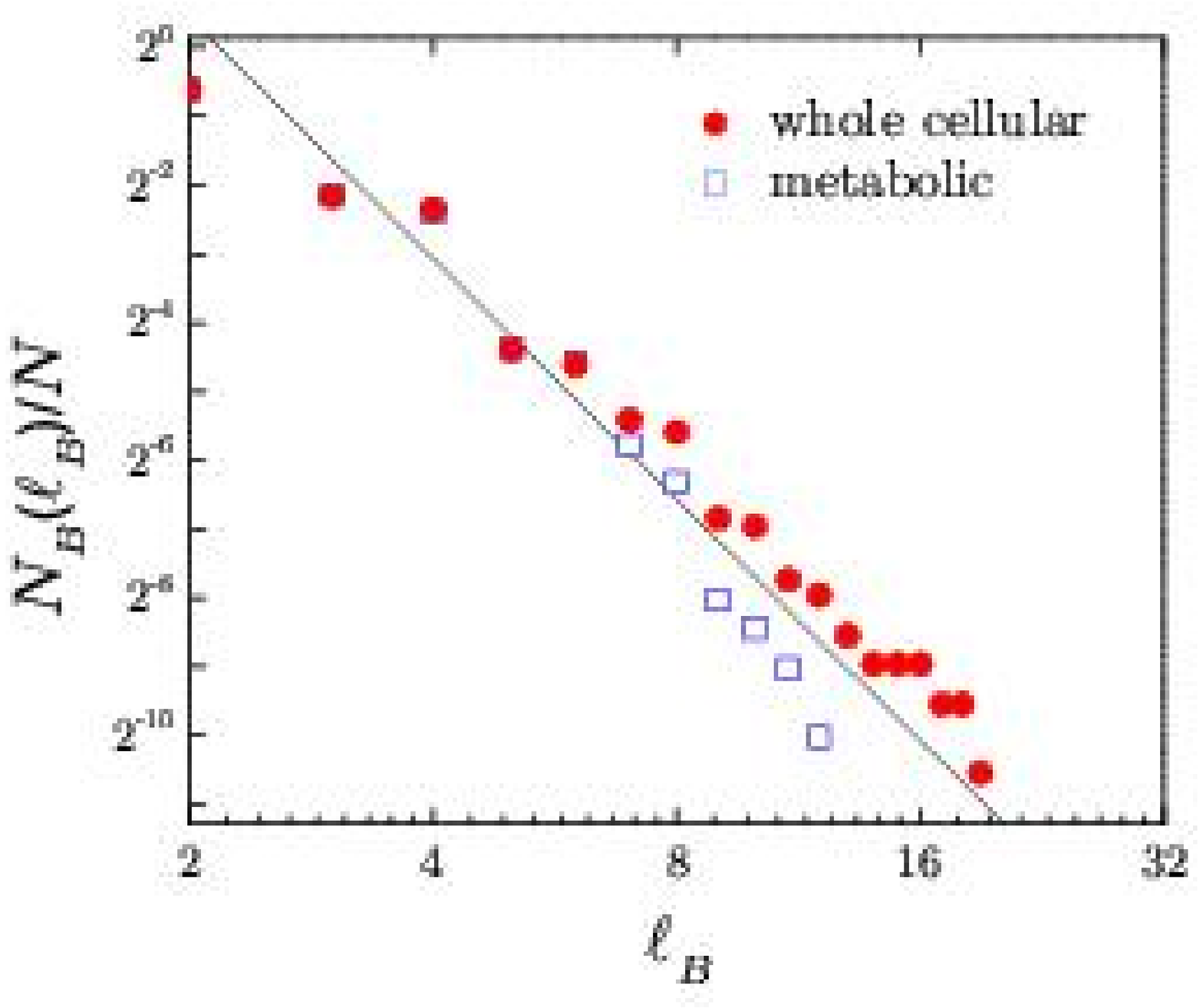}
\end{minipage}%
\end{center}

\begin{center}
\begin{minipage}[c]{0.33\textwidth}
\centering Neisseria
meningitidis\\\includegraphics[width=5.9cm]{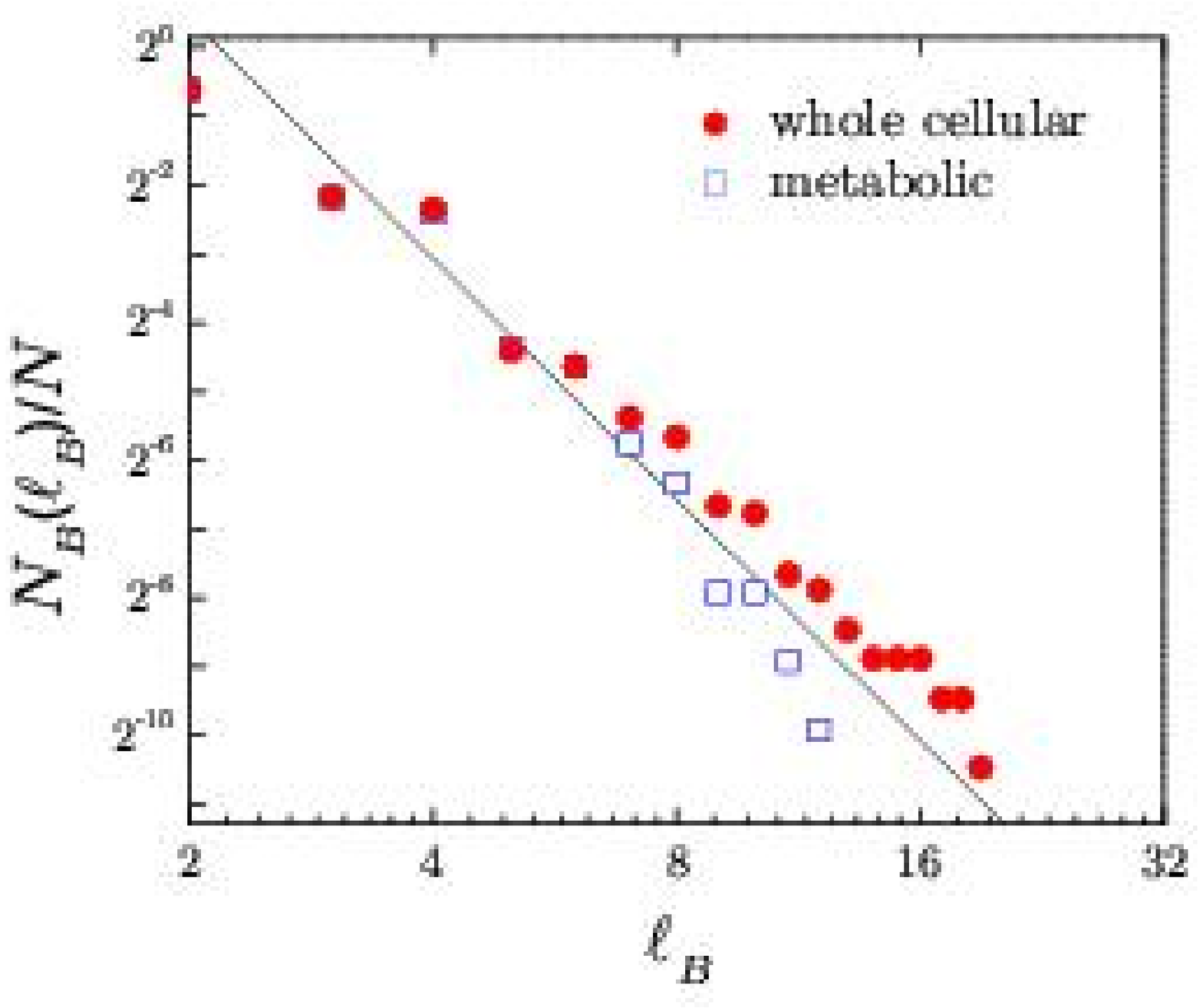}
\end{minipage}%
\begin{minipage}[c]{0.33\textwidth}
\centering Oryza sativa\\\includegraphics[width=5.9cm]{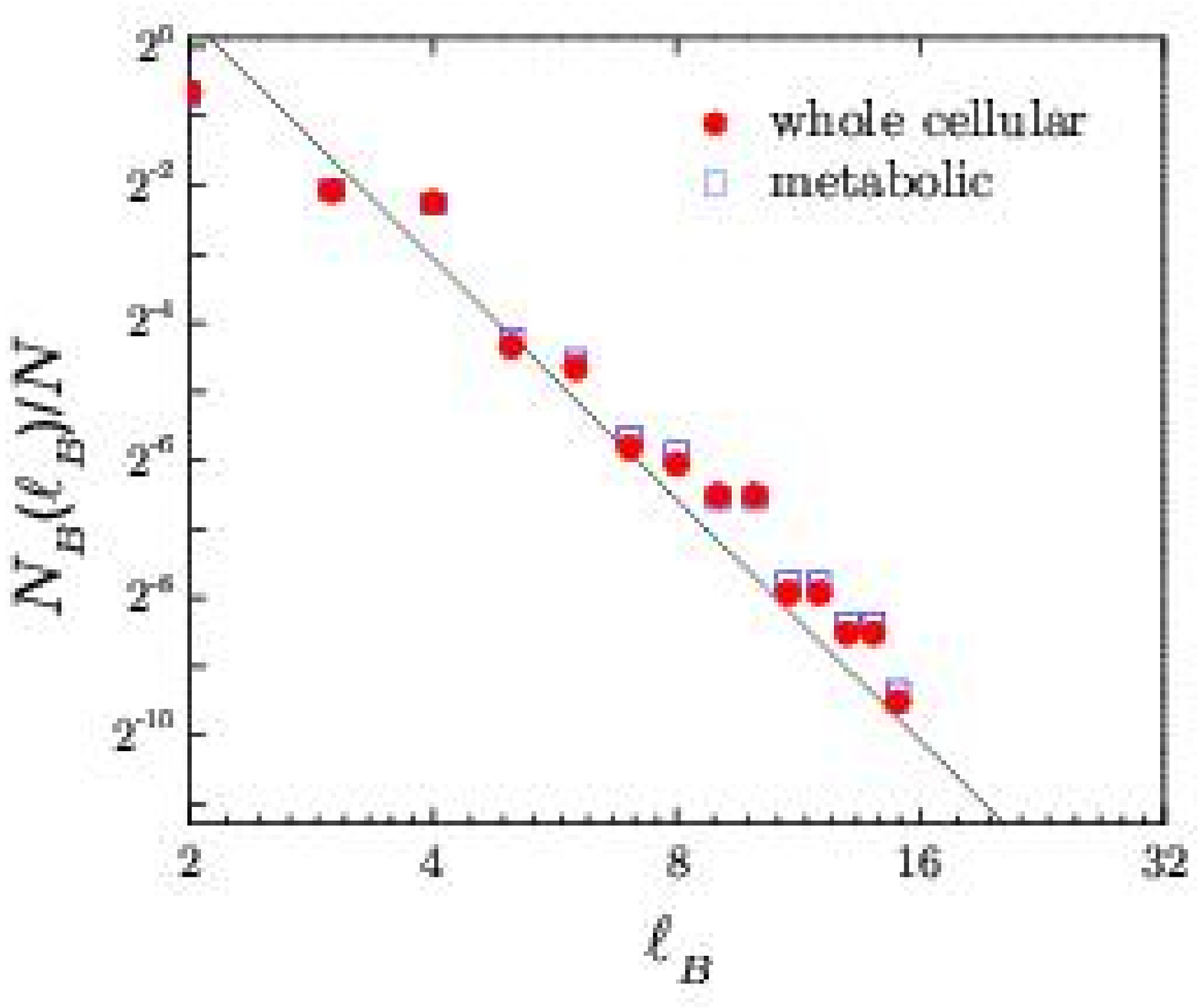}
\end{minipage}%
\begin{minipage}[c]{0.33\textwidth}
\centering Pseudomonas
aeruginosa\\\includegraphics[width=5.9cm]{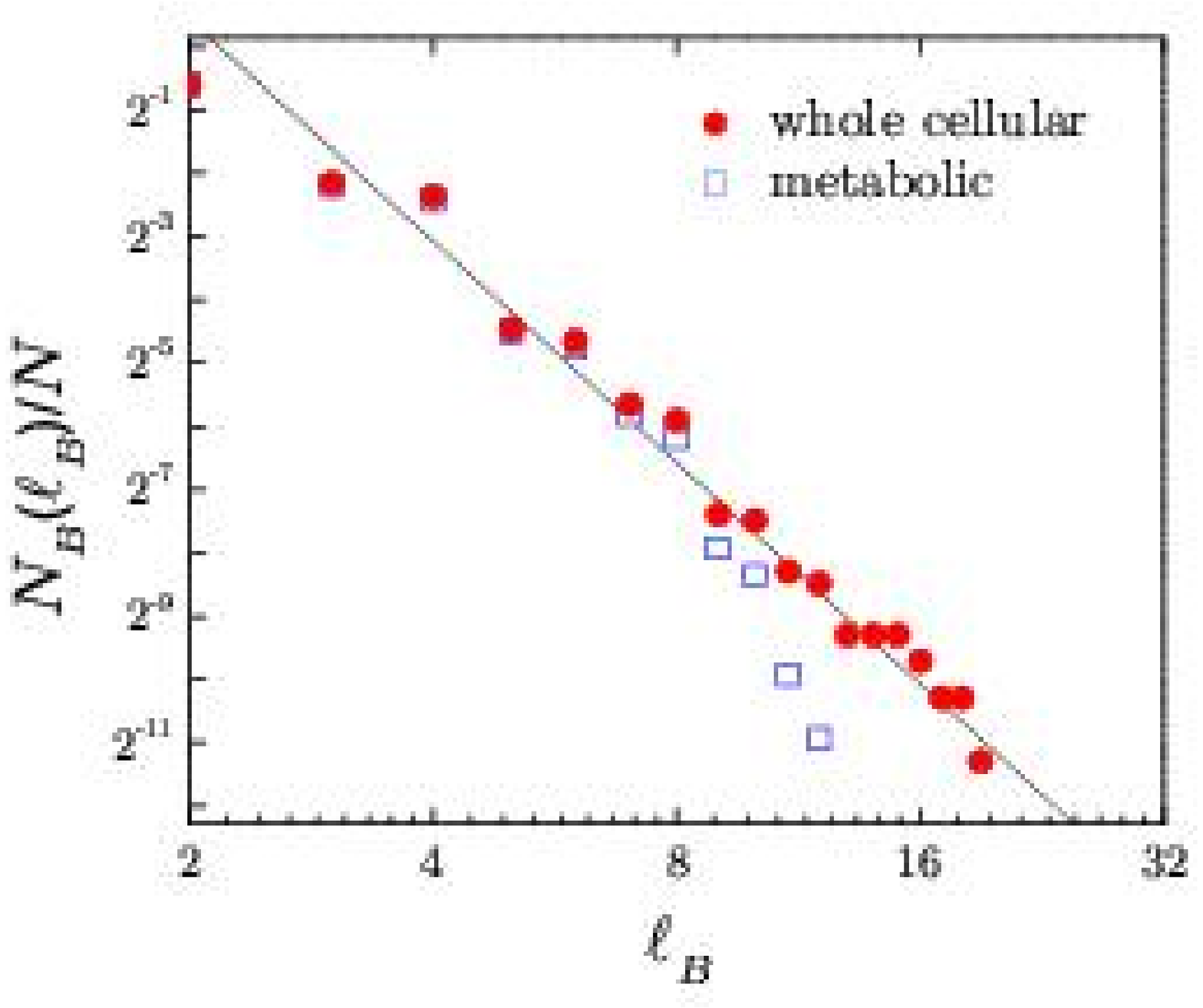}
\end{minipage}%
\end{center}

\begin{center}
\begin{minipage}[c]{0.33\textwidth}
\centering Pyrococcus
furiosus\\\includegraphics[width=5.9cm]{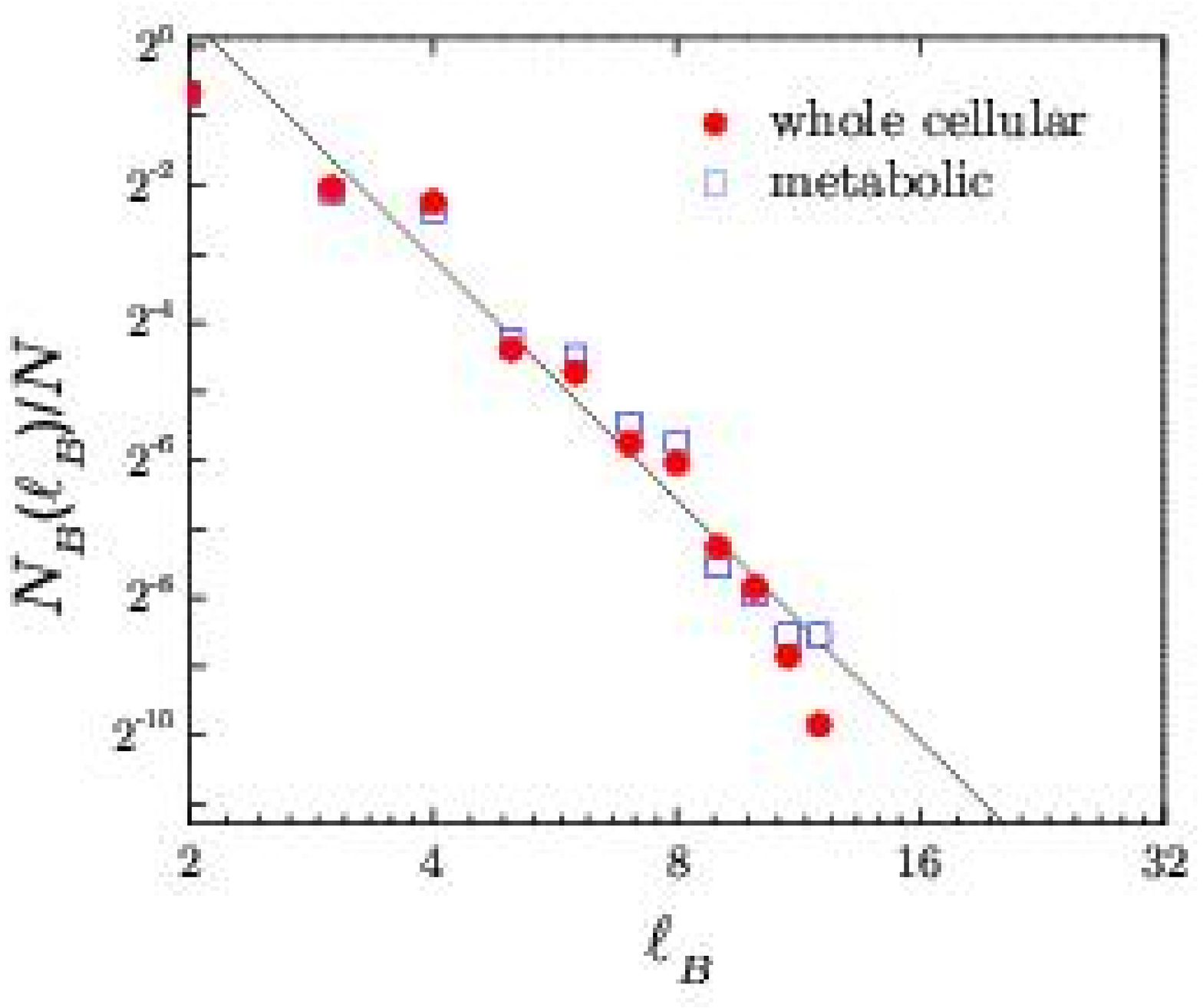}
\end{minipage}%
\begin{minipage}[c]{0.33\textwidth}
\centering Porphyromonas
gingivalis\\\includegraphics[width=5.9cm]{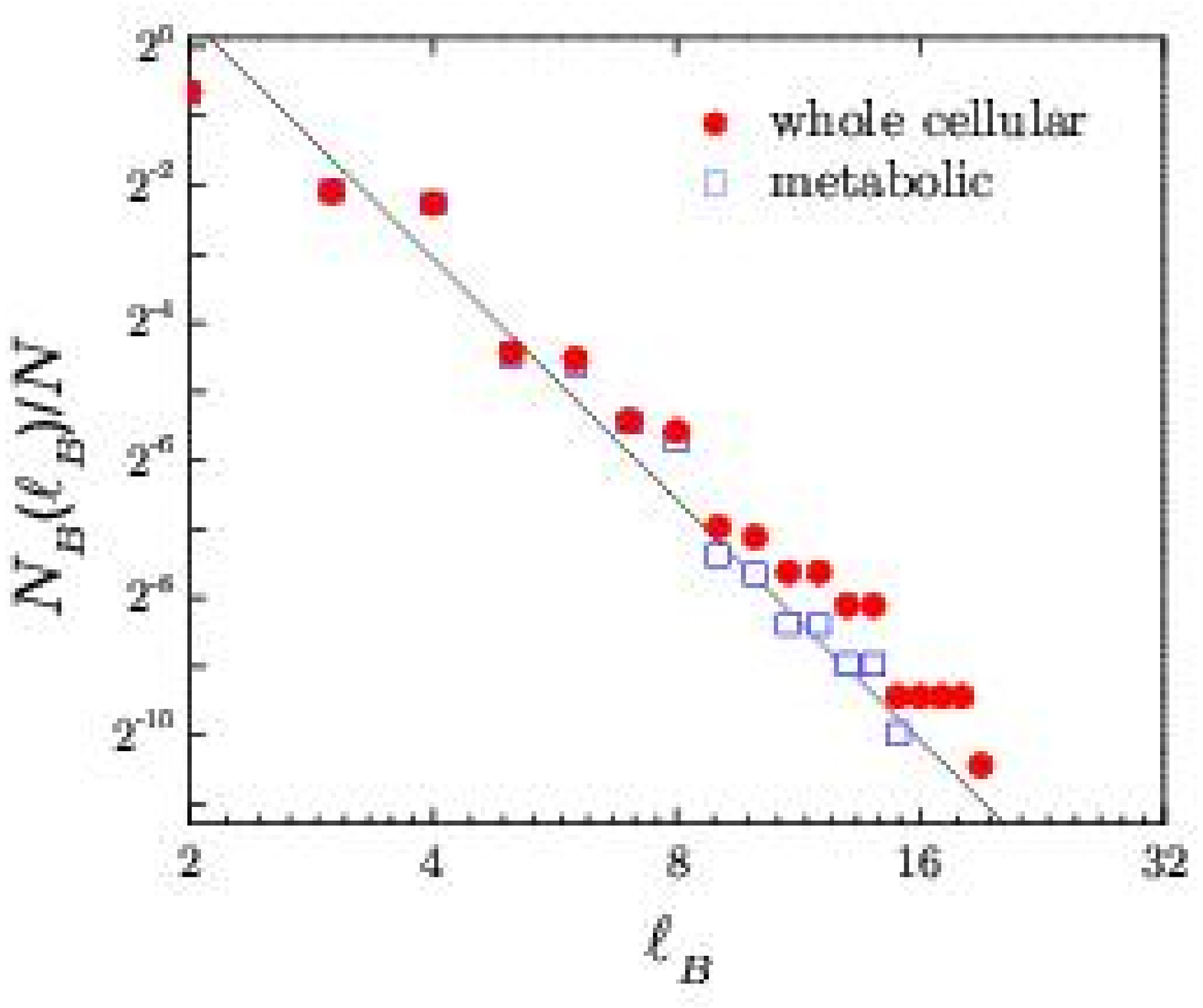}
\end{minipage}%
\begin{minipage}[c]{0.33\textwidth}
\centering Pyrococcus
horikoshii\\\includegraphics[width=5.9cm]{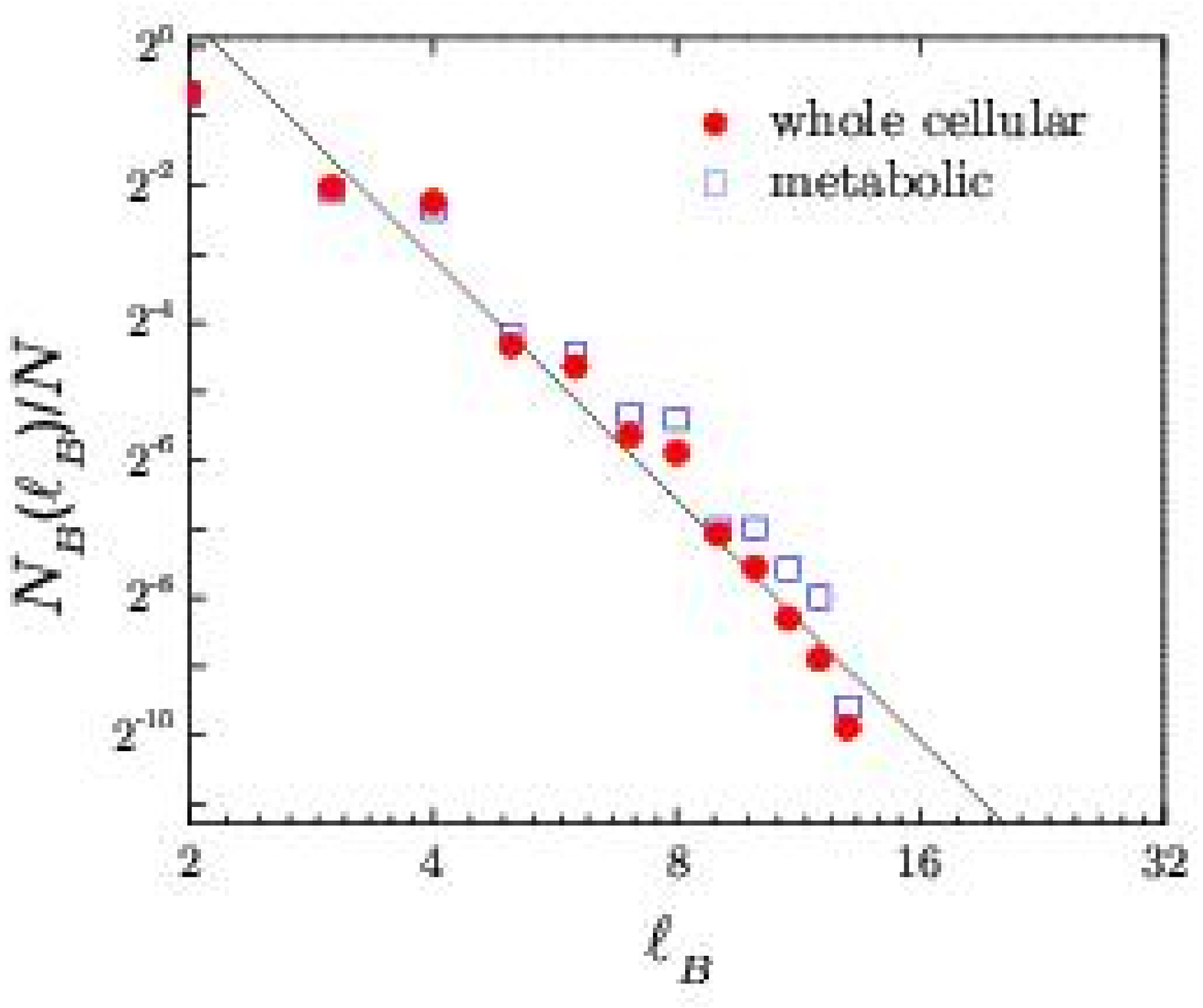}
\end{minipage}%
\end{center}

\begin{center}
\begin{minipage}[c]{0.33\textwidth}
\centering Streptococcus
pneumoniae\\\includegraphics[width=5.9cm]{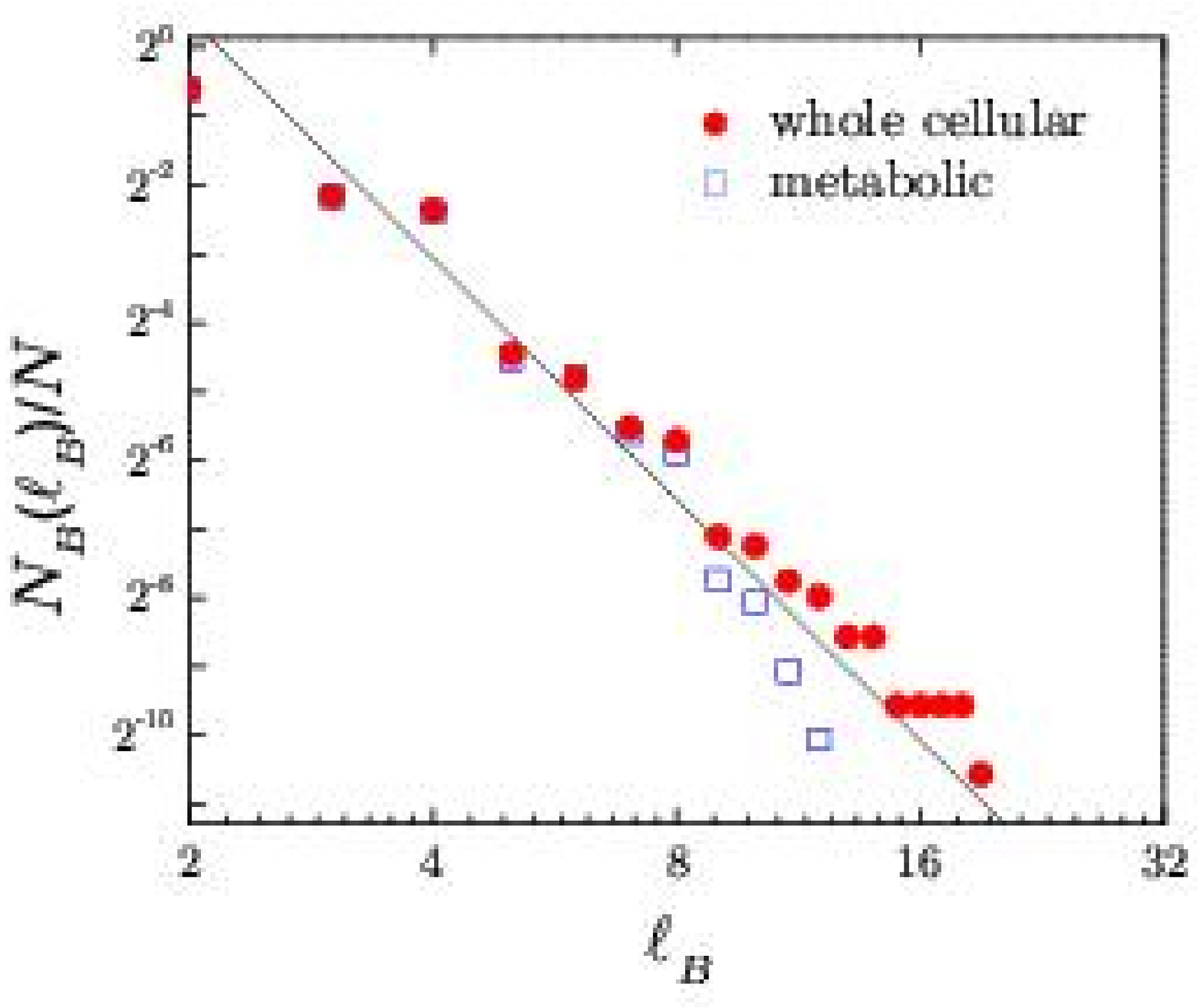}
\end{minipage}%
\begin{minipage}[c]{0.33\textwidth}
\centering Rhodobacter
capsulatus\\\includegraphics[width=5.9cm]{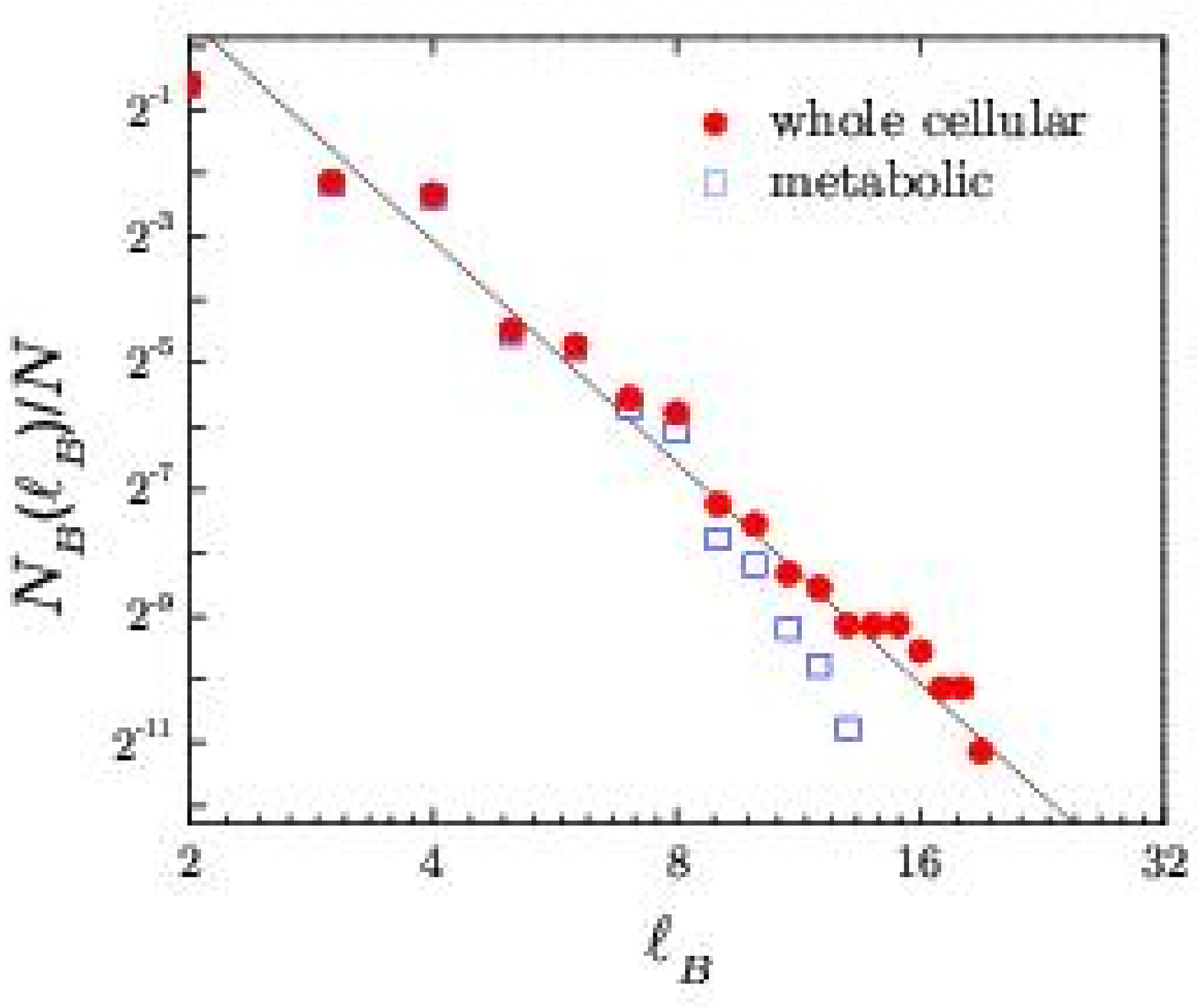}
\end{minipage}%
\begin{minipage}[c]{0.33\textwidth}
\centering Rickettsia
prowazekii\\\includegraphics[width=5.9cm]{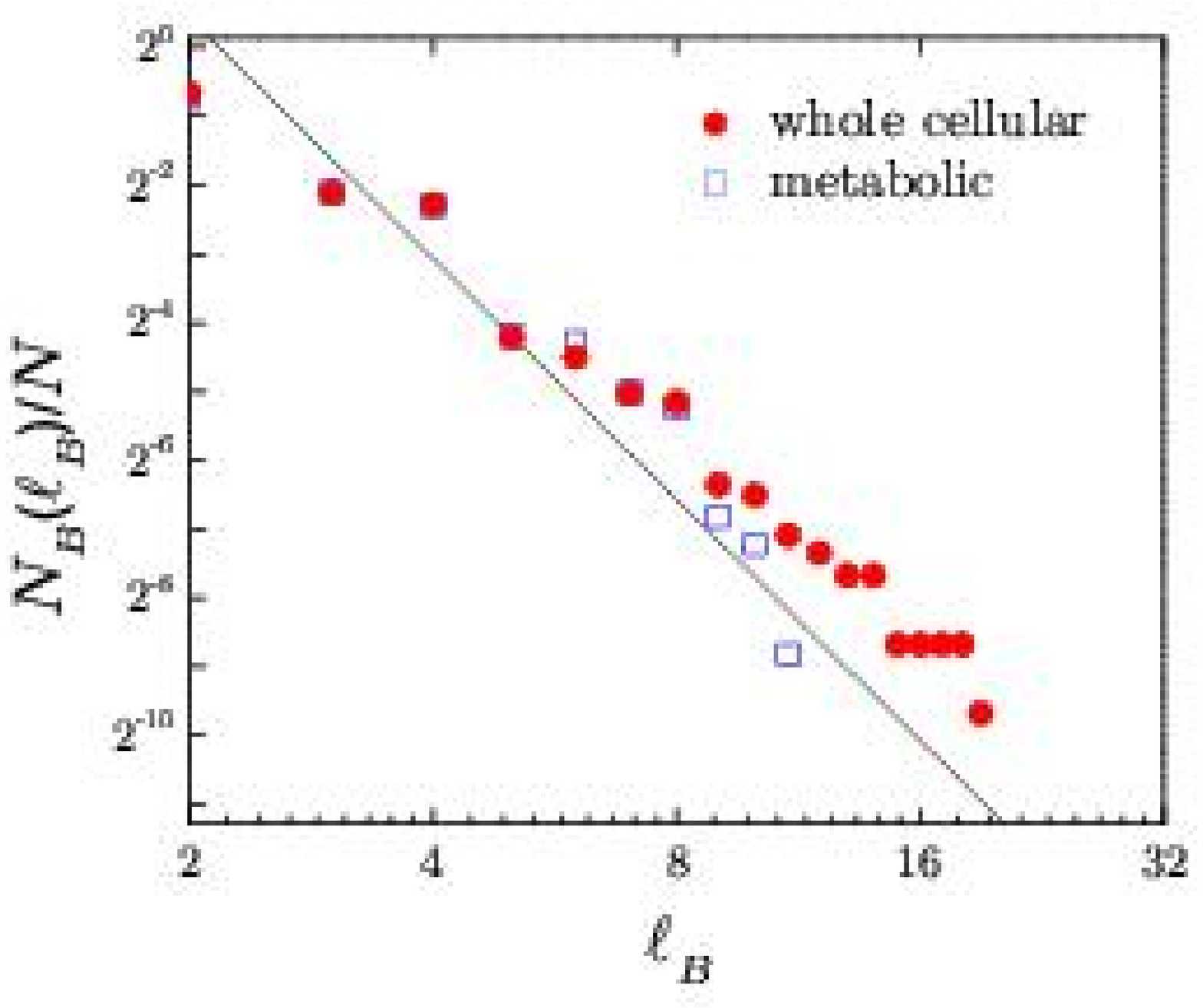}
\end{minipage}%
\end{center}

\begin{center}
\begin{minipage}[c]{0.33\textwidth}
\centering Saccharomyces
cerevisiae\\\includegraphics[width=5.9cm]{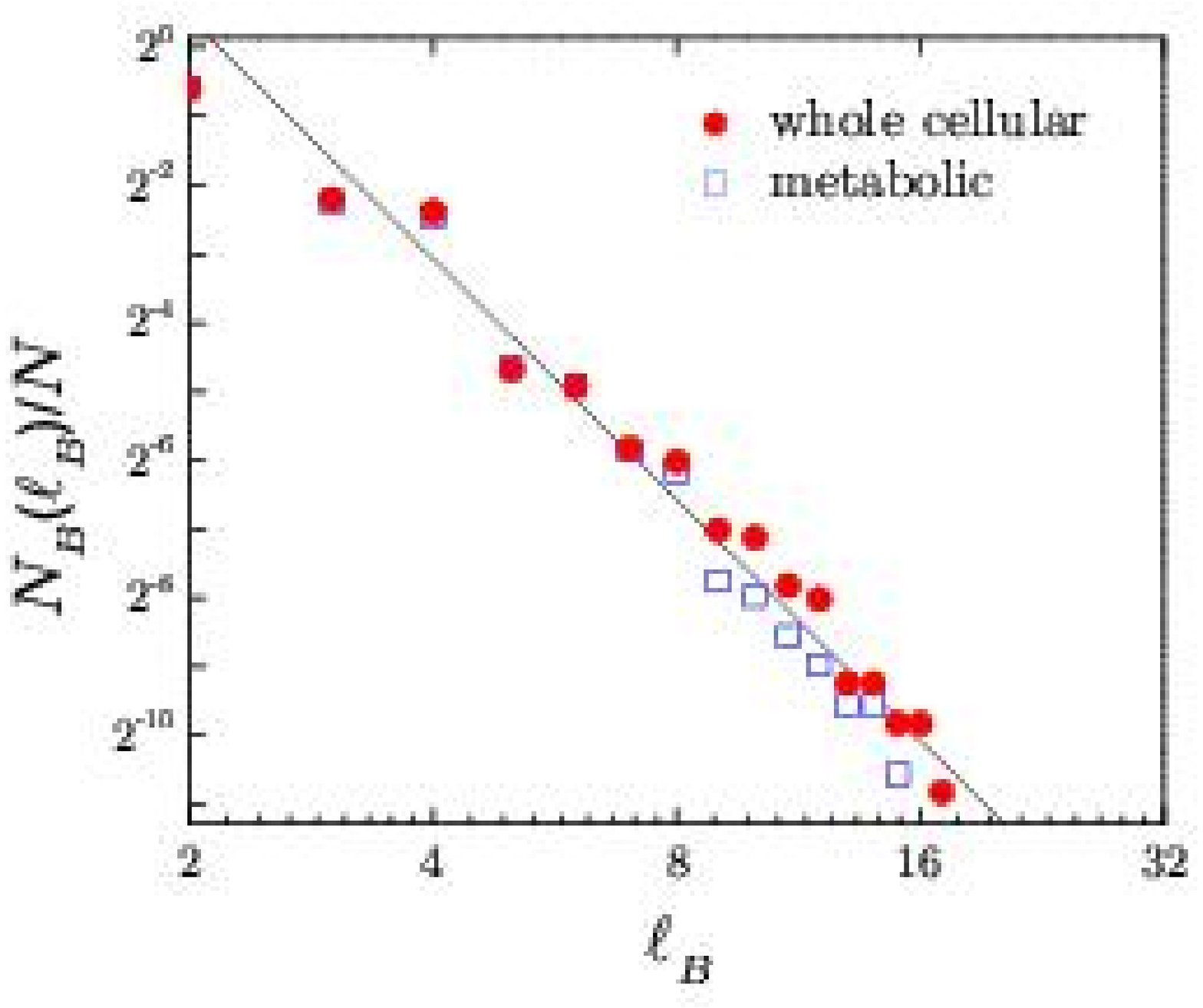}
\end{minipage}%
\begin{minipage}[c]{0.33\textwidth}
\centering Streptococcus
pyogenes\\\includegraphics[width=5.9cm]{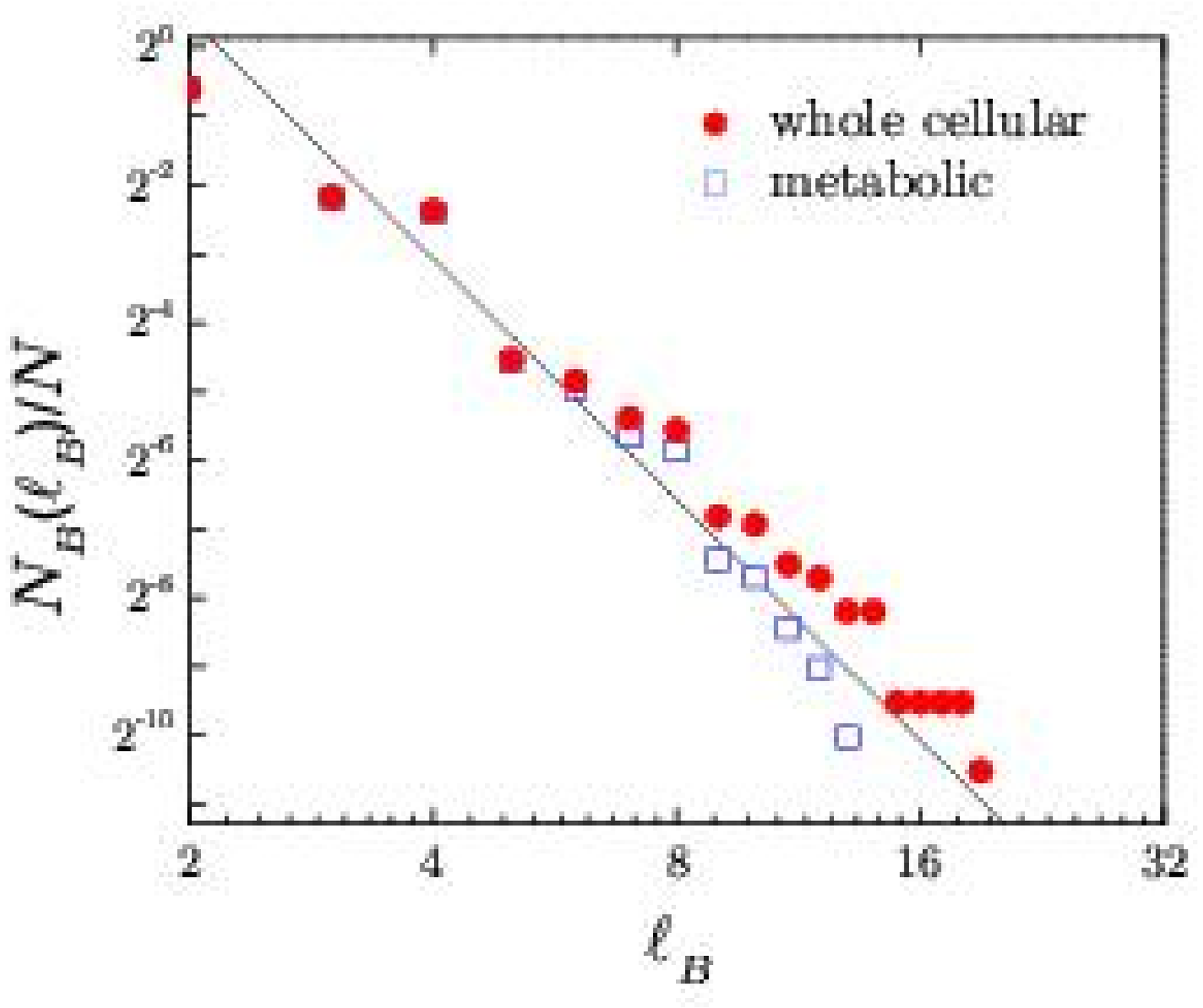}
\end{minipage}%
\begin{minipage}[c]{0.33\textwidth}
\centering Methanobacterium
thermoautotrophicum\\\includegraphics[width=5.9cm]{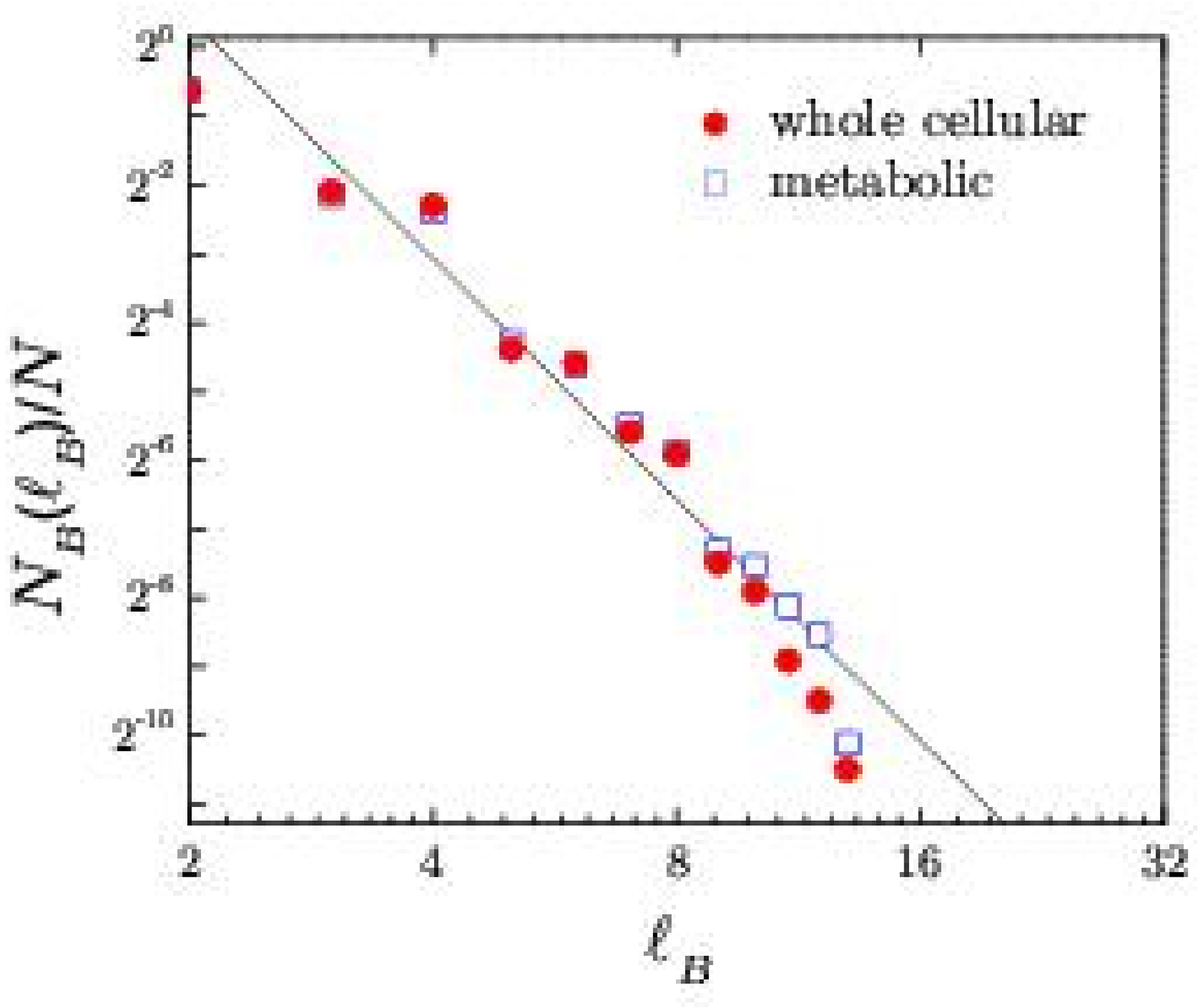}
\end{minipage}%
\end{center}

\begin{center}
\begin{minipage}[c]{0.33\textwidth}
\centering Thermotoga
maritima\\\includegraphics[width=5.9cm]{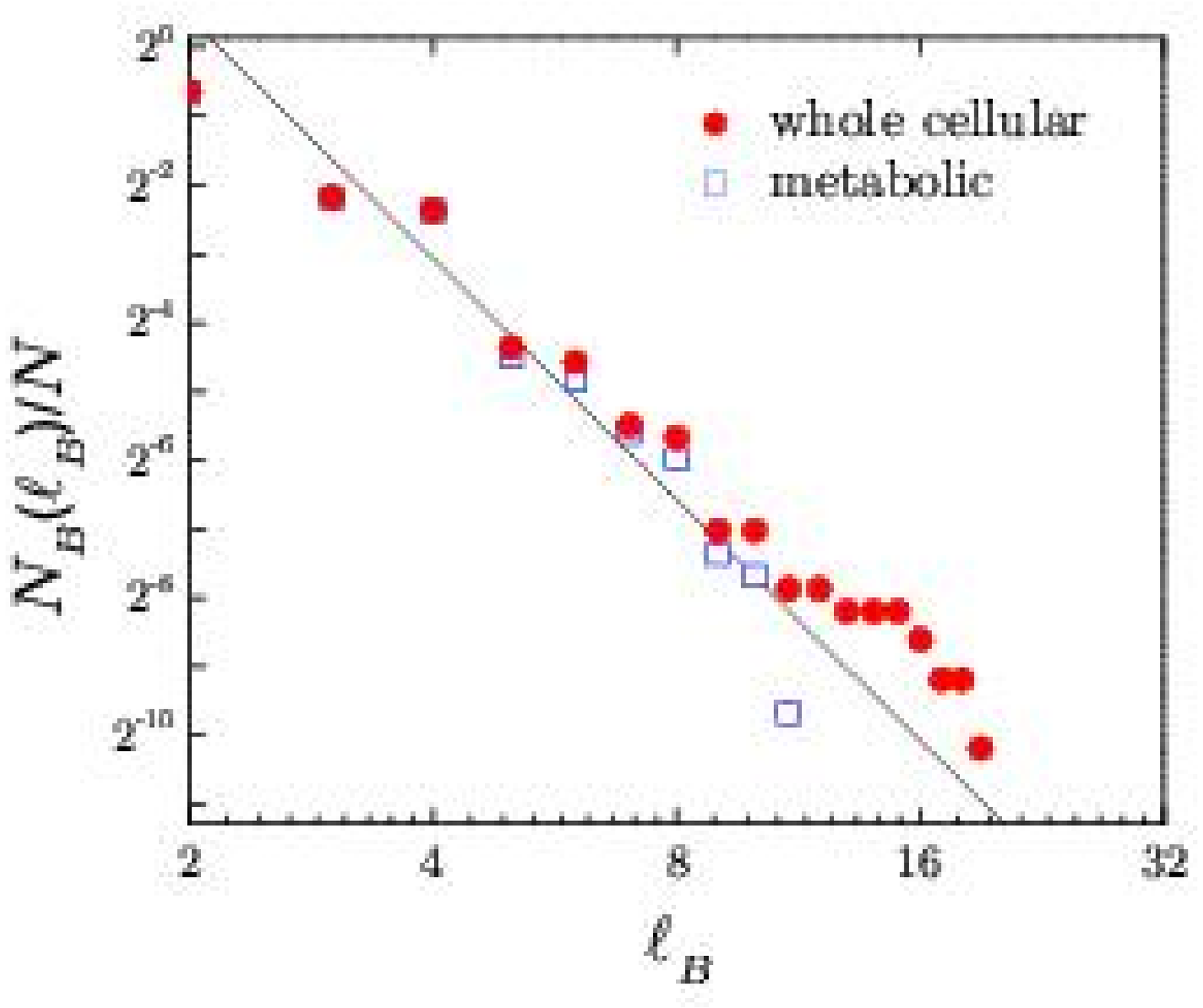}
\end{minipage}%
\begin{minipage}[c]{0.33\textwidth}
\centering Treponema
pallidum\\\includegraphics[width=5.9cm]{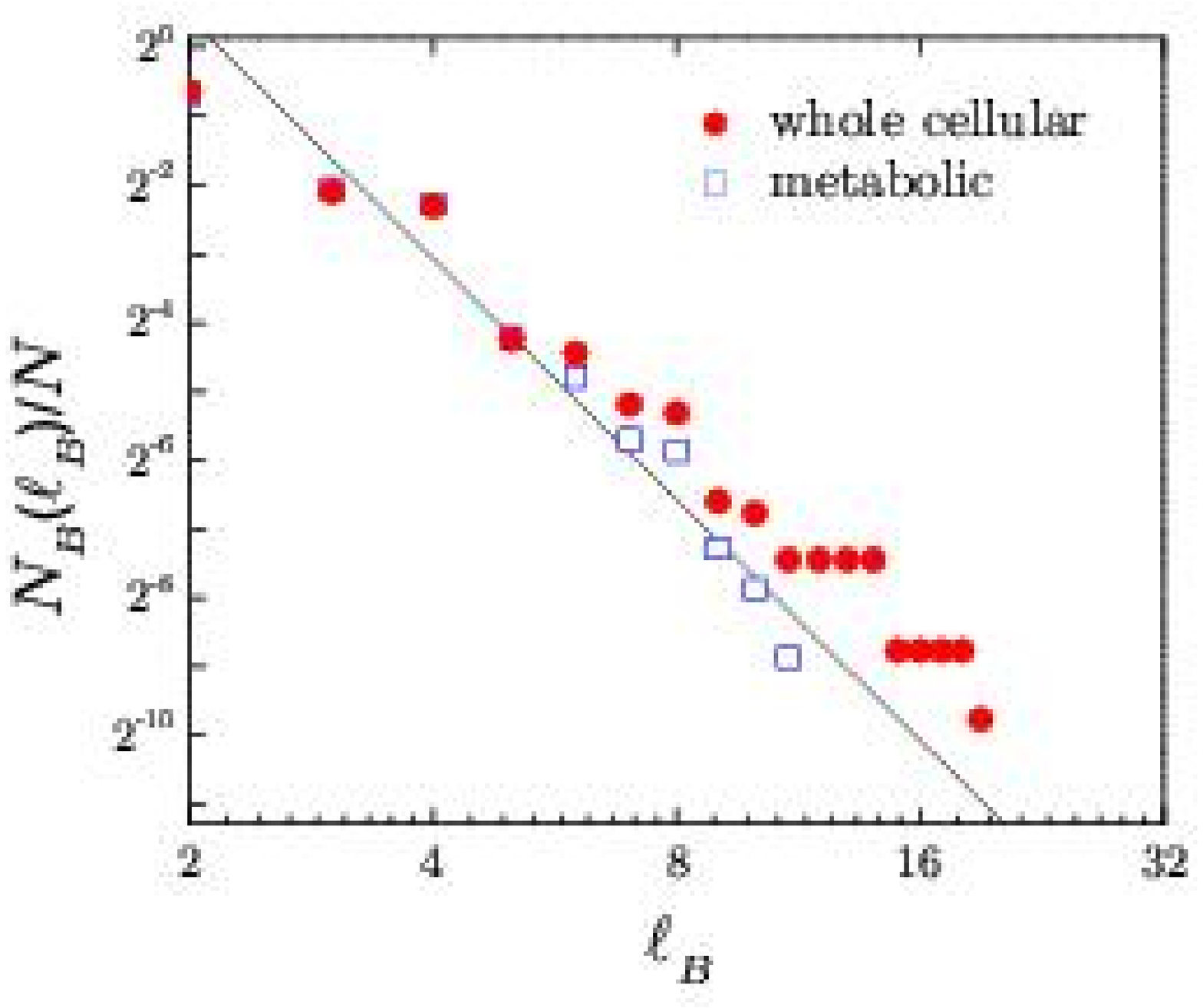}
\end{minipage}%
\begin{minipage}[c]{0.33\textwidth}
\centering Salmonella typhi\\\includegraphics[width=5.9cm]{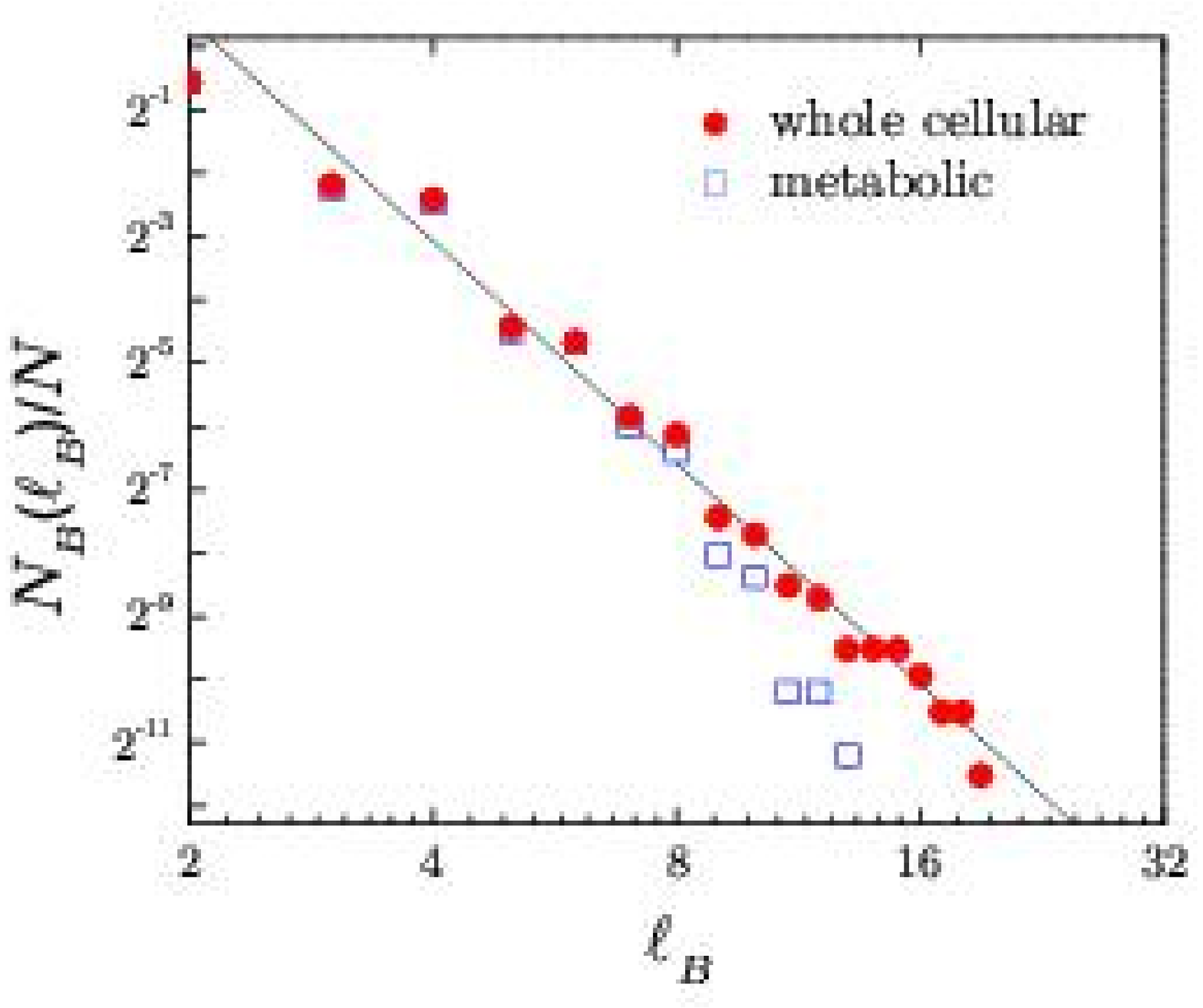}
\end{minipage}%
\end{center}

\begin{center}
\begin{minipage}[c]{0.33\textwidth}
\centering Yersinia pestis\\\includegraphics[width=5.9cm]{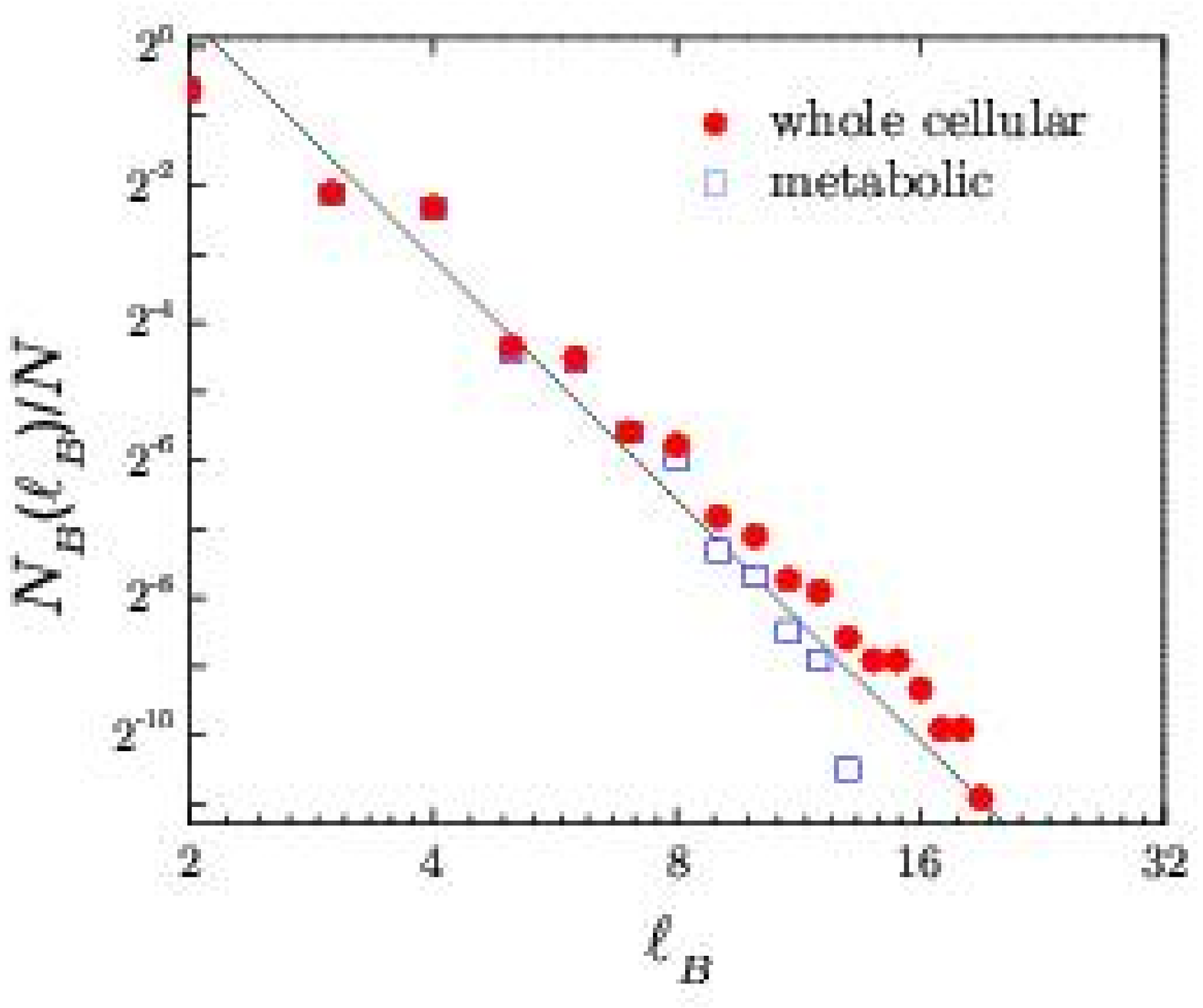}
\end{minipage}%
\end{center}

\end{document}